\documentclass[twocolumn,preprintnumbers,amsmath,amssymb]{revtex4}  

\usepackage{amsfonts}
\usepackage[T1]{fontenc}
\usepackage{amsmath,amsbsy,amssymb,graphicx}
\usepackage{times}
\usepackage{amsmath}
\usepackage{amssymb}
\usepackage{graphicx}
\usepackage{color}
\usepackage{float}
\usepackage[mathscr]{eucal}

\begin{document}

\title{One-dimensional flat bands and Dirac cones in narrow zigzag dice lattice ribbons}

\author{Lei Hao} 
 \address{School of Physics, Southeast University, Nanjing 211189, China}

\date{\today}

\begin{abstract}
We show that four narrow zigzag dice lattice ribbons, which have the minimal widths among their separate categories, constitute a unique collection of systems to study physics related to one-dimensional Dirac cones and flat bands. In zero magnetic field, all three combinations, including only Dirac cones, only flat bands, coexisting Dirac cones and flat bands, are realized in the low-energy band structures of one or two of the four ribbons. In particular, we identify flat bands and Dirac cones corresponding to the edge states of wide ribbons. In a perpendicular magnetic field that gives half a flux quantum per elementary rhombus, two of the four minimal ribbons have fully pinched spectrum, and dynamical evolutions from initially localized wave packets always lead to compact Aharonov-Bohm (AB) cages. The experimental realizations of these narrow zigzag dice lattice ribbons, and the opportunities of exploring novel single-body and many-body physics therein are discussed.
\end{abstract}

\maketitle


\section{Introduction}

Novel band structures mostly lead to unusual electronic properties. Two types of band structures, Dirac (or Weyl) cones on one hand and flat bands on the other, have aroused widespread interest. Dirac cones are solid-state simulators of massless relativistic fermions and lead to peculiar phenomena such as the Klein tunneling and Zitterbewegung motion \cite{katsnelson06,katsnelson06b,castro09,sarma11,peleg07}. The flat bands, with macroscopically degenerate states of infinite effective mass, are genuine strongly correlated systems and prone to many-body phase transitions at infinitesimal interaction strengths \cite{liu14,leykam18,parameswaran13,bergholtz13,derzhko15}.

Besides well-known two-dimensional (2D) and three-dimensional (3D) systems, such as graphene and Dirac or Weyl semimetals with Dirac or Weyl cones \cite{castro09,hosur13,weng16,armitage18}, the quantum Hall states in Landau levels \cite{thouless82,laughlin83}, and the correlated insulating states or unconventional superconducting states in the flat bands of magic-angle twisted bilayer graphene \cite{cao18a,cao18b,bistritzer11}, one-dimensional (1D) systems with Dirac cones or flat bands are also of special interest. Many interesting models of fundamental importance allow exact analytical solutions in 1D \cite{giamarchibook}. In addition, there are powerful numerical techniques that are highly effective for tackling many-body problems in general 1D systems \cite{giamarchibook}. These make the research into the 1D systems with Dirac cones or flat bands a rich and productive field.

Plenty of 1D or quasi-1D (q-1D) systems are known to harbor Dirac cones, flat bands, or both of them. The armchair carbon nanotubes have a pair of 1D Dirac cones in the low-energy band structures \cite{ajiki93,kane97,egger97,kane97m,yoshioka99}. The zigzag graphene nanoribbons are well known to have flat electronic bands on the edge \cite{klein94,fujita96,nakada96,brey06}. Quasi-1D metamaterial lattices, including photonic lattices \cite{mukherjee18,kremer20,zurita20}, optical lattices of cold atoms \cite{kang18,kang20}, and Josephson junction arrays \cite{pop08}, have also been fabricated and are shown to support Dirac cones and (or) flat bands in the band structures.
Because of the fundamental interest, and also inspired by the successful fabrication in metamaterials, 1D or q-1D lattices having Dirac cones or flat bands in their band structures are enthusiastically pursued and studied \cite{mukherjee18,kremer20,zurita20,kang18,kang20,pop08,azaria98,gulacsi07,derzhko10,molina12,molina15,lopez16,huda20,creutz99,vidal00,doucot02,hyrkas13,tovmasyan18,
cartwright18,tilleke20,roy20,orito21,bischoff17,zhang19prb,tada19,perrin20,gligoric20,he21,hung21}.

Among the 1D and q-1D systems considered, an interesting group may be regarded as the narrow ribbons of certain 2D lattices having 2D flat bands and (or) Dirac cones in the band structures. Famous examples of such 2D lattices with both flat bands and Dirac cones include the kagome lattice \cite{mielke91,bergman08}, the Lieb lattice \cite{lieb89,shen10}, and the dice lattice \cite{sutherland86,vidal98,rizzi06,bercioux09}.
These exotic 2D lattices have been experimentally realized in a variety of systems \cite{liu14,leykam18}, ranging from solid-states films and layered materials \cite{oritz19,mihalyuk22}, to various artificial networks and lattices \cite{abilio99,naud01,jo12,leung20,vicencio15,mukherjee15,xia18,mihalyuk21}, and to atomistic chemical 2D systems made of molecular frameworks \cite{springer20,jing20,jiang21}. Compared to the kagome lattice and the Lieb lattice that have a single Dirac cone in the band structures, the dice lattice is unique since its band structures contain two degenerate Dirac cones that constitute a two-component valley degree of freedom, which allows for applications in valleytronics \cite{tan21}.
By reducing the widths of the ribbons of these 2D lattices to the minimum, the influence of uninteresting subbands becomes minimal, which distinguishes the narrow ribbons from the wide ribbons. For both the kagome lattice \cite{azaria98,derzhko10,molina12,molina15,lopez16} and the Lieb lattice \cite{molina15,lopez16,huda20}, a series of narrow ribbons have been studied which all contain 1D flat bands in the band structures. The diamond chain lattice \cite{vidal00}, which has a flat band and a Dirac cone in the band structures, may be considered as the narrowest zigzag ribbon of the dice lattice. What other narrow ribbons of the dice lattice may be interesting as regards the 1D Dirac cones and flat bands are presently unknown. On the other hand, since the wide ribbons of many novel 2D lattices harbor edge states \cite{klein94,fujita96,nakada96,brey06,xu17,oriekhov18,bugaiko19,chen19,alam19,tan21,wang21b,hao22}, an important open question as regards the studies over the narrow ribbons is whether there are interesting edge states, such as Dirac cones or flat bands, that survive the limit of the smallest ribbon widths.

In this work, we try to answer the above couple of questions by studying four narrow zigzag dice lattice ribbons. These four narrow ribbons have the smallest widths among their separate categories and therefore will also be called the minimal zigzag ribbons. We study the spectra of these minimal zigzag ribbons in both zero and nonzero magnetic fields. In zero magnetic field, the four minimal zigzag ribbons have either flat bands, or Dirac cones, or coexisting flat bands and Dirac cones, in the low-energy part of (i.e., the middle of) the band structures. To our knowledge, this is the only series of narrow ribbons of a 2D lattice that have such rich varieties of behaviors as regards the 1D Dirac cones and flat bands. In particular, we identify 1D flat bands and Dirac cones corresponding to the edge states of wide ribbons. The diamond chain lattice, in comparison, do not have such states. As far as we know, these are the first explicit examples of edge states of wide ribbons that survive the extreme reduction of the ribbon width. For the Dirac cone states, of two minimal zigzag ribbons, we provide a picture based on the folding of the 1D Brillouin zone (BZ) and elucidate the connection between the evolution of the Dirac cone states from one minimal ribbon to the other and the lattice structures of the two minimal ribbons. For the flat bands, we construct the compact localized states (CLS) to characterize the states therein. The CLS also form a good starting point for studying the many-body phases of the narrow ribbons \cite{derzhko10,gulacsi07,wu07,derzhko09}. Besides the known CLS for flat bands inherited from the bulk dice lattice, the flat bands corresponding to the edge states of wide ribbons give peculiar triangular CLS.

In a perpendicular magnetic field that gives half a flux quantum to each elementary rhombus of the ribbon lattices, two of the four minimal zigzag ribbons have fully pinched spectra, in which all the bands in the whole band structures are flat bands. The flat bands in the fully pinched spectra separate into three sets, which have energies identical to those for the bulk dice lattice. By studying the dynamical evolutions from localized wave packets, we show that the fully pinched spectra lead to bounded evolutions. The compact pulsating states formed in these evolutions are usually called Aharonov-Bohm (AB) cages \cite{vidal98}. The present finding therefore add new systems where the phenomenon of extreme localization by a magnetic field is realized \cite{vidal98}. Different from the bulk dice lattice and the diamond chain lattice that have two types of AB cages \cite{vidal98,vidal00}, we find additional AB cages associated with initial states on sites of the ribbon edges. By studying the CLS for the flat band states in the fully pinched spectrum, we identify interesting changes brought to the flat band states by the magnetic field, which are otherwise hidden in the featureless dispersions of the flat bands. With all these properties, the four minimal zigzag dice lattice ribbons constitute ideal new platforms for studying physics of 1D flat bands and Dirac cones.

The rest of the paper is organized as follows. In Section II, we define the dice lattice and the four narrow zigzag dice lattice ribbons of minimal widths. We define the models for the electrons in the ribbons (the momentum-space Hamiltonian matrices are defined in the Appendix), in both zero and nonzero magnetic fields. In Sec.III, we study the electronic spectra of the four minimal zigzag ribbons in zero magnetic field. For the 1D Dirac cones and flat bands corresponding to the edge states of wide ribbons, we study their wave functions and their close relations with the geometric structures of the ribbons. We also construct the CLS for the flat bands. In Sec.IV, we study the spectra of the four minimal zigzag ribbons in the presence of a uniform magnetic field, for which each elementary rhombus has half a flux quantum. Two of the four minimal zigzag ribbons are shown to have spectra with flat bands pinched at three isolated energies. We also study the AB cages formed by the dynamical evolutions from localized initial states, and the qualitative changes in the CLS induced by the magnetic field. We then discuss in Sec.V the experimental systems where the minimal zigzag ribbons may be realized and outlooks for future theoretical tasks.

\section{lattices and models}

\begin{figure}[!htb]\label{fig1} \centering
\hspace{-2.95cm} {\textbf{(a)}} \hspace{3.8cm}{\textbf{(b)}}\\
\hspace{0cm}\includegraphics[width=5cm,height=3.47cm]{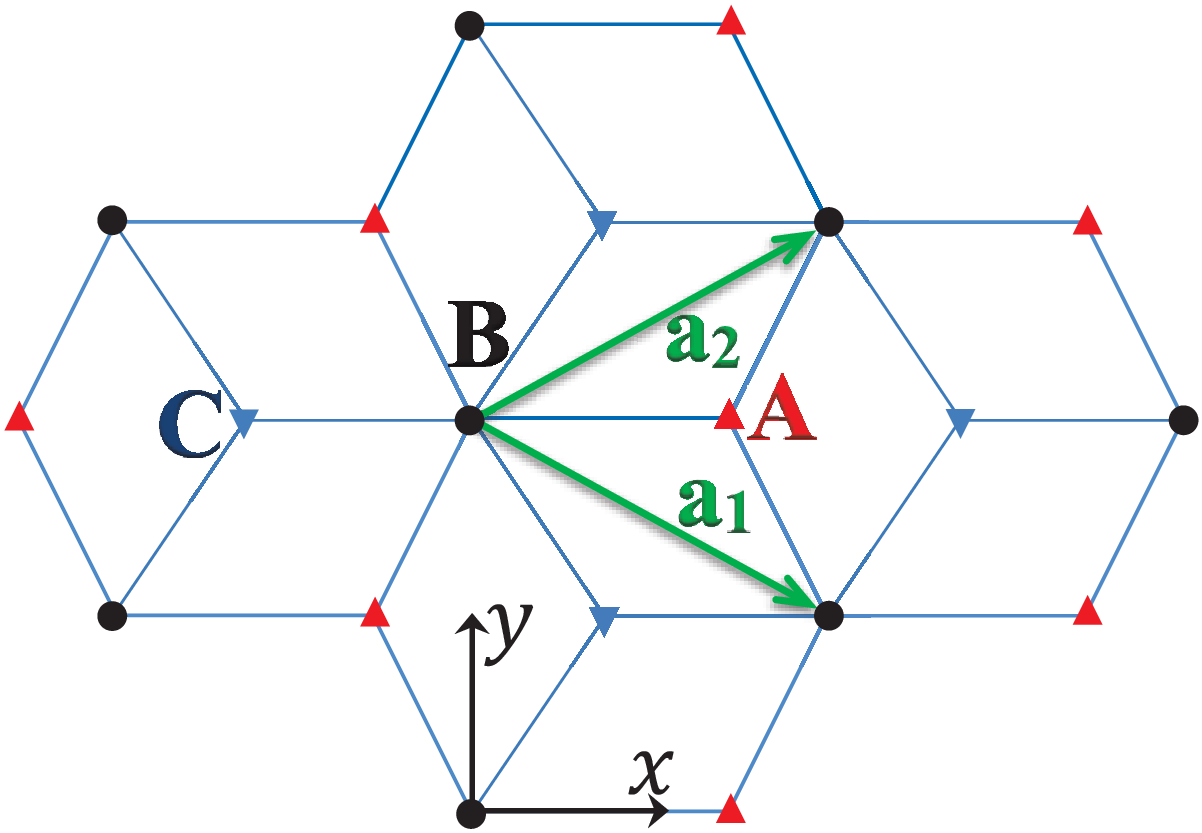}
\includegraphics[width=3.3cm,height=3.34cm]{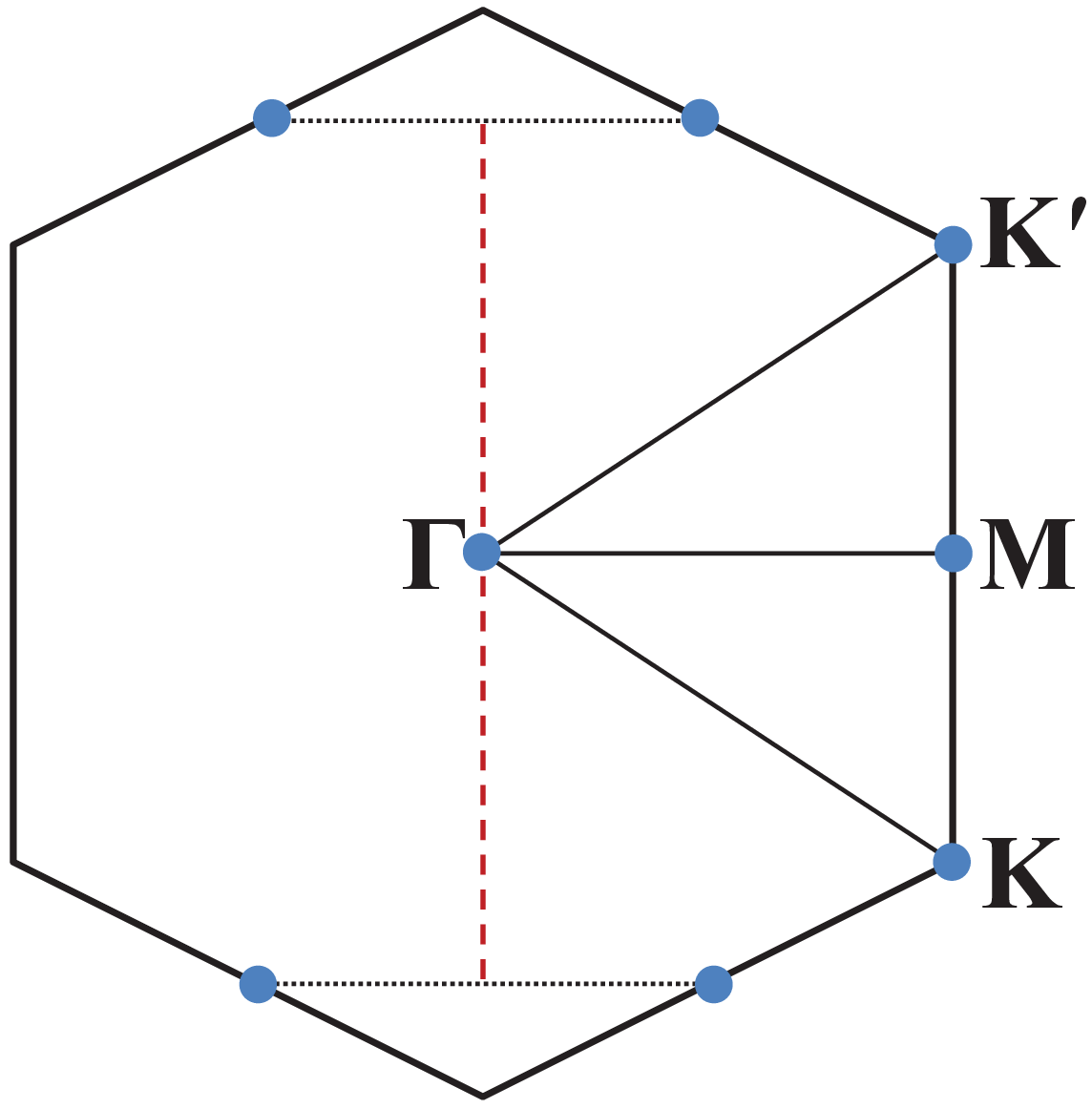}\\
\hspace{-7.2cm} {\textbf{(c)}} \\
\hspace{0cm}\includegraphics[width=8cm,height=5.74cm]{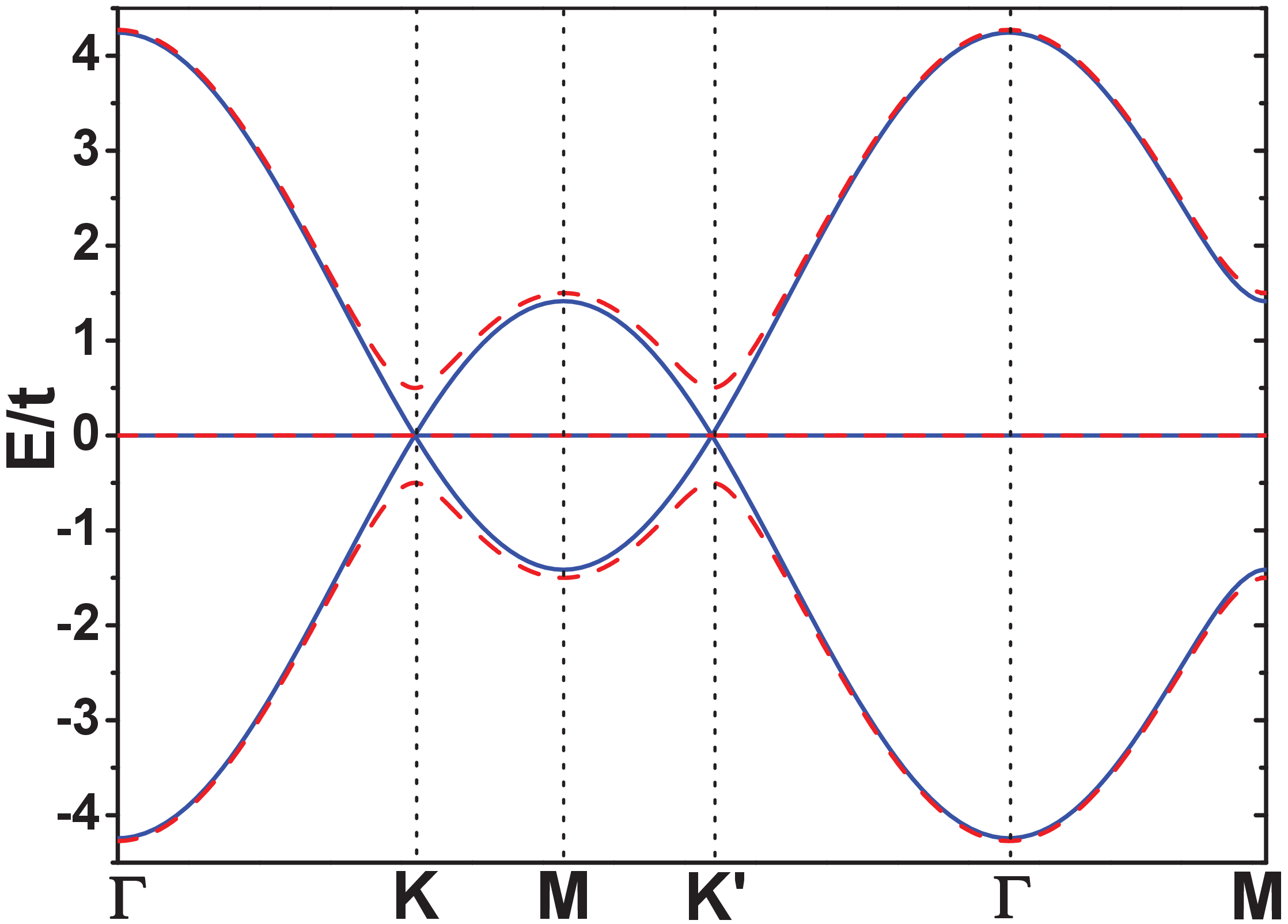} \\
\vspace{-0.10cm}
\caption{(a) The dice lattice. (b) Brillouin zone (BZ) of the dice lattice. The high symmetry points of the BZ are indicated. The 1D BZ of the zigzag ribbons are shown as the dashed vertical line crossing the $\boldsymbol{\Gamma}$ point. (c) The band structures of the bulk dice lattice along several high-symmetry lines of the 2D BZ. $t=1$ is taken as the energy unit. The solid blue curves are for $\Delta=0$. The dashed red curves are for $\Delta=0.5t$. The zero-energy flat bands for the two cases coincide exactly.}
\end{figure}

The dice lattice, as shown in Fig. 1(a), is a 2D lattice with three sublattices A, B, and C \cite{sutherland86,vidal98,rizzi06,bercioux09}. The three sublattices separate into two groups, the single hub sublattice (the B sublattice) versus two rim sublattices (the A and C sublattices). There are bonds only between the hub sublattice sites and their nearest-neighboring (NN) rim sublattices sites. For the low-energy properties of the dice lattice with a single orbital on each site, in zero magnetic field, we consider the following tight-binding model
\begin{equation}
\hat{H}=t\sum\limits_{\langle i,j\rangle}(b^{\dagger}_{i}a_{j}
+b^{\dagger}_{i}c_{j}+\text{H.c.})
+\Delta\sum\limits_{i}(a^{\dagger}_{i}a_{i}
-c^{\dagger}_{i}c_{i}).
\end{equation}
$a_{i}$, $b_{i}$, and $c_{i}$ separately annihilates a spinless electron on the A, B, and C sites of the $i$-th unit cell. H.c. means the Hermitian conjugate of the terms explicitly written out. The first term of the model is the NN hopping term, where the summation $\langle i,j\rangle$ runs over all NN bonds of the dice lattice. For the pure dice lattice, we take the hopping amplitudes between the hub sublattice sites and the two distinct rim sublattice sites the same. The second term of the model is the on-site energy term representing a symmetric bias of the A and C sublattices with respect to the B sublattice \cite{betancur17,xu17,hao21,hao21epj}. The three vectors connecting NN sites of the hub and rim sublattices include $\boldsymbol{\delta}_{1}=(-1,0)a_{0}$, $\boldsymbol{\delta}_{2}=(\frac{1}{2},-\frac{\sqrt{3}}{2})a_{0}$, and $\boldsymbol{\delta}_{3}=(\frac{1}{2},\frac{\sqrt{3}}{2})a_{0}$. The two primitive lattice vectors are $\mathbf{a}_{1}=(\frac{\sqrt{3}}{2},-\frac{1}{2})a$ and $\mathbf{a}_{2}=(\frac{\sqrt{3}}{2},\frac{1}{2})a$, where $a=\sqrt{3}a_{0}$.

As shown in Fig.1(c) are the band structures for the unbiased dice model and a typical biased (for $\Delta=0.5t$) dice model, along several high symmetry lines of the 2D BZ defined in Fig.1(b). The pure dice model is featured by a zero-energy flat band coexisting with a pairs of Dirac cones. With a nonzero symmetric bias (i.e., $\Delta\ne0$), the Dirac cones are gapped out but the zero-energy flat band remains.

Under a proper perpendicular magnetic field, the bulk dice lattice is well known to show fully pinched spectrum, in which all the single-particle states pinch at three isolated energies \cite{vidal98}. In addition, dynamical evolutions from localized initial states give peculiar pulsating states in compact regions of the lattice, which are called (compact) AB cages \cite{vidal98,abilio99,naud01,perrin20,danieli21}. The fully-pinched spectrum and compact AB cages are rare examples of the extreme localization phenomenon induced by the interplay of quantum interference and lattice geometry.

The impact of the perpendicular magnetic field is included through the Pierls' substitution of replacing the hopping amplitude $t_{ij}$ in $t_{ij}c_{i}^{\dagger}c_{j}$ with $t_{ij}e^{i\gamma_{ij}}$, where the phase factor
\begin{equation}
\gamma_{ij}=\frac{2\pi}{\phi_{0}}\int_{i}^{j}\mathbf{A}\cdot d\mathbf{l}.
\end{equation}
$\phi_{0}=hc/e$ is the magnetic flux quantum. The magnitude of the magnetic field is measured by
\begin{equation}
\phi=H\frac{a^{2}}{2\sqrt{3}},
\end{equation}
which is the magnetic flux through an elementary rhombus of the dice lattice or its ribbons. It is also convenient to introduce the reduced flux $f=\phi/\phi_{0}$. The fully-pinched spectrum and compact AB cages of the bulk dice lattice are obtained for $f=\frac{1}{2}$ and $\Delta=0$ \cite{vidal98}.

The four narrow zigzag dice lattice ribbons that we will study are defined in Figs.2(a)-2(d). For comparison, we also show the diamond chain lattice in Fig.2(e).
These narrow ribbons are obtained by cutting the bulk dice lattice along two straight lines parallel to the $y$ axis and retaining the lattice in between the two cutting lines. They have a pair of edges along the zigzag direction of the dice lattice. In zero magnetic field, the ribbons are translation invariant along $y$, with the unit cell of each ribbon encircled by a dashed rectangular box. $k_{y}\in(-\pi/a,\pi/a]$, which is shown in Fig. 1(b) as the vertical dashed line crossing the $\boldsymbol{\Gamma}$ point, defines the 1D BZ of all the five narrow zigzag dice lattice ribbons.

\begin{figure}[H]\label{fig2}
\centering
\hspace{-6.5cm} {\textbf{(a)}}\\
\includegraphics[width=7.0cm,height=2.435cm]{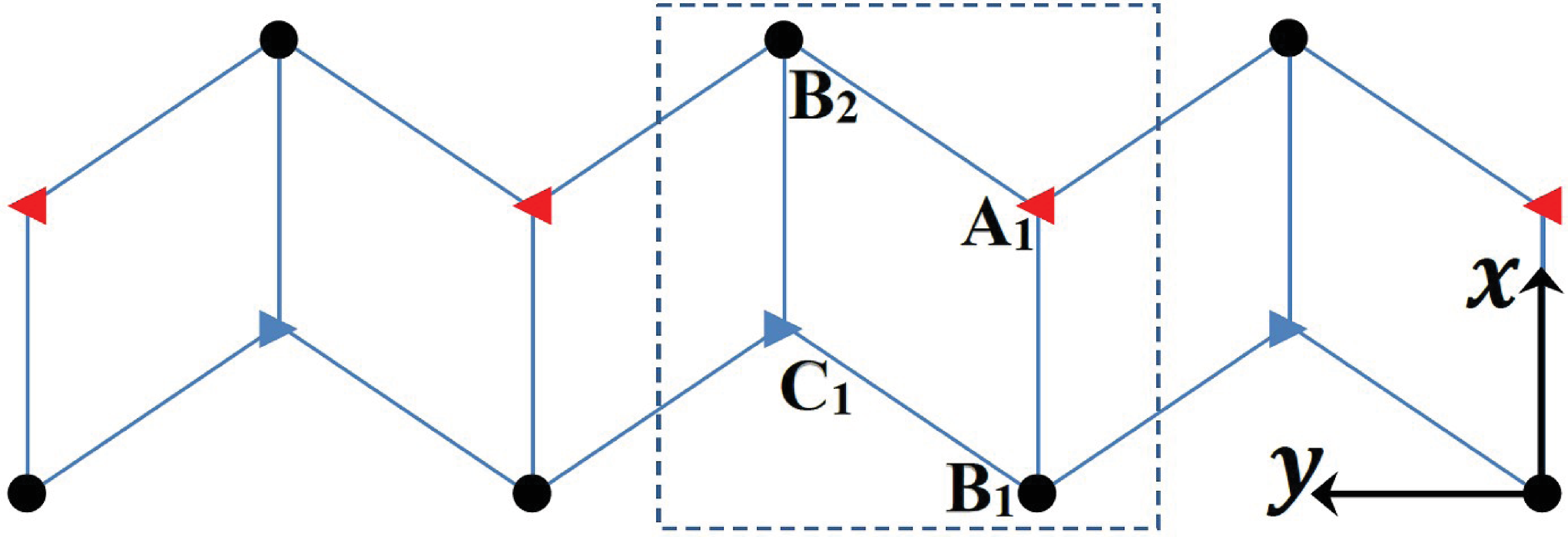} \\ \vspace{-0.05cm}
\hspace{-6.5cm} {\textbf{(b)}}\\
\includegraphics[width=7.0cm,height=4.492cm]{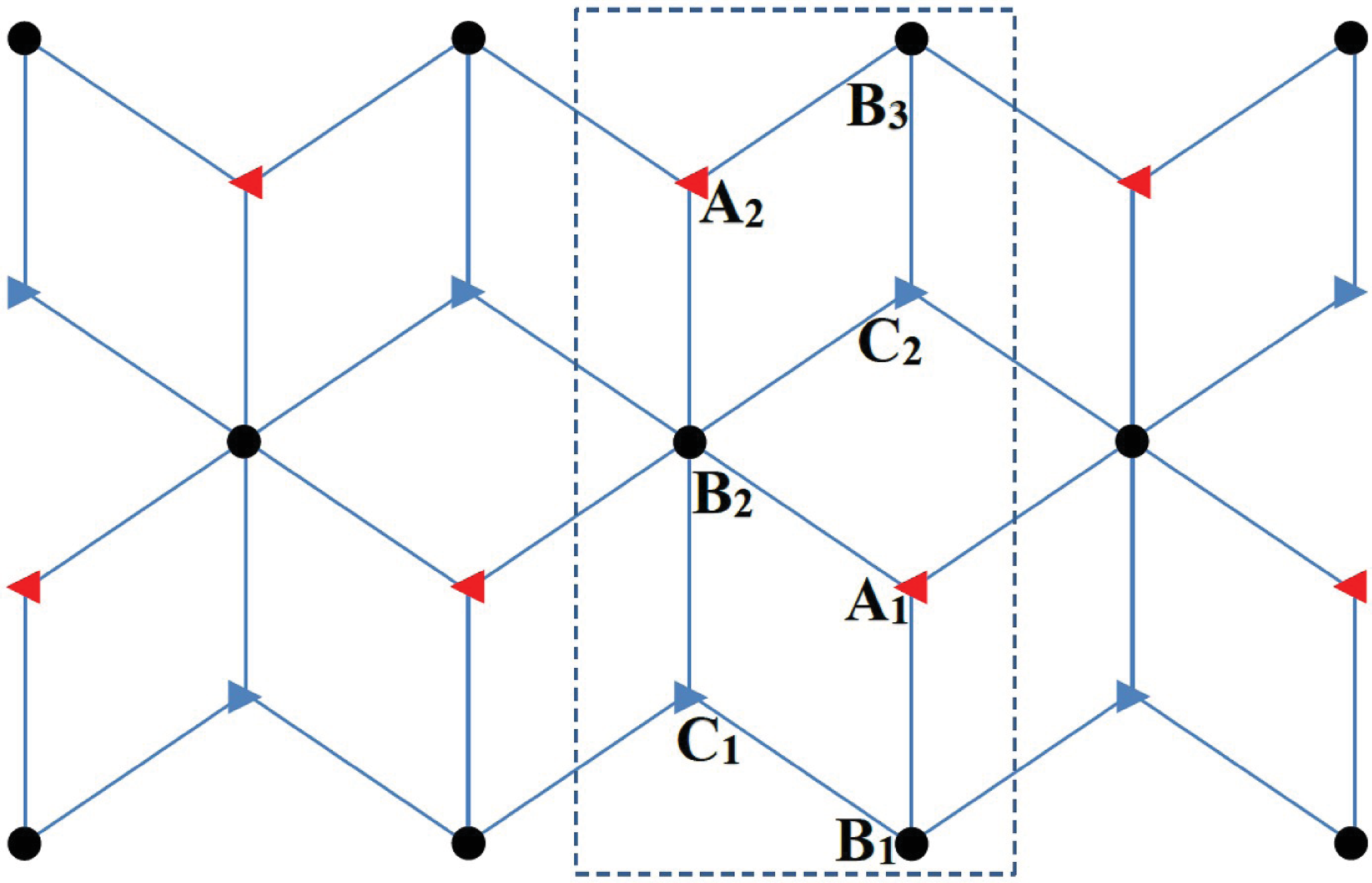}  \\ \vspace{-0.05cm}
\hspace{-6.5cm} {\textbf{(c)}}\\
\includegraphics[width=7.0cm,height=3.075cm]{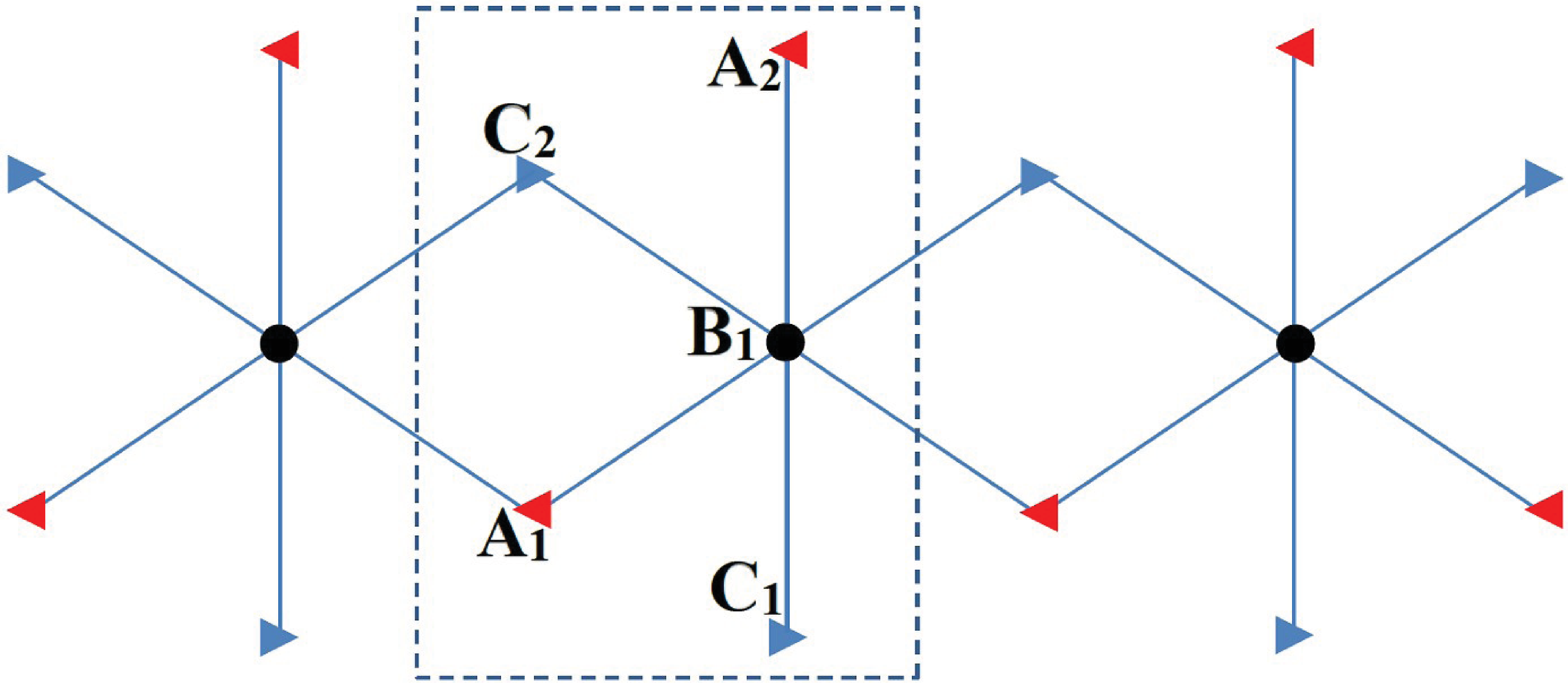} \\ \vspace{-0.05cm}
\hspace{-6.5cm} {\textbf{(d)}}\\
\includegraphics[width=7.0cm,height=5.143cm]{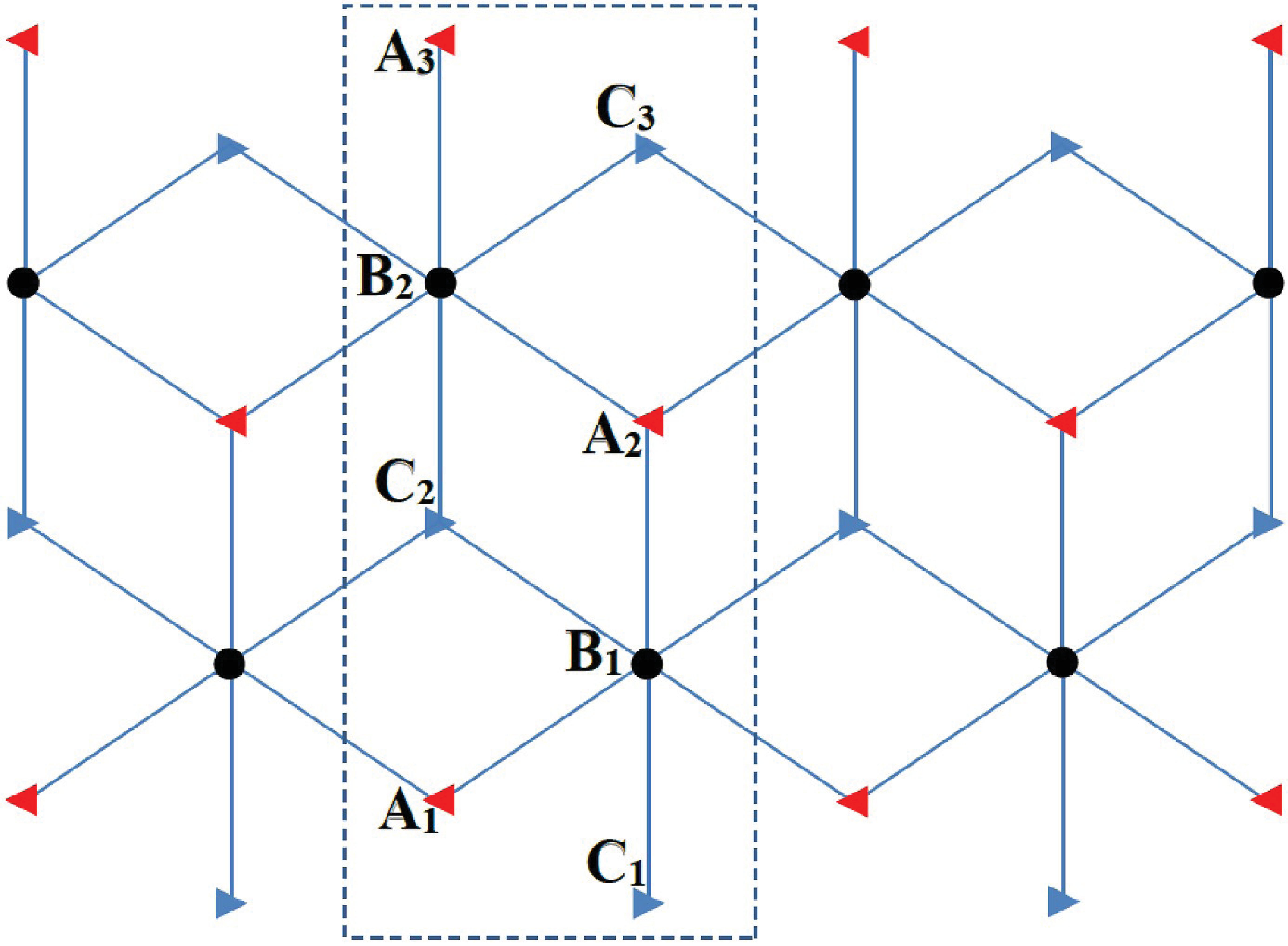} \\ \vspace{-0.05cm}
\hspace{-6.5cm} {\textbf{(e)}}\\
\includegraphics[width=7.0cm,height=1.967cm]{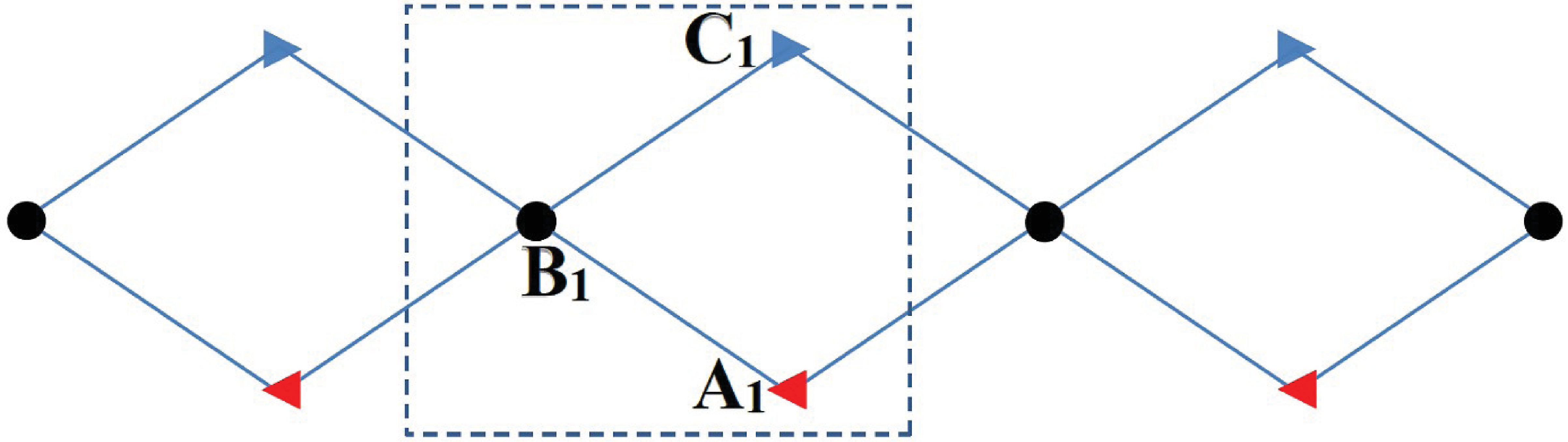} \\
\caption{The lattices of five minimal zigzag dice lattice ribbons. (a) The minimal BB-off ribbon. (b) The minimal BB-in ribbon. (c) The minimal CA-in ribbon. (d) The minimal CA-off ribbon. (e) The minimal AC-in ribbon (i.e., the diamond chain lattice). As shown in (a), the lattices and the coordinate systems have been rotated counterclockwise by 90$^{\circ}$, so that the horizontal axis from right to left is the $y$ axis and the vertical axis from bottom to top is the $x$ axis. A unit cell of each lattice is encircled by a dashed rectangular box.}
\end{figure}

According to the two chains of lattice sites terminating the two edges of the ribbons, we denote the ribbons as $\alpha\beta$-in and $\alpha\beta$-off \cite{hao22}. $\alpha$ and $\beta$ take values among the three sublattices (i.e., A, B, and C) and represent the types of sites terminating the left and right edges of the ribbon. The left edge has a smaller $x$ coordinate than the right edge has. The suffix ``in'' (``off'') indicates that the leftmost and rightmost chains of sites of the ribbon will (will not) coincide if they are projected to the same line parallel to the $y$ axis. In this notation, the four narrow zigzag dice lattice ribbons are separately called the minimal BB-off ribbon (Fig.2a), the minimal BB-in ribbon (Fig.2b), the minimal CA-in ribbon (Fig.2c), and the minimal CA-off ribbon (Fig.2d). The diamond chain lattice (Fig.2e) is the minimal AC-in ribbon in this notation.

To describe the single-particle states in these minimal zigzag ribbons, we retain the terms of the model defined in Eqs.(1)-(3) that involve only the degrees of freedom contained in the minimal ribbons.
In the presence of a nonzero magnetic field (i.e., $f\ne0$), we take the Landau gauge $\mathbf{A}=(0,x,0)H$, so that translational invariance along the $y$ direction is retained and the wave vector $k_{y}$ is still a good quantum number.
From the definition in Eq.(2) for the phase factor, and the vector potential in the Landau gauge, there is a phase factor for each non-horizontal bond with a finite projection along the $y$ axis, but no phase factor for the horizontal bonds (i.e., bonds parallel to the $x$ axis). For example, for the minimal BB-off ribbon, as shown in Fig.2(a), there are two types of non-horizontal bonds in the unit cell. Taking the $x$ coordinate of the $B_{1}$ sites as 0, the phase factor for the hopping terms $tb_{1Y}^{\dagger}c_{1,Y+\frac{a}{2}}$ and $tc_{1,Y-\frac{a}{2}}^{\dagger}b_{1Y}$ is $\gamma_{1}=\frac{\pi}{2}f$, while the phase factor for the hopping terms $ta_{1Y}^{\dagger}b_{2,Y+\frac{a}{2}}$ and $tb_{2,Y-\frac{a}{2}}^{\dagger}a_{1Y}$ is $\gamma_{2}=\frac{5\pi}{2}f$. $Y$ represents the $y$ coordinate of the corresponding site in the unit cell. The phase factors for the other minimal ribbons are obtained in the same way and are shown in Appendix A.

As is well known, the minimal AC-in ribbon (i.e., the diamond chain lattice) has a single Dirac cone coexisting with a zero-energy flat band for $f=0$ and $\Delta=0$, and a fully-pinched spectrum for $f=\frac{1}{2}$ and $\Delta=0$ \cite{vidal00,doucot02}. We show in what follows that the four minimal zigzag ribbons defined in Figs.2(a)-2(d) have more interesting and diverse properties as regards 1D flat bands and Dirac cones.

\section{electronic spectra in zero magnetic field}

For each minimal zigzag ribbon defined in Fig. 2, we show in Fig.3 the band structures (in zero magnetic field) for the unbiased model (i.e., $\Delta=0$) and a symmetrically biased model with $\Delta=0.5t$. For both the unbiased and symmetrically biased dice models, the low-energy band structures have a pair of Dirac cones for the minimal BB-off ribbon, a pair of Dirac cones coexisting with a zero-energy flat band for the minimal BB-in ribbon, and isolated flat bands for the minimal CA-in and minimal CA-off ribbons. In contrast, the single Dirac cone of the minimal AC-in ribbon is gapped out by a finite $\Delta$.

\begin{figure}[H]\label{fig3}
\centering
\hspace{-2.95cm} {\textbf{(a)}} \hspace{3.8cm}{\textbf{(b)}}\\
\hspace{0cm}\includegraphics[width=4.08cm,height=3.4cm]{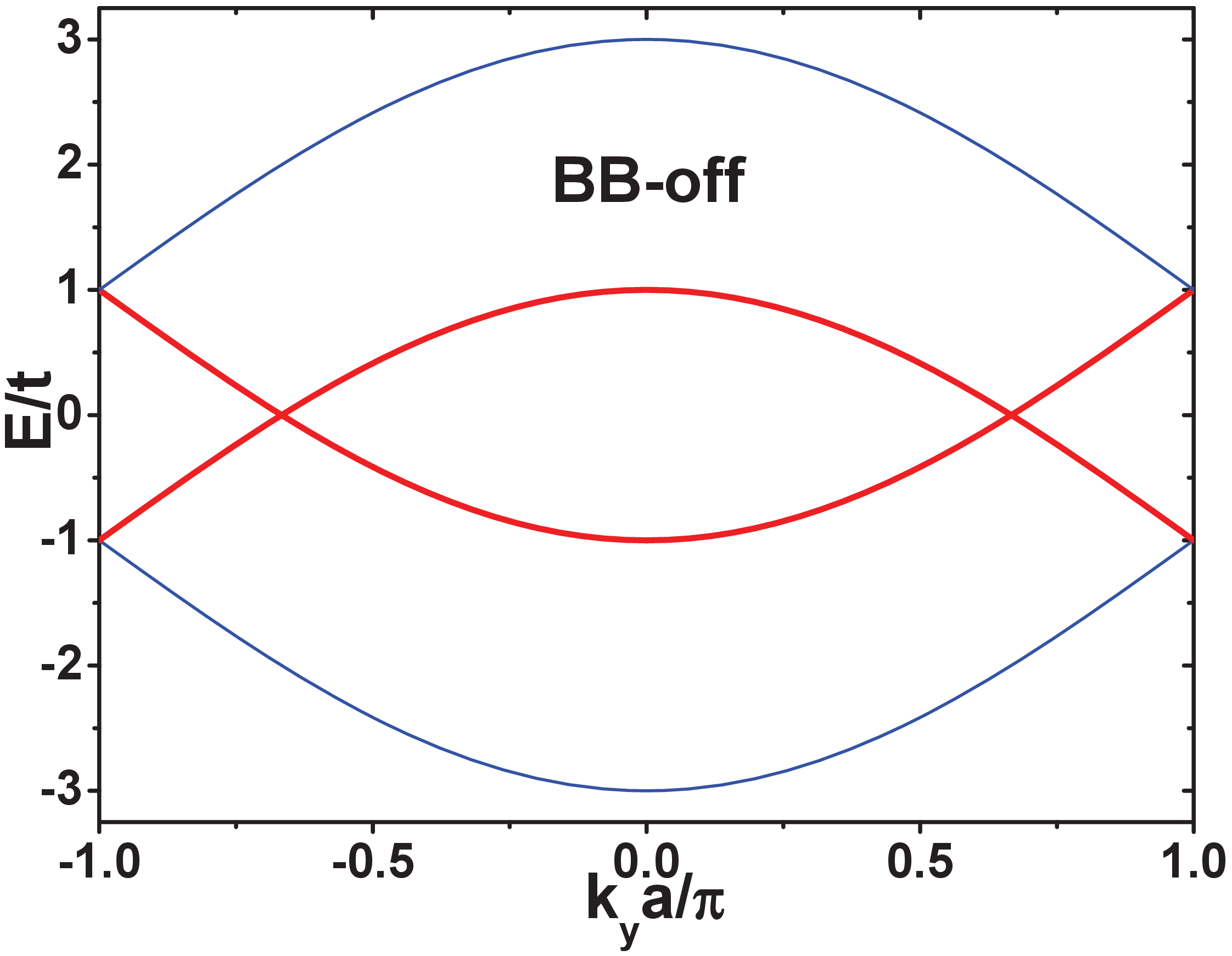}
\includegraphics[width=4.08cm,height=3.4cm]{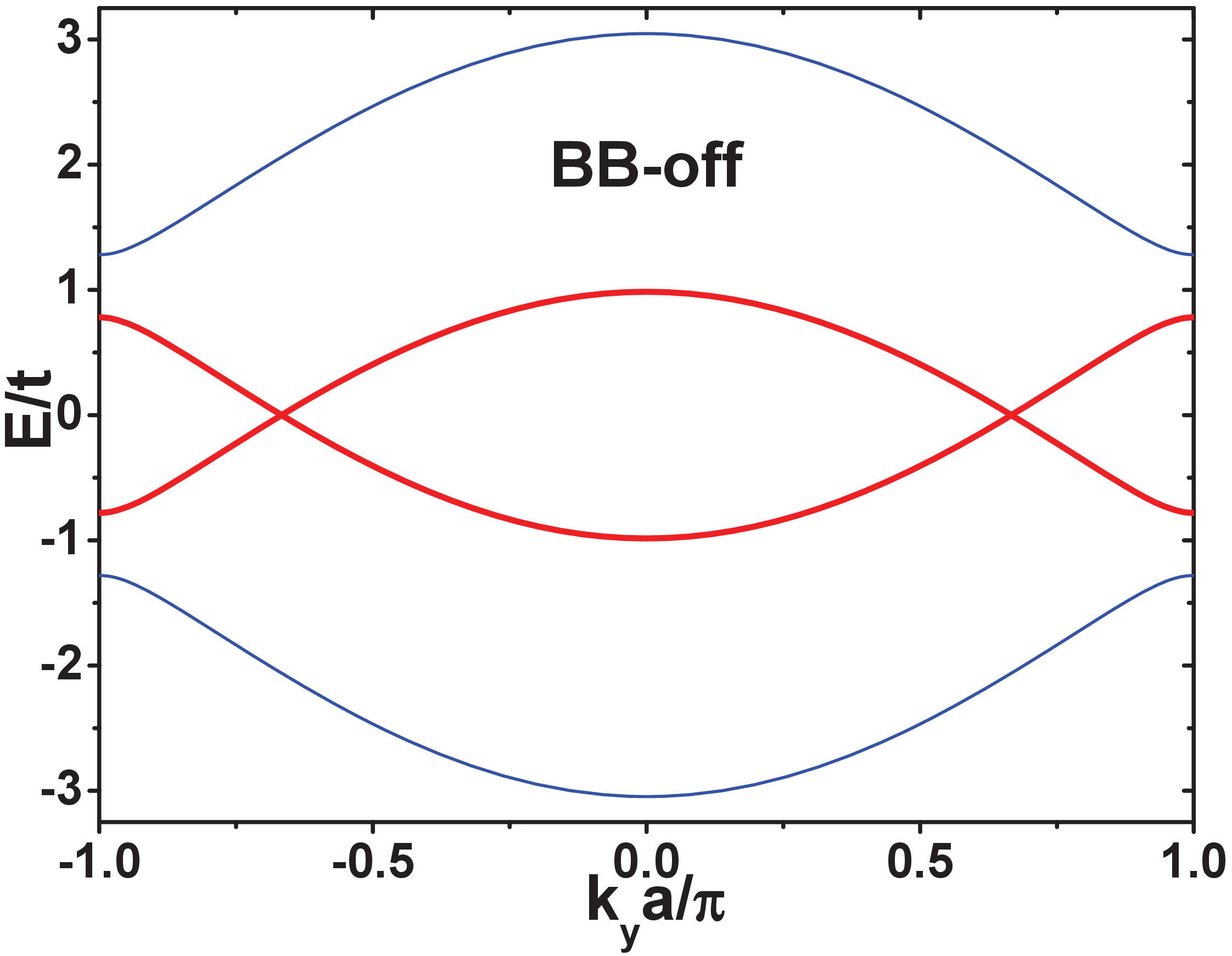} \\
\vspace{-0.10cm}
\hspace{-2.95cm} {\textbf{(c)}} \hspace{3.8cm}{\textbf{(d)}}\\
\hspace{0cm}\includegraphics[width=4.08cm,height=3.4cm]{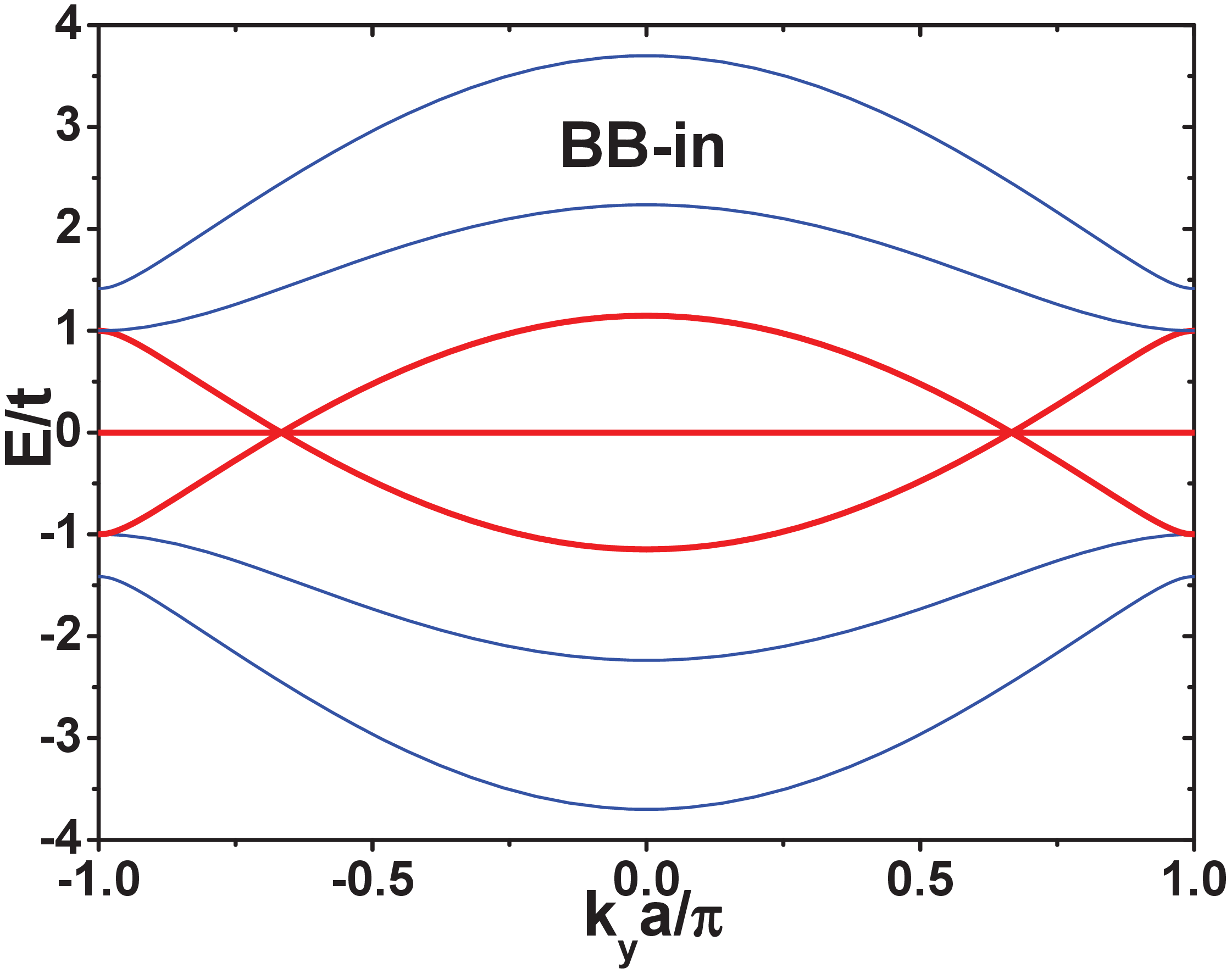}
\includegraphics[width=4.08cm,height=3.4cm]{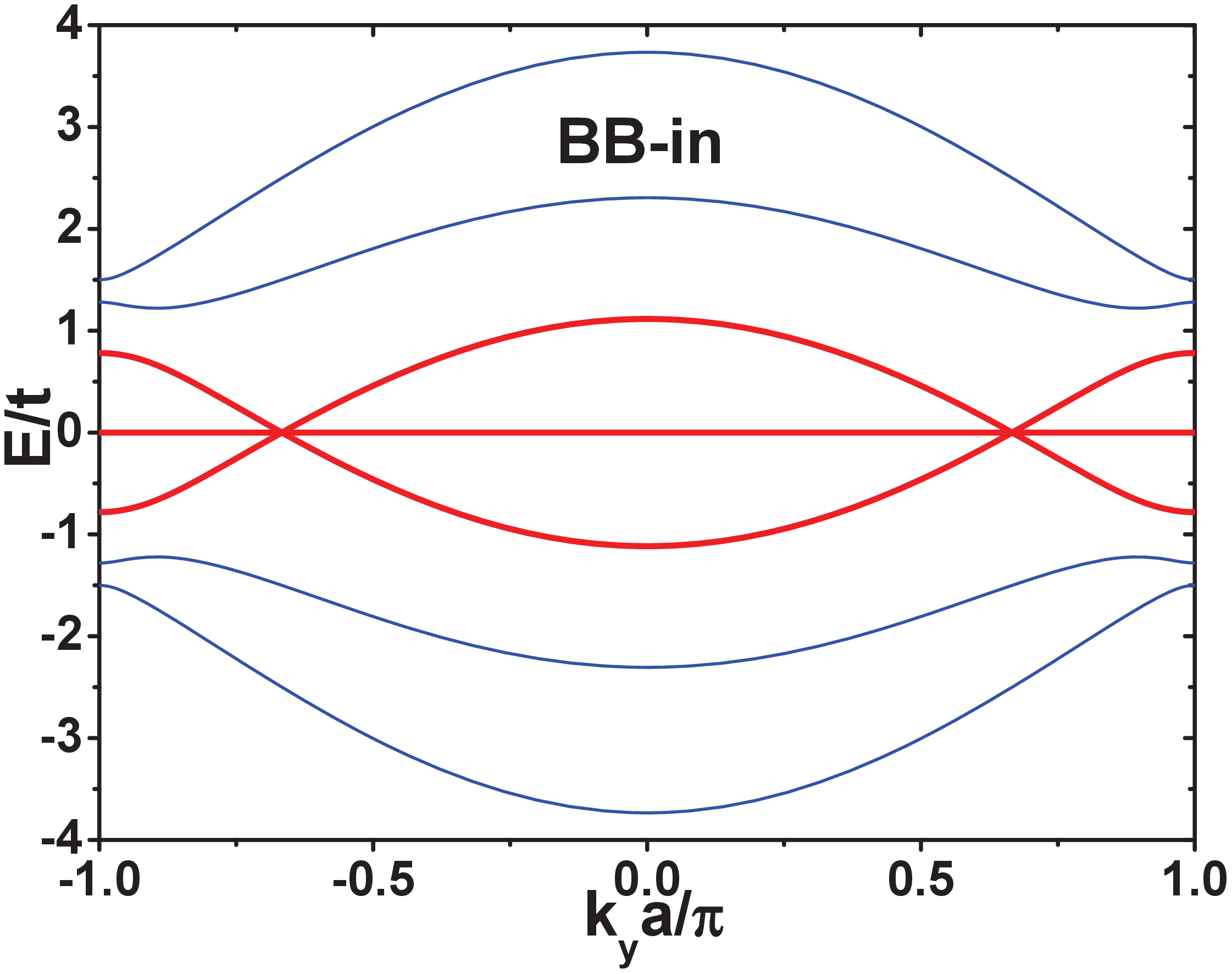}\\
\vspace{-0.10cm}
\hspace{-2.95cm} {\textbf{(e)}} \hspace{3.8cm}{\textbf{(f)}}\\
\hspace{0cm}\includegraphics[width=4.08cm,height=3.4cm]{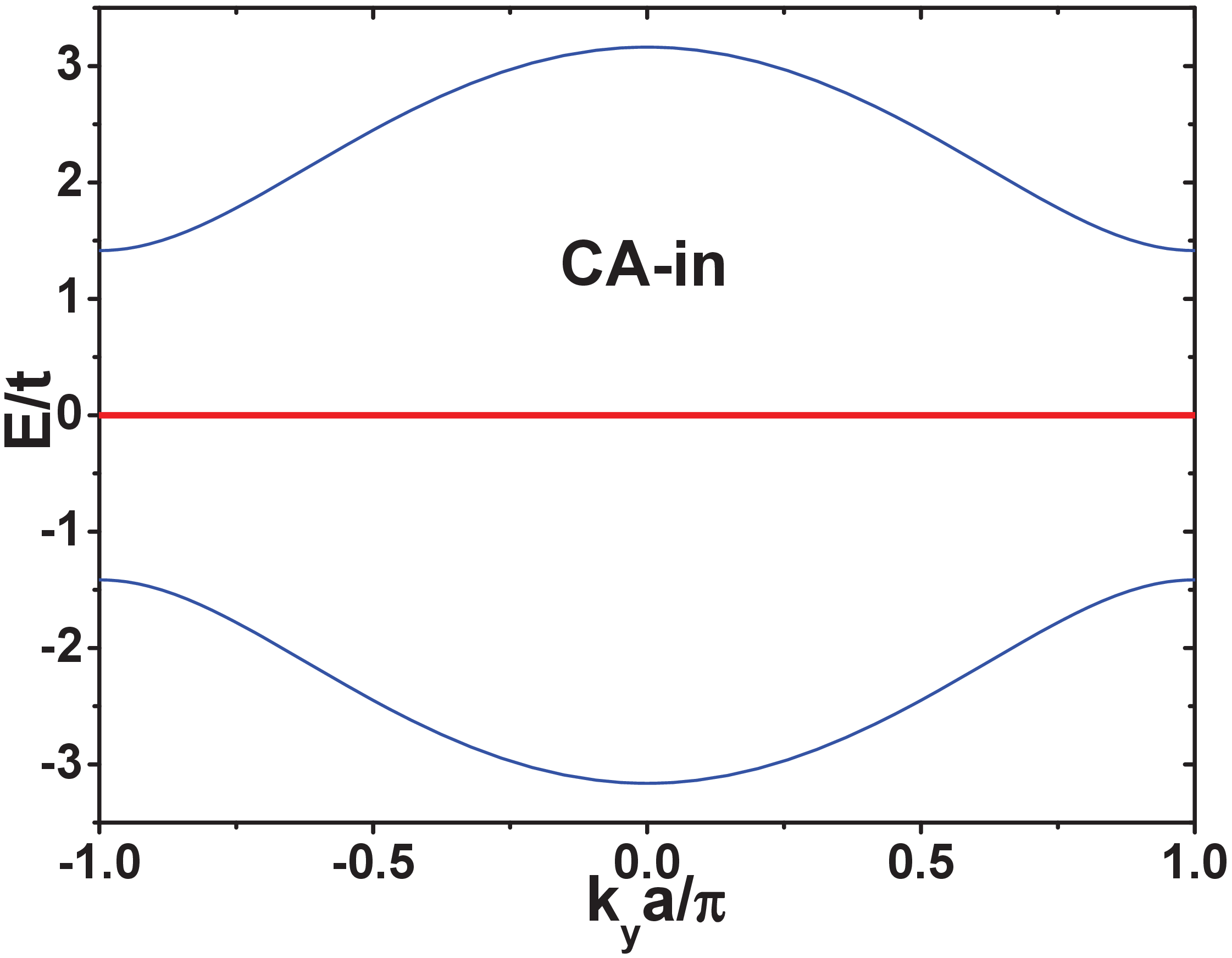}
\includegraphics[width=4.08cm,height=3.4cm]{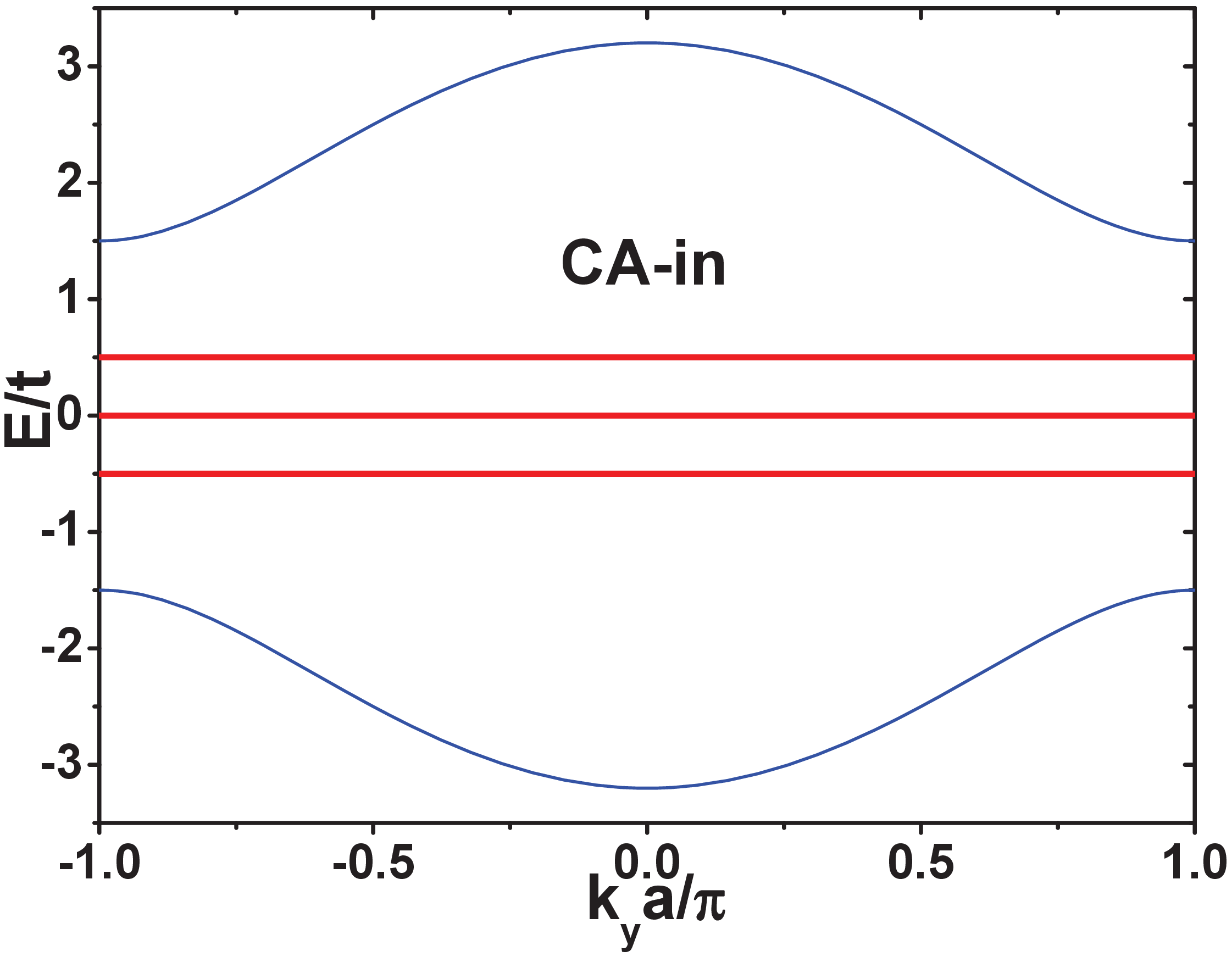} \\
\vspace{-0.10cm}
\hspace{-2.95cm} {\textbf{(g)}} \hspace{3.8cm}{\textbf{(h)}}\\
\hspace{0cm}\includegraphics[width=4.08cm,height=3.4cm]{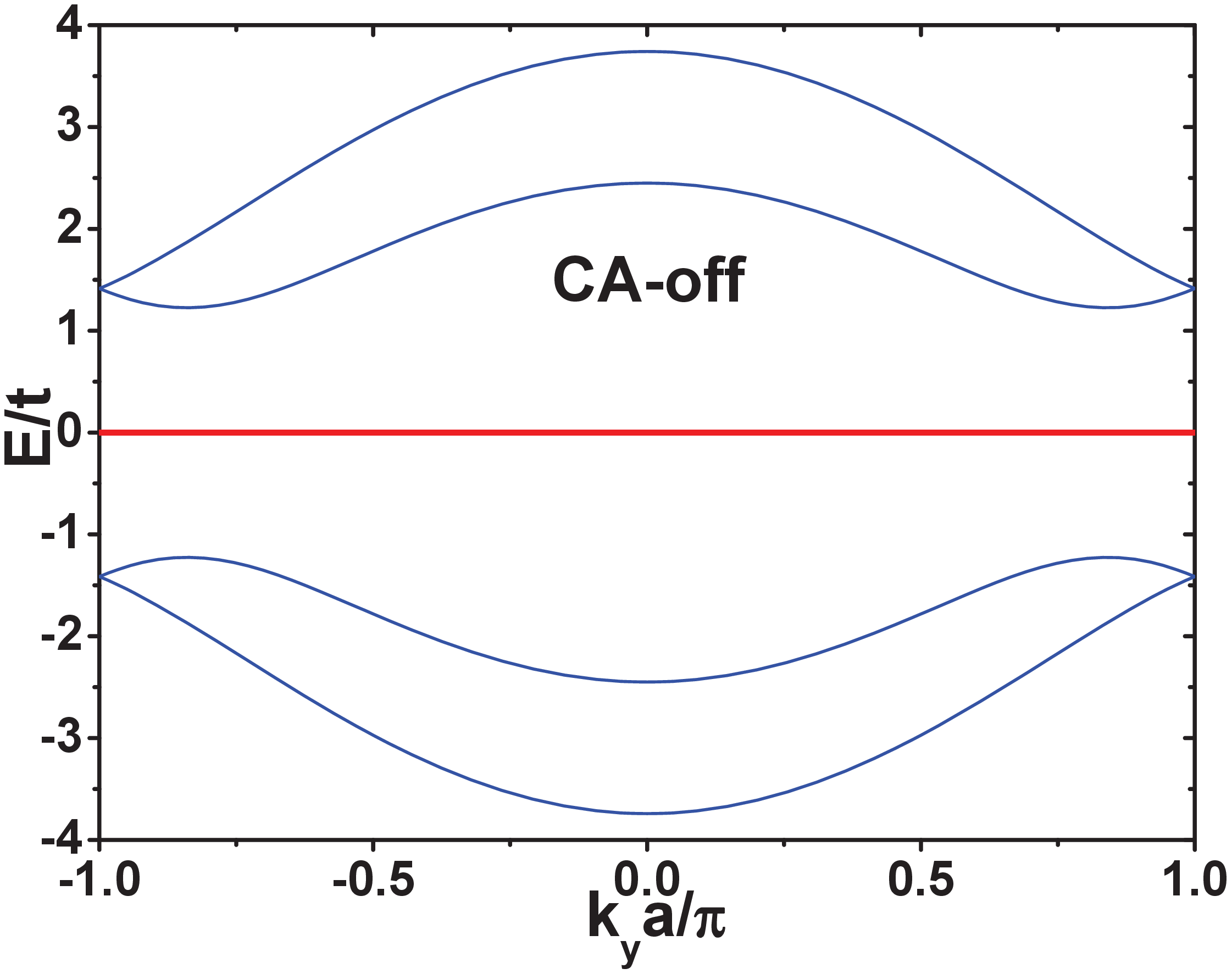}
\includegraphics[width=4.08cm,height=3.4cm]{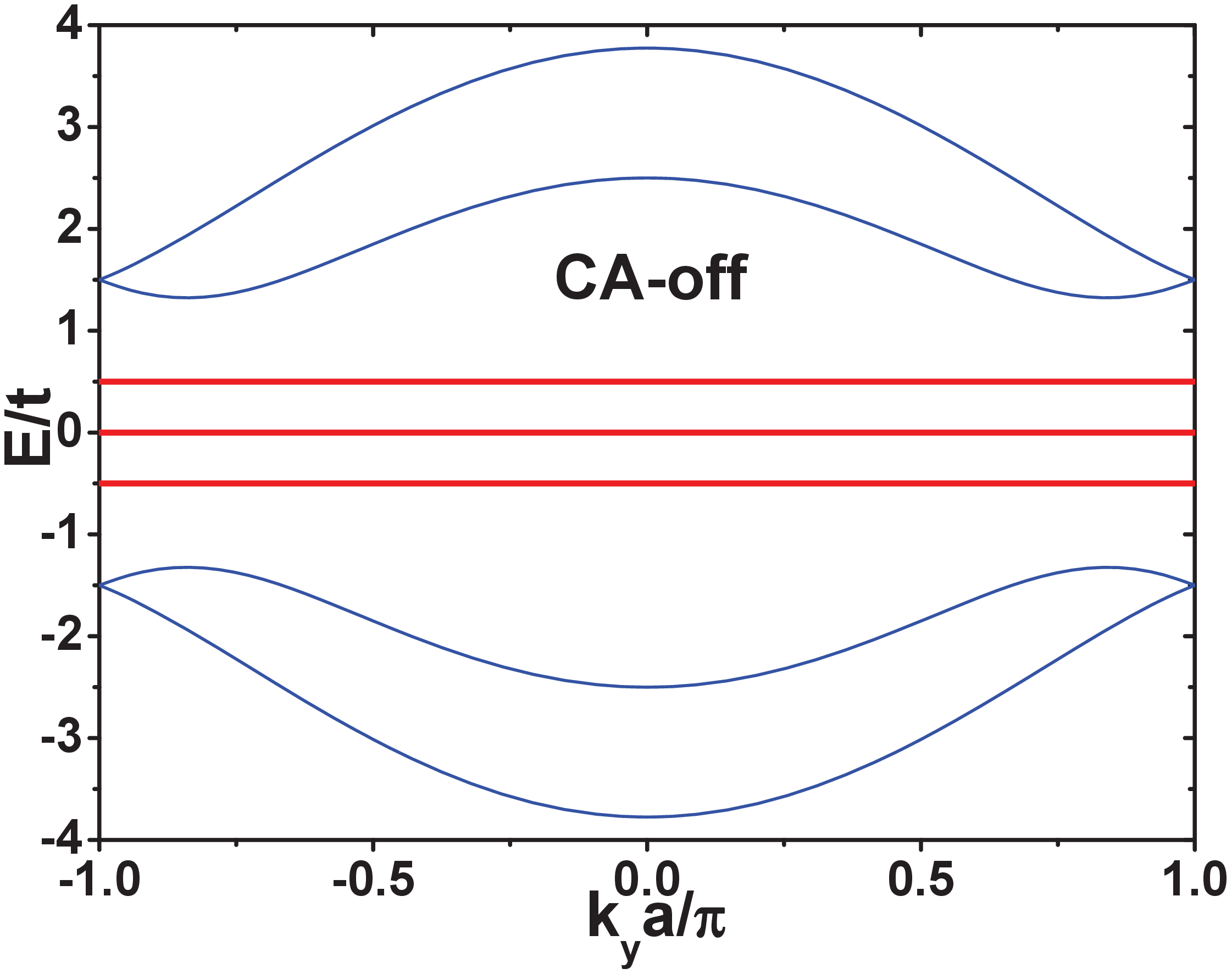} \\
\vspace{-0.10cm}
\hspace{-2.95cm} {\textbf{(i)}} \hspace{3.8cm}{\textbf{(j)}}\\
\hspace{0cm}\includegraphics[width=4.08cm,height=3.4cm]{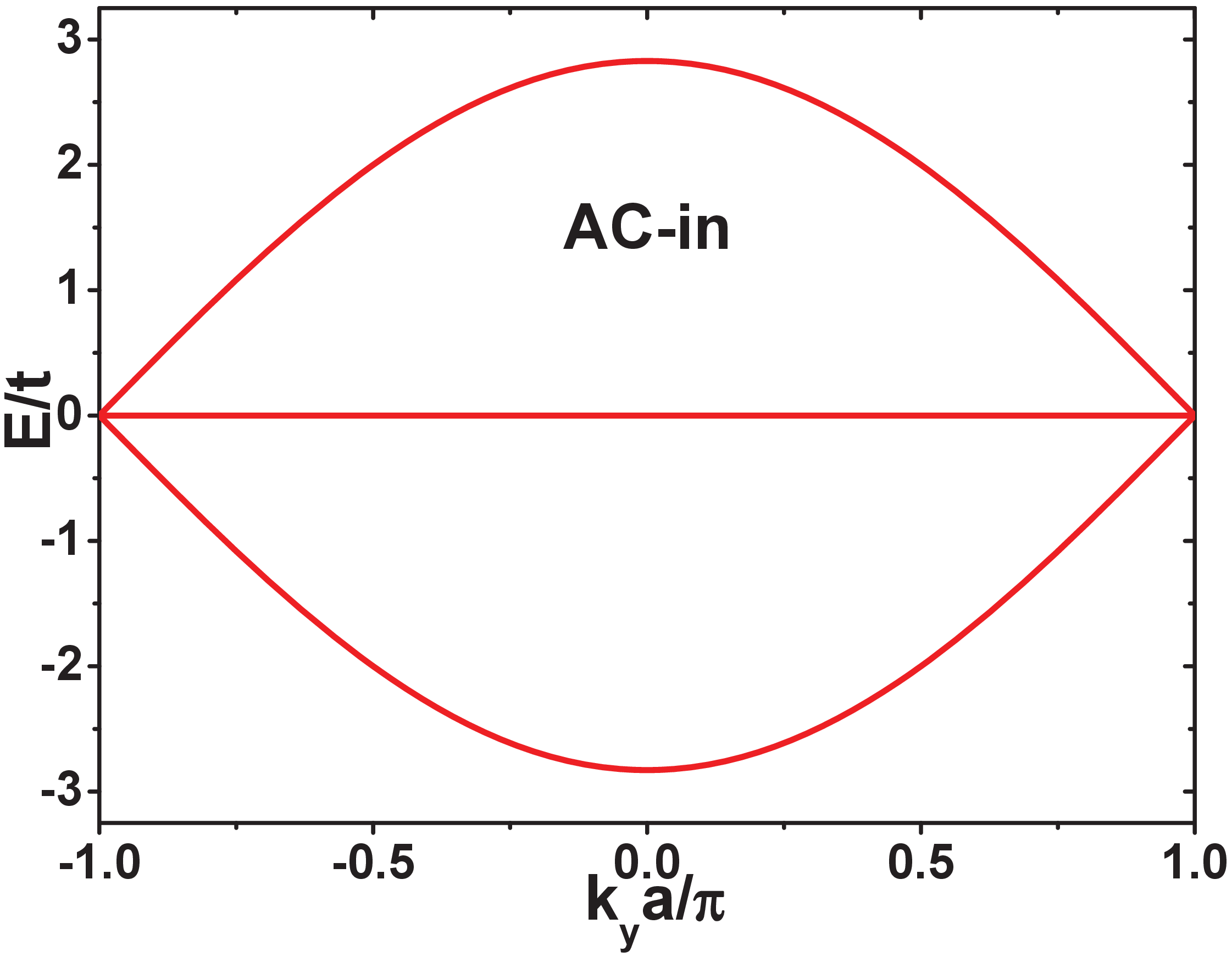}
\includegraphics[width=4.08cm,height=3.4cm]{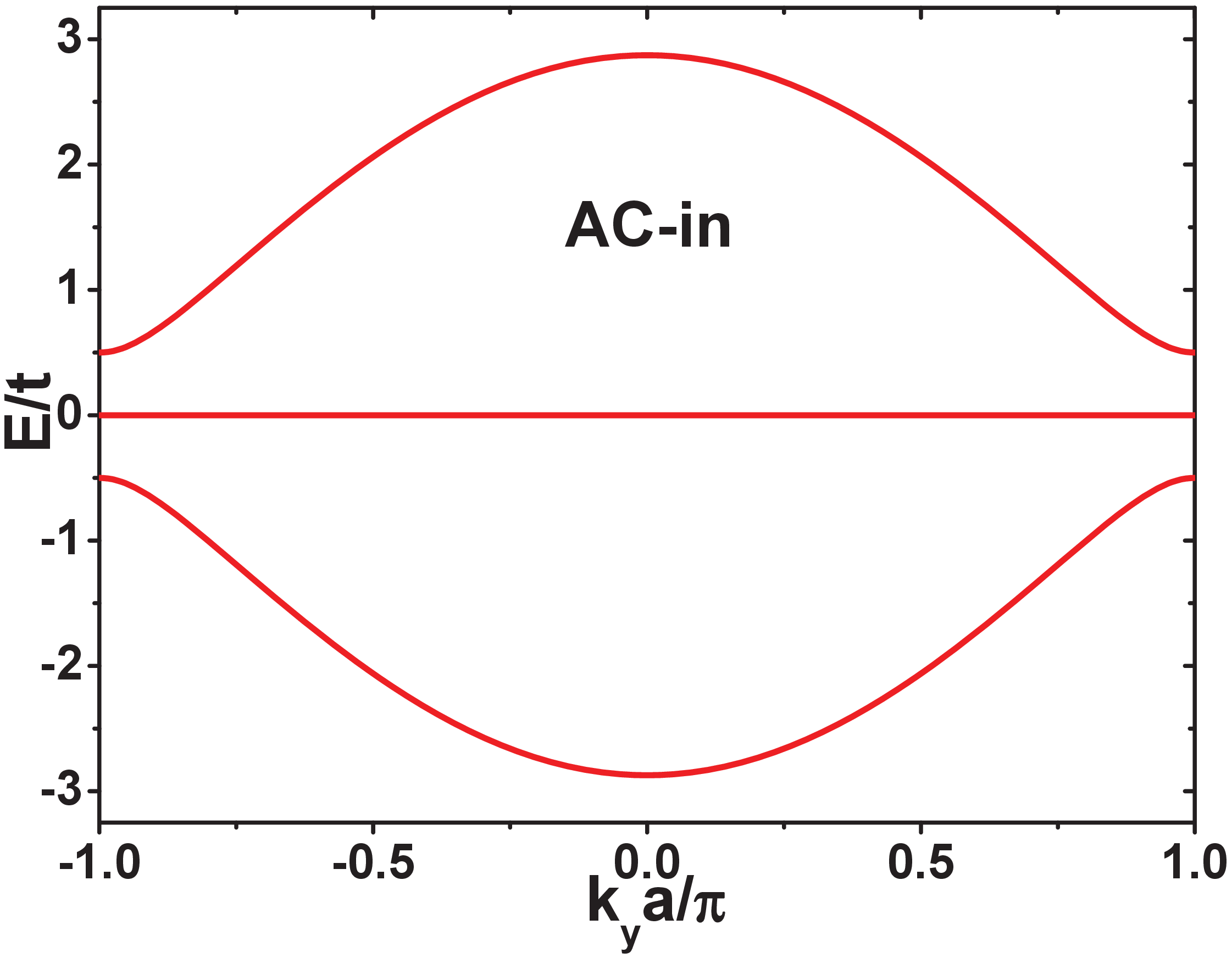}
\caption{Band structures of the five minimal zigzag dice lattice ribbons, for the unbiased and symmetrically biased dice models. The left column (a, c, e, g, i) and the right column (b, d, f, h, j) are separately the band structures for $\Delta=0$ and for $\Delta=0.5t$, of the five ribbons defined in Fig.2 in sequence. The flat bands and the low-energy bands containing the (gapped) Dirac cones are highlighted in bold and plotted in red.}
\end{figure}

To our knowledge, the five minimal zigzag dice lattice ribbons, in particular the four new ribbons, constitute the only series of narrow ribbons for the same 2D lattice that have such rich varieties of combinations of 1D Dirac cones and flat bands as those shown in Fig.3. Note that the band structures for the minimal zigzag ribbons are clearly not a simple combination of those for the narrowest minimal ribbons, the minimal AC-in and minimal BB-off ribbons.

In what follows, we study in more details the Dirac cones and flat bands in the four minimal zigzag dice lattice ribbons. In particular, we determine whether the Dirac cones and flat bands are inherited from the bulk bands or related to the edge states of corresponding wide ribbons. We also study the connection of these states with the geometric structures of the ribbon lattices, and identify new characters of the flat bands in terms of peculiar CLS.

For notational simplicity, we denote the wave vector (wavenumber) $k_{y}$ in the 1D BZ as $k$. The Hamiltonian matrices in the $k$ space and the corresponding eigenequations are put to Appendix A.

\subsection{Dirac cones of the minimal BB-off and BB-in ribbons}

According to Appendix A.1, the four bands of the minimal BB-off ribbon are
\begin{equation}
E_{\alpha\beta}(k)=\alpha\left|\sqrt{t^{2}+\frac{\Delta^{2}}{4}}+\beta\sqrt{t'^{2}+\frac{\Delta^{2}}{4}}\right|,
\end{equation}
where $t'=2t\cos\frac{ka}{2}$, $\alpha=\pm$ and $\beta=\pm$. The two low-energy bands $E_{--}(k)$ and $E_{+-}(k)$ touch at $k=\pm\frac{2\pi}{3a}$, which give two valleys of 1D Dirac cones. In the neighborhood of the Dirac point $k_{\tau}=\tau\frac{2\pi}{3a}$ ($\tau=\pm$), we may expand the Dirac cone band $E_{\alpha-}(k)$ ($\alpha=\pm$) into the leading order polynomial of the deviation $q=k-k_{\tau}$,
\begin{equation}
E_{\alpha-}\simeq\alpha\frac{\sqrt{3}t^{2}a|q|}{2\sqrt{t^{2}+\frac{\Delta^{2}}{4}}}.
\end{equation}
The group velocity of the 1D Dirac fermions is
\begin{equation}
v=\frac{\sqrt{3}t^{2}a}{2\sqrt{t^{2}+\frac{\Delta^{2}}{4}}},
\end{equation}
which tends to $\sqrt{3}a|t|/2$ in the limit of $\Delta=0$. In the close neighborhood of the Dirac point $k_{\tau}$ and up to the leading order of $q$, the eigenvector for the state in the $E_{\alpha-}$ ($\alpha=\pm$) band is
\begin{equation}
\frac{1}{2\sqrt{4t^{2}+\Delta^{2}}}\begin{pmatrix} \sqrt{4t^{2}+\Delta^{2}}-\alpha\tau\text{sgn}(q)\Delta \\
2t\alpha\tau\text{sgn}(q) \\
-2t\alpha\tau\text{sgn}(q) \\
-\sqrt{4t^{2}+\Delta^{2}}-\alpha\tau\text{sgn}(q)\Delta   \end{pmatrix}.
\end{equation}
$\text{sgn}(q)=q/|q|$ is the sign of $q$. The basis is taken as $(b_{1k}^{\dagger},a_{1k}^{\dagger},c_{1k}^{\dagger},b_{2k}^{\dagger})$. As $\Delta$ tends to zero, the above eigenvector evolves continuously to
\begin{equation}
\frac{1}{2}\begin{pmatrix} 1 \\
\alpha\tau\text{sgn}(qt) \\
-\alpha\tau\text{sgn}(qt) \\
-1   \end{pmatrix}.
\end{equation}

The Dirac cones of the minimal BB-off ribbon can be understood from BZ folding. We may consider the lattice structure of Fig.2(a) for the minimal BB-off ribbon as a crumpled two-leg ladder. In this picture, we consider a continuously variable bond angle $\theta$ between each NN pair of bonds along the two parallel zigzag chains. $\theta\in(0,\pi)$, and $\theta=\frac{2\pi}{3}$ corresponds to the minimal BB-off ribbon of Fig.2(a). We fix the lengths of the NN bonds as $a_{0}$ and so the lattice parameter of the crumpled two-leg ladder becomes $a'=2a_{0}\sin(\frac{\theta}{2})$. The wave vectors of the 1D BZ are in the range of $k'\in(-\frac{\pi}{a'},\frac{\pi}{a'}]$. By making the substitutions of $k'$ for $k$ and $a'$ for $a$, the model and spectrum in zero magnetic field are still given by Eq.(A1) and Eq.(4). The two Dirac cones at $k'a'=\pm\frac{2\pi}{3}$ are therefore robust features of the crumpled two-leg ladder with arbitrary $\theta\in(0,\pi)$ and $\Delta$.
The BZ folding picture for the Dirac cones is understood by considering the limit of $\theta=\pi$ and $\Delta=0$, for which the crumpling is removed. In this case, the band structures for the two-leg ladder with 2 sites in a primitive unit cell have no Dirac cones. By artificially doubling the unit cell to have 4 sites, the folding of the BZ and the bands give two Dirac cones, which correspond directly to the pair of Dirac cones for $0<\theta<\pi$.

According to Appendix A.2, the band structures of the minimal BB-in ribbon have a single zero-energy flat band and two valleys of Dirac cones. In the close neighborhood of the Dirac point $k_{\tau}=\tau\frac{2\pi}{3a}$ ($\tau=\pm$), we may expand the Dirac cone bands into a polynomial of $q=k-k_{\tau}$. To the leading linear order of $q$, we find
\begin{equation}
E_{2\alpha}\simeq\alpha|qt|a\sqrt{\frac{3t^{2}(4t^{2}+\Delta^{2})}{(6t^{2}+\Delta^{2})(2t^{2}+\Delta^{2})}},
\end{equation}
where $\alpha=\pm$. The group velocity of the Dirac fermions is
\begin{equation}
v=a|t|\sqrt{\frac{3t^{2}(4t^{2}+\Delta^{2})}{(6t^{2}+\Delta^{2})(2t^{2}+\Delta^{2})}}.
\end{equation}
In the limit of $\Delta=0$, we have $E_{2\alpha}\simeq\alpha|qt|a$ and $v=a|t|$.

To the leading order of $q$, we find the eigenvector $\psi_{\alpha}(k_{\tau}+q)$ ($\alpha=\pm$, $\tau=\pm$) of the Dirac cone states at $\Delta=0$ to be
\begin{equation}
\psi_{\alpha}(k_{\tau}+q)\simeq\frac{1}{\sqrt{6}}\begin{pmatrix} 1 \\ \frac{\sqrt{3}}{2}\alpha\tau\text{sgn}(qt) \\ -\frac{\sqrt{3}}{2}\alpha\tau\text{sgn}(qt) \\ -1 \\ -\frac{\sqrt{3}}{2}\alpha\tau\text{sgn}(qt) \\ \frac{\sqrt{3}}{2}\alpha\tau\text{sgn}(qt) \\ 1 \end{pmatrix},
\end{equation}
in the basis of $(b_{1k}^{\dagger},a_{1k}^{\dagger},c_{1k}^{\dagger} ,b_{2k}^{\dagger},a_{2k}^{\dagger},c_{2k}^{\dagger},b_{3k}^{\dagger})$.
It is interesting that $\psi_{\alpha}(k_{\tau}+q)$ is in the form of a symmetric merging of the two sections B$_{1}$A$_{1}$C$_{1}$B$_{2}$ and $B_{3}$C$_{2}$A$_{2}$B$_{2}$ via the common orbital of the B$_{2}$ site. This corresponds to the geometrical relation between the minimal BB-in ribbon and the minimal BB-off ribbon. Namely, the minimal BB-in ribbon may be considered as the merging of two minimal BB-off ribbons via the common $y$-chain of B$_{2}$ sites.

In addition, a comparison of Eq.(11) with Eq.(8) suggests a particularly regular evolution of the Dirac cone states with the increase of the ribbon width. Comparison with the known eigenvector and eigenspectrum for the Dirac cone states of very wide BB-in or BB-off ribbons strengthens this expectation \cite{hao22}. Briefly summarizing the trend, for $\Delta=0$, the group velocity of the Dirac fermions changes from $\frac{\sqrt{3}}{2}a|t|$ for the minimal BB-off ribbon to $a|t|$ for the minimal BB-in ribbon, and finally to $\frac{\sqrt{6}}{2}a|t|$ for very wide BB ribbons \cite{hao22}. For the eigenvector, the total weight on the A and C sites is equal to the weight on the B sites, which drives the ratio between the amplitudes on an A or C site and a B site to evolve from 1 for the minimal BB-off ribbon to $\frac{\sqrt{3}}{2}$ for the minimal BB-in ribbon, and finally to $\frac{\sqrt{2}}{2}$ for very wide BB ribbons \cite{hao22}.

Are the Dirac cones of the minimal BB-in and BB-off ribbons inherited from the bulk states or related to the in-gap states (i.e., edge states) of wide ribbons? For wide zigzag dice lattice ribbons, it has been found that the various edge states can be indicated by the in-gap states at the boundary of the 1D BZ, $ka=\pi$ \cite{hao22}. At $ka=\pi$, $t'=0$, the Hamiltonian matrices describe a collection of disconnected CBA trimers, CB or BA dimers, and A or C monomers \cite{hao22}. The eigenstates of the CBA trimers correspond to the projection of the bulk bands. The eigenstates of the dimers and monomers give in-gap states, which connect continuously to the edge states for general wave vectors \cite{hao22}. Therefore, the states for $ka=\pi$ can be taken as indicators to tell whether a band of the ribbon corresponds to bulk-like bands or edge states of wide ribbons.

For the minimal BB-off ribbon defined in Fig. 2(a), the Hamiltonian matrix of Eq.(A1) describes two isolated dimers, B$_1$A$_1$ and B$_2$C$_1$, at $ka=\pi$ and for $f=0$. In this respect, all the four bands in Figs.3(a) and 3(b) for the minimal BB-off ribbon correspond to the edge states of wide BB-off ribbons. At $ka=\pi$ and for $f=0$, the Hamiltonian matrix Eq.(A5) of the minimal BB-in ribbon describes an isolated trimer C$_1$B$_2$A$_2$ and two isolated dimers B$_1$A$_1$ and B$_3$C$_2$. The three eigenstates of the isolated trimer connect to the zero-energy flat band and the two highest energy dispersive bands. The single flat band therefore inherits directly from the bulk zero-energy flat band. The eigenstates of the two isolated dimers connect to the remaining four bands that have the same qualitative structure as the four bands of the minimal BB-off ribbon. The Dirac cones of the minimal BB-in ribbon therefore correspond also to the edge states of wide BB-in ribbons.

We highlight in passing two qualitative differences between the Dirac cones of the minimal BB-off and minimal BB-in ribbons and the Dirac cone of the minimal AC-in ribbon. Firstly, the single Dirac cone of the minimal AC-in ribbon is a unique property of the unbiased minimal AC-in ribbon itself and does not correspond to the edge states of wider AC-in ribbons, nor does it correspond to the Dirac cones of the bulk dice lattice. To our knowledge, the Dirac cones of the minimal BB-off and minimal BB-in ribbons are the only known cases of Dirac cone edge states that survive the extreme reduction of the ribbon width. Secondly, the two Dirac cones of the minimal BB-off and minimal BB-in ribbons constitute a two-component valley degree of freedom, akin to the bulk dice model. The minimal BB-off and minimal BB-in ribbons therefore have alluring application prospects in valleytronics.

\subsection{Isolated flat bands in minimal CA-in and CA-off ribbons}

According to Appendix A.3, the band structures of the minimal CA-in ribbon have one zero-energy flat band $E_{0}(k)=0$, two other flat bands at $E_{1\alpha}(k)=\alpha|\Delta|$, and two dispersive bands with dispersions $E_{2\alpha}(k)=\alpha\sqrt{\Delta^{2}+2(t^{2}+t'^{2})}$ ($\alpha=\pm$). When $\Delta=0$, we have a triply degenerate zero-energy flat band, which splits into three nondegenerate flat bands for $\Delta\neq0$. The flat bands are always separated from the dispersive bands by at least $\sqrt{\Delta^{2}+2t^{2}}-|\Delta|$.

In the basis $(c_{1k}^{\dagger},a_{1k}^{\dagger},b_{1k}^{\dagger},c_{2k}^{\dagger},a_{2k}^{\dagger})$, the eigenvector for the zero-energy flat band of the minimal CA-in ribbon is
\begin{equation}
\frac{1}{\sqrt{2t^{2}+2t'^{2}+\Delta^{2}}}\begin{pmatrix} t, & -t', & \Delta, & t', & -t \\ \end{pmatrix}^{\text{T}}.
\end{equation}
The superscript $``\text{T}"$ means taking the transpose of the array. The flat band at $\Delta$ locates completely on the A$_{1}$ and A$_{2}$ sublattices, with the eigenvector
\begin{equation}
\frac{1}{\sqrt{t^{2}+t'^{2}}}\begin{pmatrix} 0, & -t, & 0, & 0, & t' \\ \end{pmatrix}^{\text{T}}.
\end{equation}
The flat band at $-\Delta$ resides completely on the C$_{1}$ and C$_{2}$ sublattices, with the eigenvector
\begin{equation}
\frac{1}{\sqrt{t^{2}+t'^{2}}}\begin{pmatrix} t', & 0, & 0, & -t, & 0 \\ \end{pmatrix}^{\text{T}}.
\end{equation}

The band structures of the minimal CA-off ribbon for a general $\Delta$, according to Appendix A.4, have two zero-energy flat bands and two flat bands at $\pm\Delta$. These flat bands are all isolated from the four higher energy dispersive bands. Taking the basis as $(c_{1k}^{\dagger},a_{1k}^{\dagger},b_{1k}^{\dagger},c_{2k}^{\dagger},a_{2k}^{\dagger} ,b_{2k}^{\dagger},c_{3k}^{\dagger},a_{3k}^{\dagger})$, the eigenvectors for the two zero-energy flat bands are
\begin{equation}
\begin{cases}
\frac{1}{\sqrt{2t^{2}+2t'^{2}+\Delta^{2}}} (t,-t',\Delta,t',-t,0,0,0)^{\text{T}}, \\
\frac{1}{\sqrt{2t^{2}+2t'^{2}+\Delta^{2}}} (0,0,0,t,-t',\Delta,t',-t)^{\text{T}}. \\
\end{cases}
\end{equation}
While these two vectors are not orthogonal to each other, they constitute a complete basis set for the two-fold degenerate zero-energy flat bands. In addition, if we compare the 8 sites in the unit cell of the minimal CA-off ribbon in Fig.2(d) and the 5 sites in the unit cell of the minimal CA-in ribbon in Fig.2(c), we see that the above two eigenvectors have the same form as Eq.(12) for the single zero-energy flat band of the minimal CA-in ribbon. These two flat bands clearly correspond to the two chains of completely coordinated B sublattice sites, which are the B$_{1}$ and B$_{2}$ sites of the unit cell.
This is an explicit illustration of the conclusion drawn for the wide zigzag dice lattice ribbons that each $y$-chain of fully coordinated B sublattice sites gives such a zero-energy flat band that is robust to $\Delta\ne0$ \cite{hao22}.

The eigenvector for the flat band at $E=\Delta$ is
\begin{equation}
\frac{1}{\sqrt{t^{4}+t^{2}t'^{2}+t'^{4}}}\begin{pmatrix} 0, & t^{2}, & 0, & 0, & -tt', & 0, & 0, & t'^{2} \\ \end{pmatrix}^{\text{T}}.
\end{equation}
The eigenvector for the flat band at $E=-\Delta$ is
\begin{equation}
\frac{1}{\sqrt{t^{4}+t^{2}t'^{2}+t'^{4}}}\begin{pmatrix} t'^{2}, & 0, & 0, & -tt', & 0, & 0, & t^{2}, & 0 \\ \end{pmatrix}^{\text{T}}.
\end{equation}
Eqs.(16) and (17) are direct generalizations of Eqs.(13) and (14) for the minimal CA-in ribbon. On one hand, the eigenstate of energy $\Delta$ ($-\Delta$) resides completely on the A (C) sublattice sites. On the other hand, the wave function amplitudes of the eigenstate of energy $\Delta$ ($-\Delta$) decay or increase in proportion of $-t'/t$ from left to right (from right to left). Geometrically, the minimal CA-off ribbon may be considered as the merging of two CA-in ribbons via the common C$_{2}$ and A$_{2}$ sublattices. The evolution of the eigenvectors from Eqs.(13) and (14) to Eqs.(16) and (17) illustrates this geometrical connection between the two minimal CA ribbons.

Are the flat bands of the minimal CA-in and CA-off ribbons related to the bulk bands or the edge states (i.e., in-gap states) of wide ribbons?
At $ka=\pi$ and for $f=0$, Eq.(A23) for the Hamiltonian matrix of the minimal CA-in ribbon describes an isolated trimer C$_1$B$_1$A$_2$ and two isolated monomers A$_1$ and C$_2$. The three eigenstates of the isolated trimer connect to one zero-energy flat band and two dispersive bands. The eigenstates of the two isolated monomers A$_1$ and C$_2$ connect separately to the two flat bands with energy $\Delta$ and $-\Delta$. Therefore, we both have a flat band corresponding to the zero-energy flat band of the bulk dice lattice and have two additional flat bands corresponding to the edge states of wide CA-in ribbons \cite{hao22}.

At the BZ boundary $ka=\pi$ and for $f=0$, Eq.(A27) for the $k$-space Hamiltonian matrix of the minimal CA-off ribbon describes two isolated trimers C$_1$B$_1$A$_2$ and C$_2$B$_2$A$_3$ and two isolated monomers A$_{1}$ and C$_3$. The zero-energy eigenstates of the two trimers connect to the two zero-energy flat bands corresponding to the zero-energy flat band of the bulk dice lattice. The eigenstates of the two monomers A$_{1}$ and C$_3$ connect to the two flat bands at $\Delta$ and $-\Delta$, which correspond to the edge states of wide CA-off ribbons \cite{hao22}.

\subsection{CLS in the flat bands}

States of the flat bands may be represented in terms of CLS, which are distributed within restricted regions of the lattice \cite{flach14,sutherland86,vidal98,bergman08}. The CLS reflect the localized nature of the flat band states. Different spatial distributions of the CLS show the different natures of the distinct flat bands. The CLS may also find interesting applications in information technology, such as the diffractionless transmission of information demonstrated in photonic lattices \cite{vicencio15,mukherjee15}.

Here, we construct the CLS for the flat bands shown in Fig.3, except for the flat band of the minimal AC-in ribbon which are well known \cite{vidal00}. Because of the complete degeneracy of the eigenstates in a flat band, arbitrary linear combinations of the eigenstates are still an eigenstate of the system with the flat-band energy. For the single flat band of the minimal BB-in ribbon, according to Eq.(A8) for the eigenvector, the CLS associated with a $y$ coordinate of $Y$ is constructed as \cite{bergman08}
\begin{equation}
A_{Y}^{\dagger}=N_{0}\sum_{k}e^{-ikY}[-t'a_{1k}^{\dagger}+tc_{1k}^{\dagger}+\Delta b_{2k}^{\dagger}-ta_{2k}^{\dagger}+t'c_{2k}^{\dagger}],
\end{equation}
where $N_{0}$ is a normalization constant, the summation over $k$ runs over the $N$ (number of unit cells of the ribbon) wave vectors in the 1D BZ, $t'=2t\cos(ka/2)$. For each $Y$ equal to the $y$ coordinate of a B$_{2}$ site of the ribbon lattice, we have
\begin{eqnarray}
A_{Y}^{\dagger}&=&\frac{1}{\sqrt{6+(\frac{\Delta}{t})^{2}}}[c_{1Y}^{\dagger}-a_{1,Y+\frac{a}{2}}^{\dagger} -a_{1,Y-\frac{a}{2}}^{\dagger}      \notag \\
&&+\frac{\Delta}{t}b_{2Y}^{\dagger}+c_{2,Y+\frac{a}{2}}^{\dagger}+c_{2,Y-\frac{a}{2}}^{\dagger}-a_{2Y}^{\dagger}].
\end{eqnarray}
$A_{Y}^{\dagger}$ creates a CLS within a cluster consisting of seven sites (i.e., a 7-site star) centering at the B$_{2}$ site whose $y$ coordinate is $Y$. The unnormalized wave function of this CLS is shown in Fig.4(a). For each B$_{2}$ site, there is such a CLS. For both zero and nonzero $\Delta$, each 7-site star contains sites [e.g., one C$_{1}$ site and one A$_{2}$ site, according to Fig.2(b)] that are not contained in any other 7-site stars. We therefore have $N$ linearly independent $A_{Y}^{\dagger}$, which form a complete basis set for the flat band shown in Figs.3(c) and 3(d). Although the flat band touches the Dirac cone bands at two Dirac points, there are no intrinsic loop states (or, line states) in the flat band as the bulk dice lattice \cite{bergman08,xia18}. This indicates that the touching of the flat band with the Dirac cone is accidental.

\begin{figure}\label{fig4} \centering
\hspace{-2.0cm}{\textbf{(a)}} \hspace{3.0cm}{\textbf{(b)}}  \hspace{2.3cm}{\textbf{(c)}} \\
\hspace{-0.5cm}\includegraphics[width=3cm,height=3.276cm]{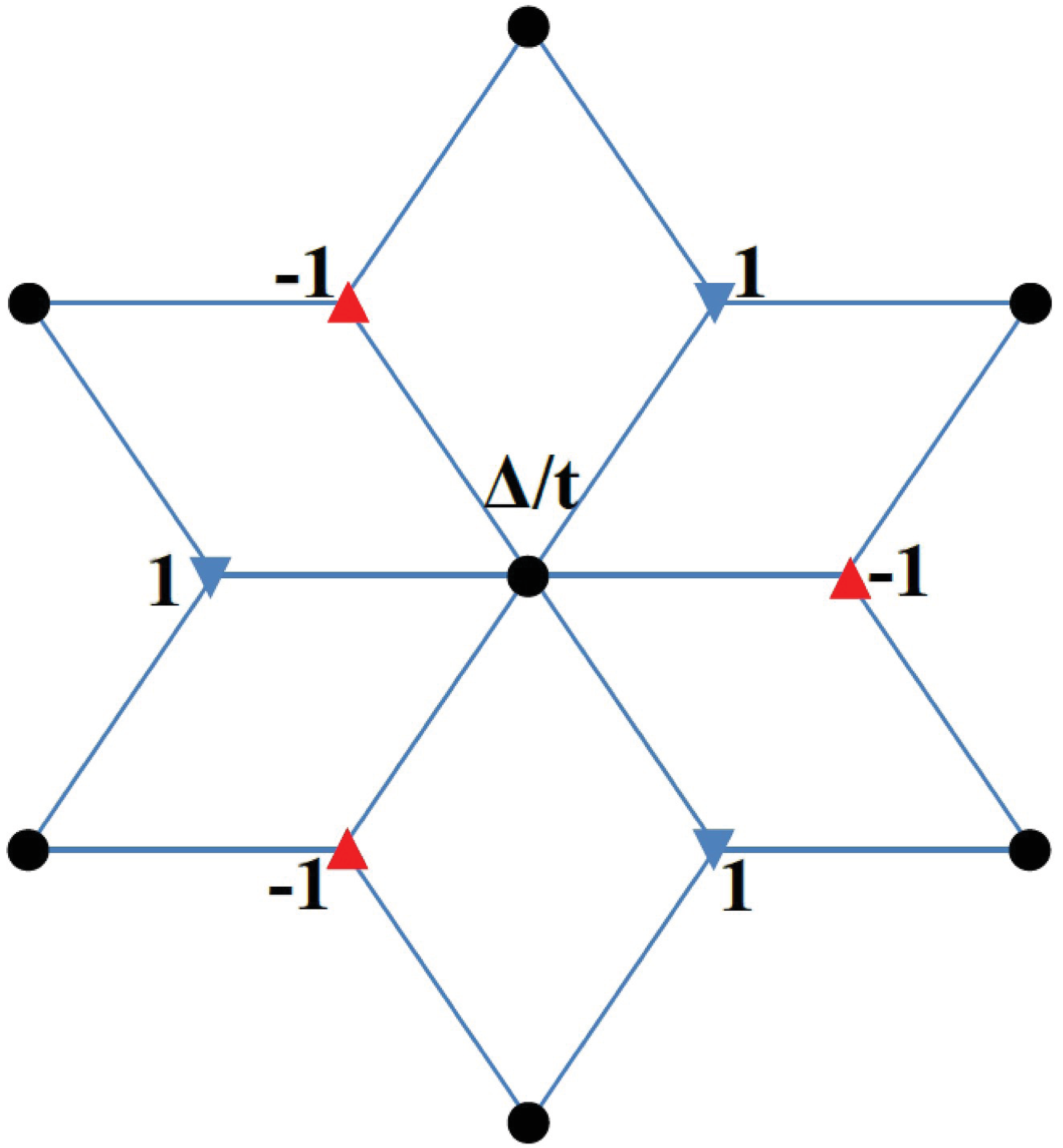}
\hspace{0.3cm}\includegraphics[width=2.3cm,height=1.955cm]{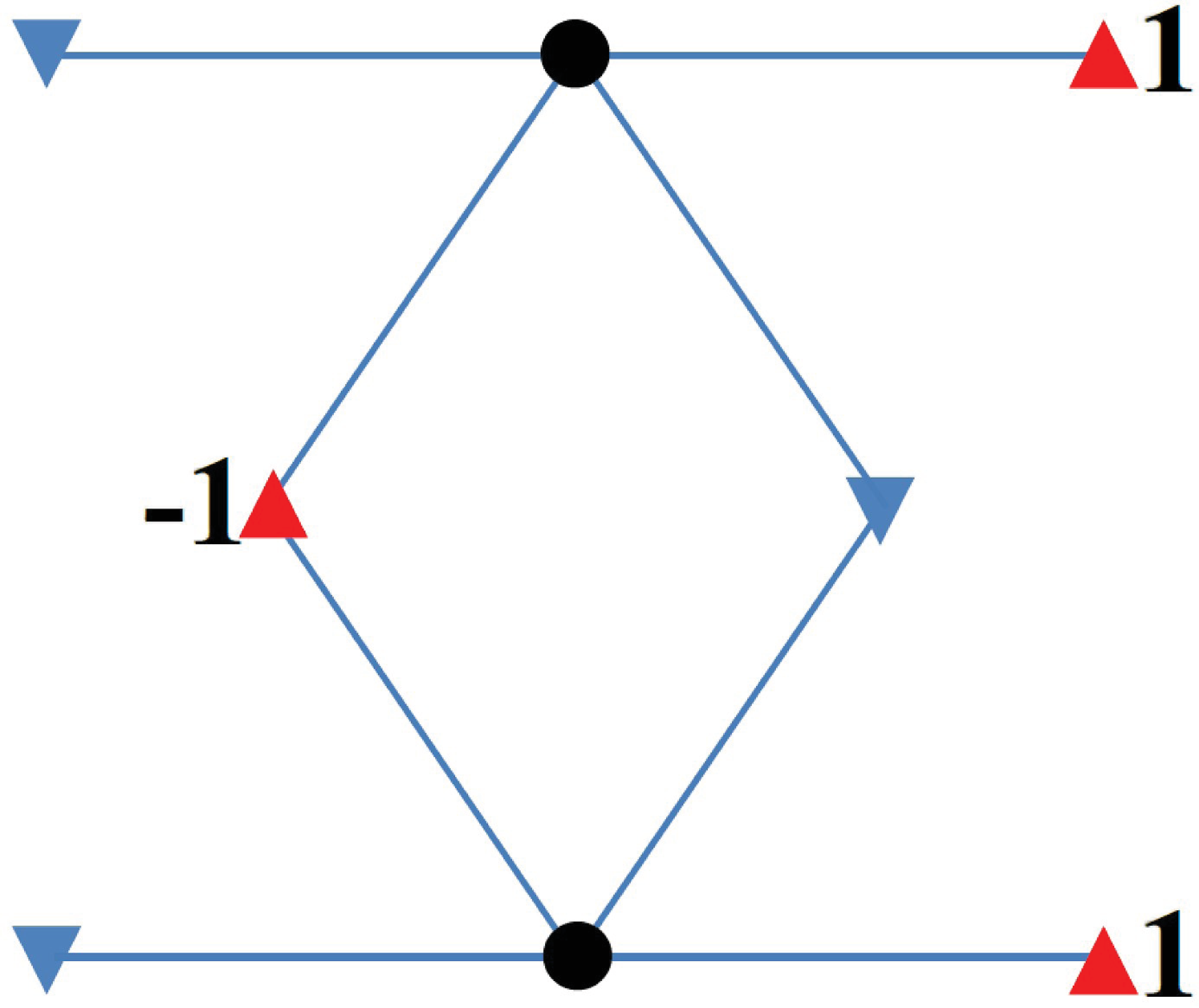}
\hspace{0.3cm}\includegraphics[width=2.3cm,height=1.976cm]{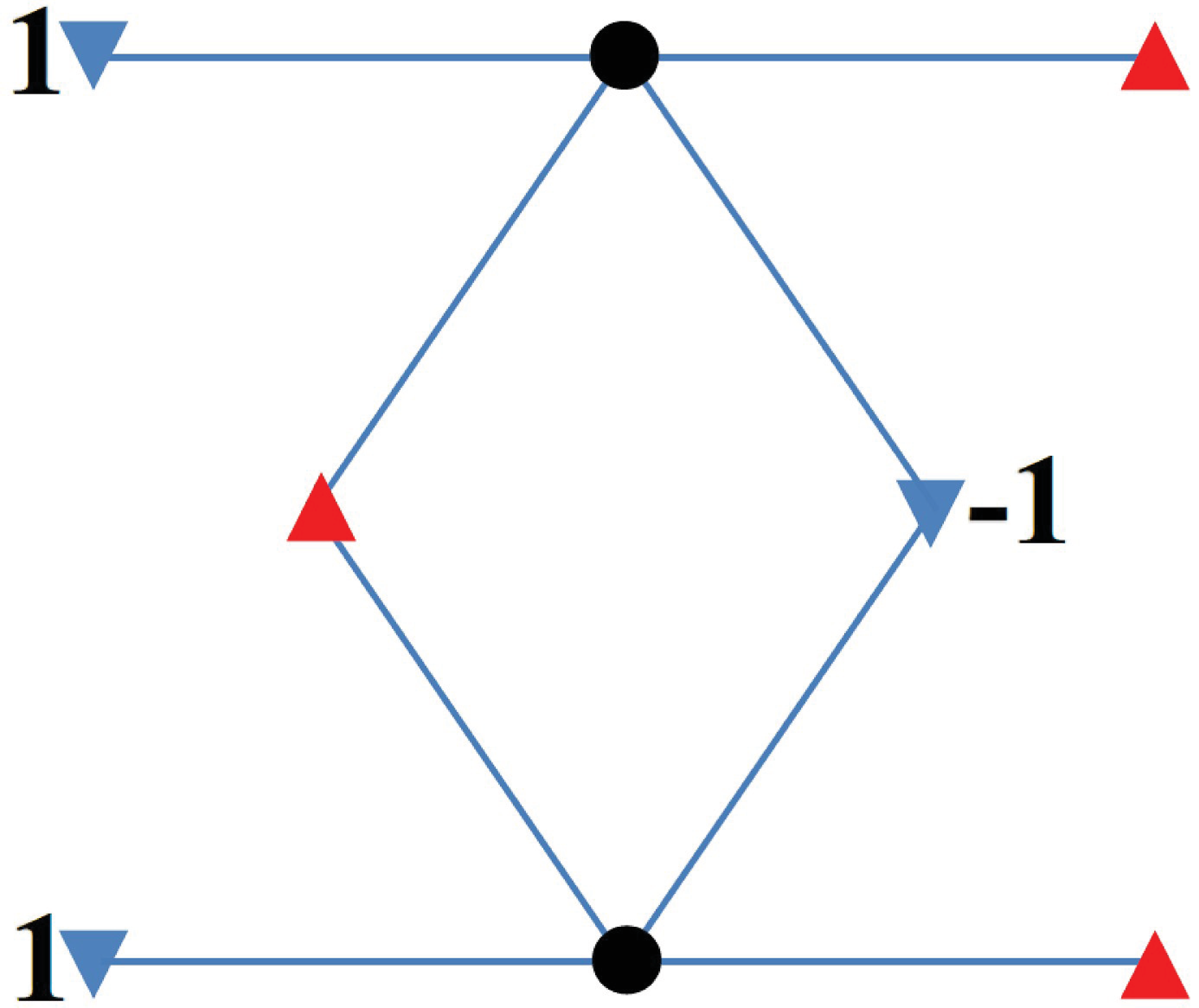}\\
\vspace{0.50cm}
\hspace{-3.6cm} {\textbf{(d)}} \hspace{4.2cm}{\textbf{(e)}}\\
\hspace{0cm}\includegraphics[width=4.0cm,height=3.763cm]{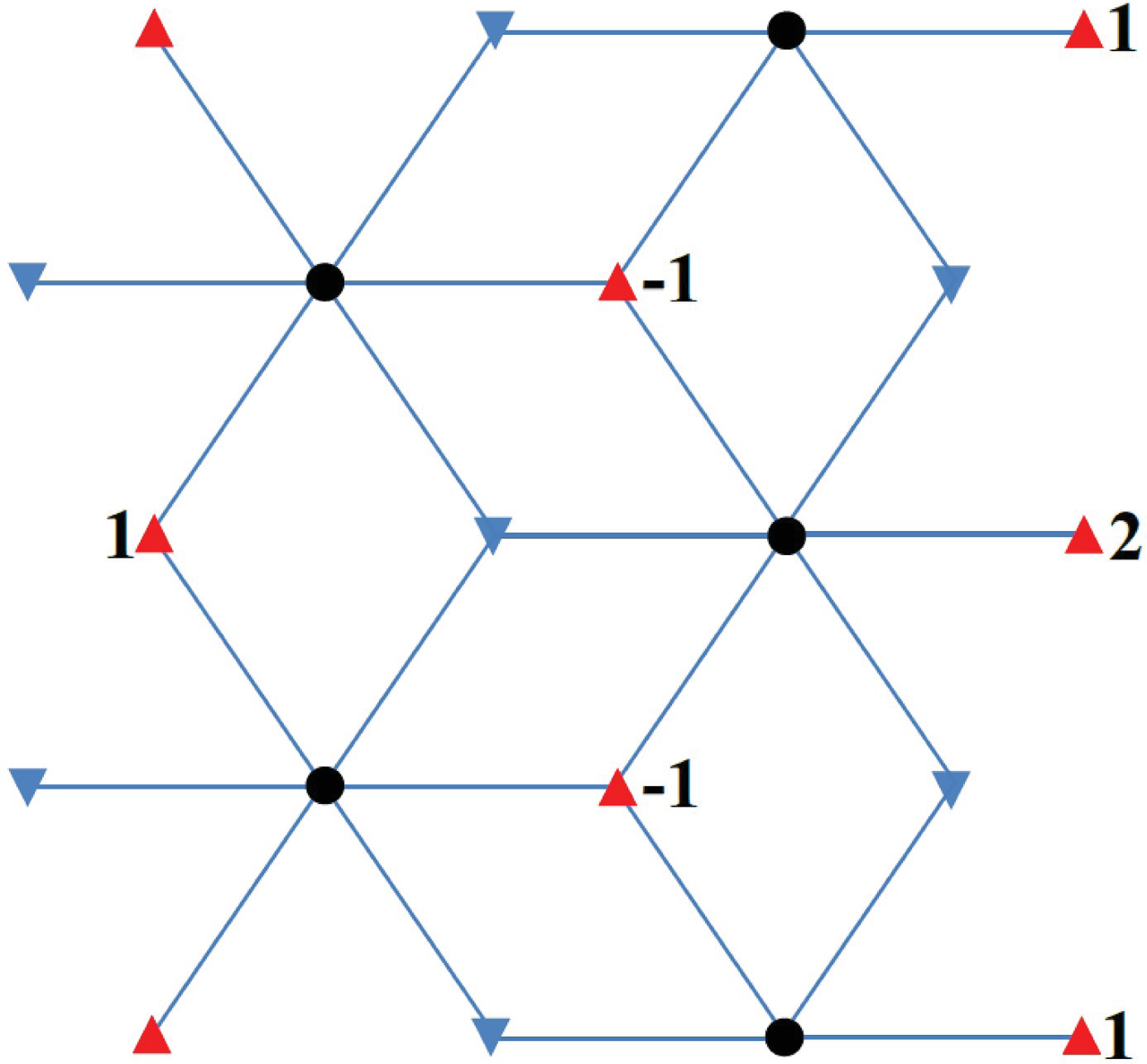}
\hspace{0.5cm}\includegraphics[width=4.0cm,height=3.801cm]{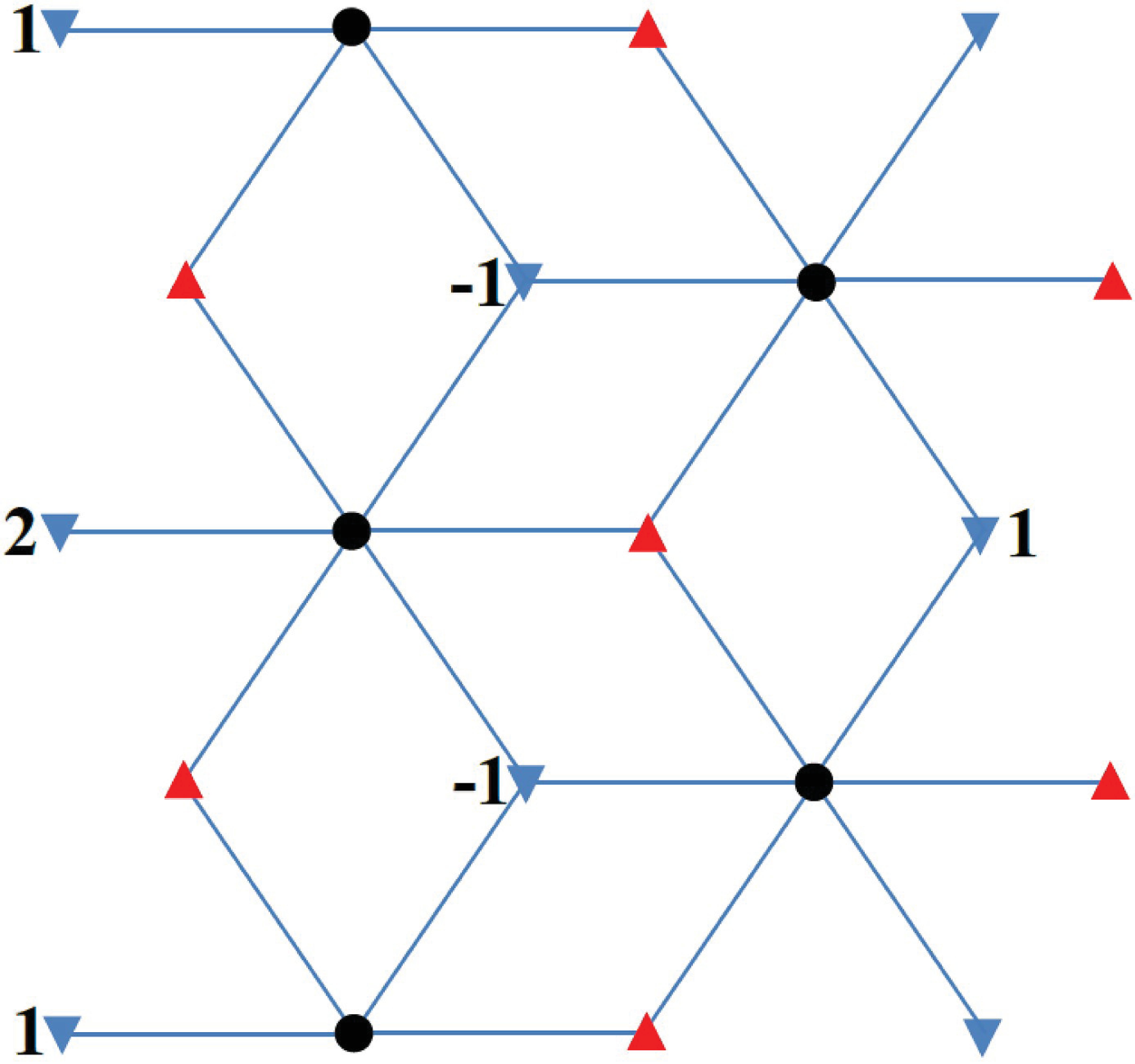}
\vspace{-0.10cm}
\caption{CLS for flat bands of the minimal zigzag dice lattice ribbons in zero magnetic field, up to normalization factors. These CLS separately correspond to (a) the $E=0$ flat band associated with a fully coordinated B sublattice site of the unit cell, (b) [(c)] the $E=\Delta$ [$E=-\Delta$] flat band of the minimal CA-in ribbon, and (d) [(e)] the $E=\Delta$ [$E=-\Delta$] flat band of the minimal CA-off ribbon. The nonzero number beside a site represents the amplitude of the CLS on that site. The sites of the ribbon lattice not shown and the sites without a number beside them contribute zero weight to the CLS.}
\end{figure}

The CLS for the flat bands of the minimal CA-in and minimal CA-off ribbons are constructed in the same manner. Since Eq.(12) for the zero-energy flat band of the minimal CA-in ribbon and Eq.(15) for the two zero-energy flat bands of the minimal CA-off ribbon are identical to Eq.(A8), the CLS shown in Fig.4(a) applies to all these flat bands.

Both the minimal CA-in and the minimal CA-off ribbons have two additional flat bands at $\pm\Delta$. From Eq.(13) for the flat band at $E=\Delta$ of the minimal CA-in ribbon, the CLS associated to an A$_{1}$ site at $y=Y$ is
\begin{eqnarray}
B_{Y}^{\dagger}&=&N_{0}\sum_{k}e^{-ikY}[-ta_{1k}^{\dagger}+t'a_{2k}^{\dagger}]   \notag \\
&=&\frac{1}{\sqrt{3}}(-a_{1Y}^{\dagger}+a_{2,Y+\frac{a}{2}}^{\dagger}+a_{2,Y-\frac{a}{2}}^{\dagger}).
\end{eqnarray}
From Eq.(14) for the flat band at $E=-\Delta$, the CLS associated to a C$_{2}$ site at $y=Y$ is
\begin{eqnarray}
C_{Y}^{\dagger}&=&N_{0}\sum_{k}e^{-ikY}[t'c_{1k}^{\dagger}-tc_{2k}^{\dagger}]   \notag \\
&=&\frac{1}{\sqrt{3}}(c_{1,Y+\frac{a}{2}}^{\dagger}+c_{1,Y-\frac{a}{2}}^{\dagger}-c_{2Y}^{\dagger}).
\end{eqnarray}
As shown in Figs. 4(b) and 4(c), $B_{Y}^{\dagger}$ and $C_{Y}^{\dagger}$ create 3-site stars (equilateral triangles, ignoring the enclosed sites of zero weight) on three nearest sites of the same rim sublattice.

From Eq.(16), we get the CLS for the $E=\Delta$ flat band of the minimal CA-off ribbon
\begin{eqnarray}
D_{Y}^{\dagger}&=&N_{0}\sum_{k}e^{-ikY}[t^{2}a_{1k}^{\dagger}-tt'a_{2k}^{\dagger}+t'^{2}a_{3k}^{\dagger}]   \notag \\
&=&\frac{1}{3}(a_{1Y}^{\dagger}-a_{2,Y+\frac{a}{2}}^{\dagger}-a_{2,Y-\frac{a}{2}}^{\dagger}   \notag \\ &&+a_{3,Y+a}^{\dagger}+a_{3,Y-a}^{\dagger}+2a_{3,Y}^{\dagger}).
\end{eqnarray}
For the CLS of the $E=-\Delta$ flat band, we get from Eq.(17)
\begin{eqnarray}
E_{Y}^{\dagger}&=&N_{0}\sum_{k}e^{-ikY}[t'^{2}c_{1k}^{\dagger}-tt'c_{2k}^{\dagger}+t^{2}c_{3k}^{\dagger}]  \notag \\
&=&\frac{1}{3}(c_{1,Y+a}^{\dagger}+c_{1,Y-a}^{\dagger}+2c_{1,Y}^{\dagger}  \notag \\
&&-c_{2,Y+\frac{a}{2}}^{\dagger}-c_{2,Y-\frac{a}{2}}^{\dagger}+c_{3Y}^{\dagger}).
\end{eqnarray}
As shown in Figs. 4(d) and 4(e), $D_{Y}^{\dagger}$ and $E_{Y}^{\dagger}$ create 6-site equilateral triangular clusters (3 sites along each edge, ignoring the enclosed sites of zero weight) with a rim sublattice site on an edge as a vertex and all the sites belong to the same rim sublattice.

Each CLS for the minimal CA-in ribbon and minimal CA-off ribbon contains sites not contained in other CLS of the same flat band. The CLS constructed for each flat band are therefore linearly independent and constitute a complete basis set.

The CLS of 7-site stars shown in Fig.4(a) are the same as the CLS in the bulk dice lattice. They are well-defined CLS for each fully coordinated B sublattice site, for both $\Delta=0$ and $\Delta\ne0$. The remaining four groups of CLS, as shown in Figs. 4(b)-4(e), are localized on A or C rim sublattice sites and are related to the edge states of corresponding wide ribbons \cite{hao22}. As is clear from the evolution from Figs. 4(b) and 4(c) for the minimal CA-in ribbon to Figs. 4(d) and 4(e) for the minimal CA-off ribbon, if we increase further the widths of the CA ribbons, the spatial extension of the corresponding CLS will grow alongside and lead to equilateral triangular CLS with three edges much longer than the lattice parameter. Therefore, these triangular CLS are well-defined CLS only for narrow CA ribbons, such as the minimal CA-in and CA-off ribbons studied here. This distinguishes the minimal CA-in and CA-off ribbons from wide CA ribbons.

While the flat band states are peculiar by themselves, the intense interest on flat bands also come from the prospective novel many-body phases realizable therein, such as intrinsic magnetism, superconductivity or superfluidity, and phases with nontrivial topological order. For example, by including both electronic correlation and spin-orbit coupling, the dice model was shown to lead to spontaneous ferromagnetic or ferrimagnetic order and quantum anomalous Hall phases \cite{wang11,soni21}. Studies on the non-minimal AC-in and AC-off ribbons also found ferrimagnetic order \cite{soni20}. It is highly intriguing to study the many-body phases that may be realized in the flat bands of the minimal CA-in and CA-off ribbons.

\section{electronic spectra in nonzero magnetic field}

As shown in Appendix A and Fig.5, two of the four minimal zigzag ribbons, the minimal CA-in and CA-off ribbons, have fully pinched spectrum at $f=1/2$ for $\Delta=0$. For the minimal CA-in ribbon, there are three zero-energy flat bands and two additional flat bands, at $\sqrt{6}t$ and $-\sqrt{6}t$. For the minimal CA-off ribbon, there are four zero-energy flat bands, two flat bands at $\sqrt{6}t$ and another pair of flat bands at $-\sqrt{6}t$. While the fully pinched spectrum in Fig.5(i) for the minimal AC-in ribbon is well known \cite{vidal00}, it is the first time that other narrow dice lattice ribbons are found to host fully pinched spectrum.

A nonzero $\Delta$ makes most of the flat bands for $\Delta=0$ nonflat, as shown in the right column of Fig.5, except for one flat band at $\Delta$ and another at $-\Delta$ for the minimal CA-in and minimal CA-off ribbons, which are split off from the zero-energy flat band for $\Delta=0$. For the minimal BB-in ribbon and the minimal BB-off ribbon, the spectra are never fully pinched at completely flat bands, for any $\Delta$.

\begin{figure}[H]\label{fig5}
\centering
\hspace{-2.95cm} {\textbf{(a)}} \hspace{3.8cm}{\textbf{(b)}}\\
\hspace{0cm}\includegraphics[width=4.2cm,height=3.5cm]{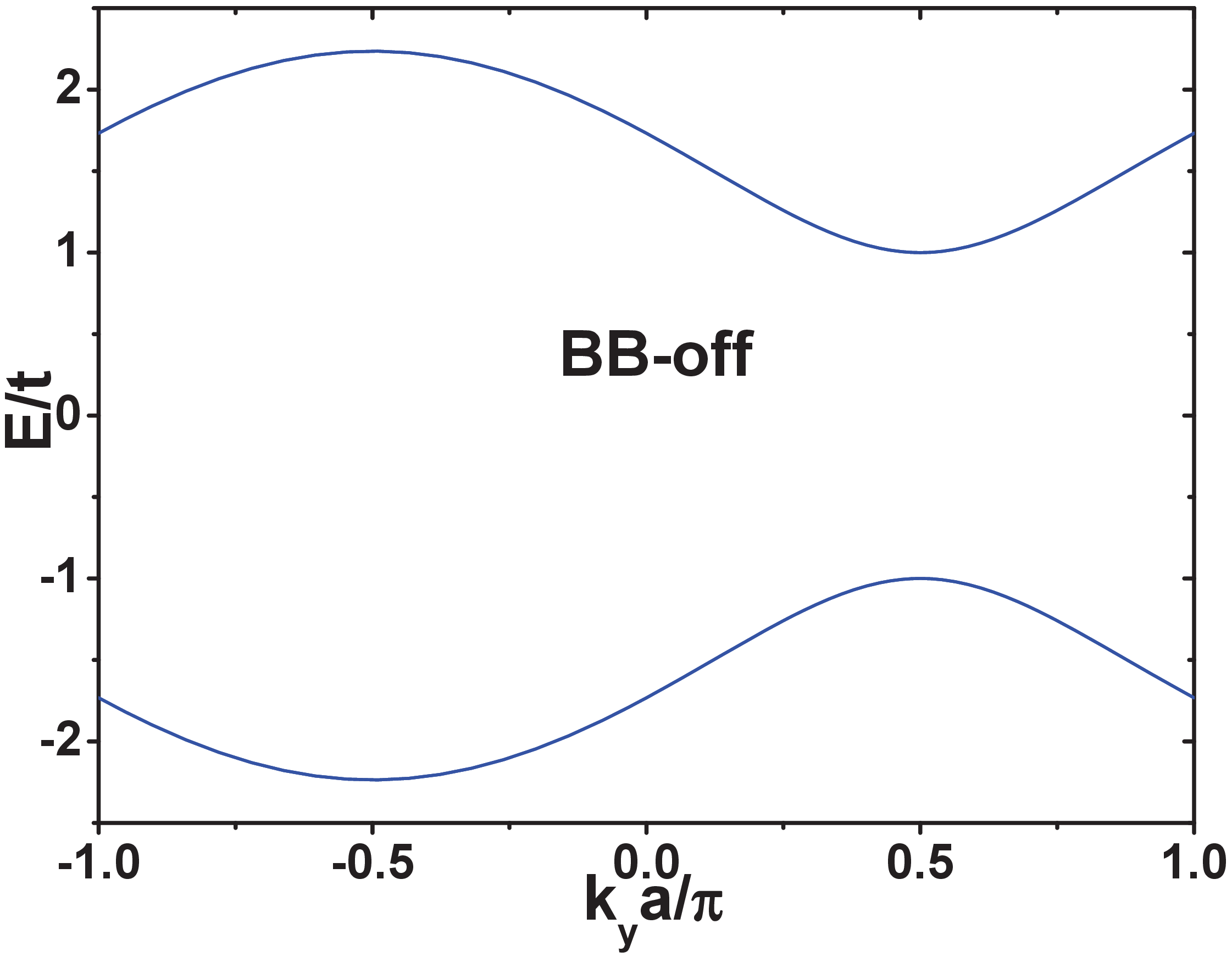}
\includegraphics[width=4.2cm,height=3.5cm]{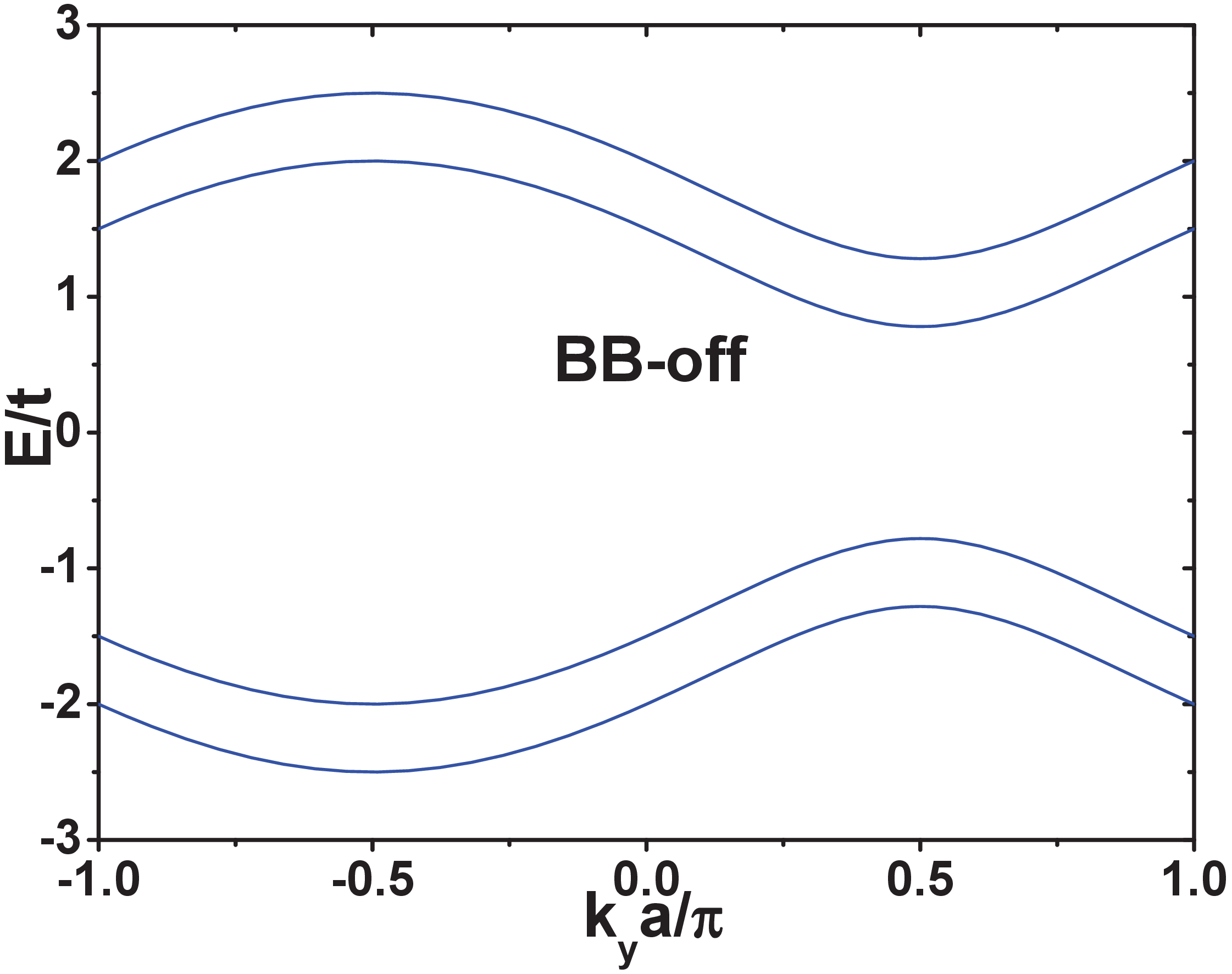} \\
\vspace{-0.10cm}
\hspace{-2.95cm} {\textbf{(c)}} \hspace{3.8cm}{\textbf{(d)}}\\
\hspace{0cm}\includegraphics[width=4.2cm,height=3.5cm]{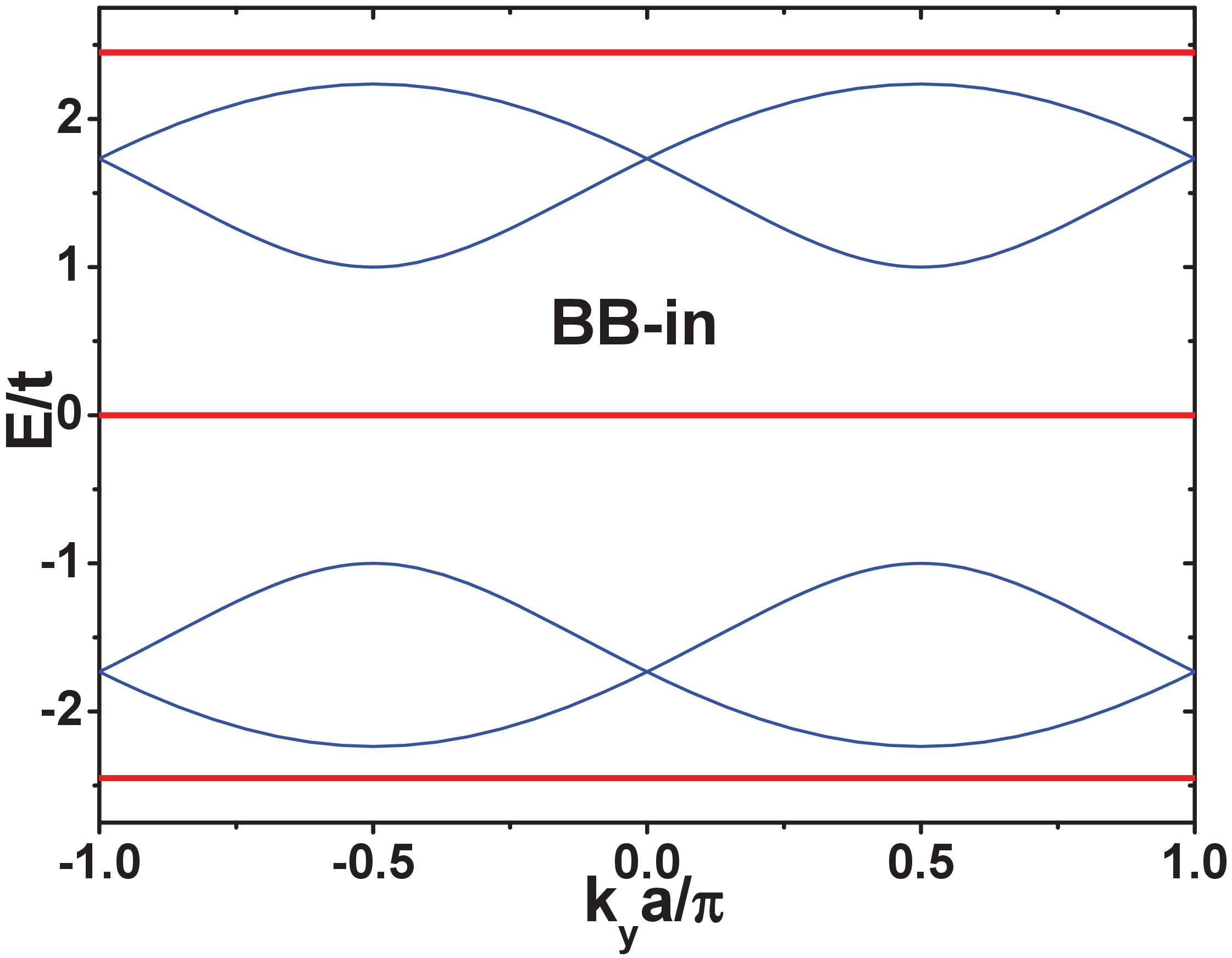}
\includegraphics[width=4.2cm,height=3.5cm]{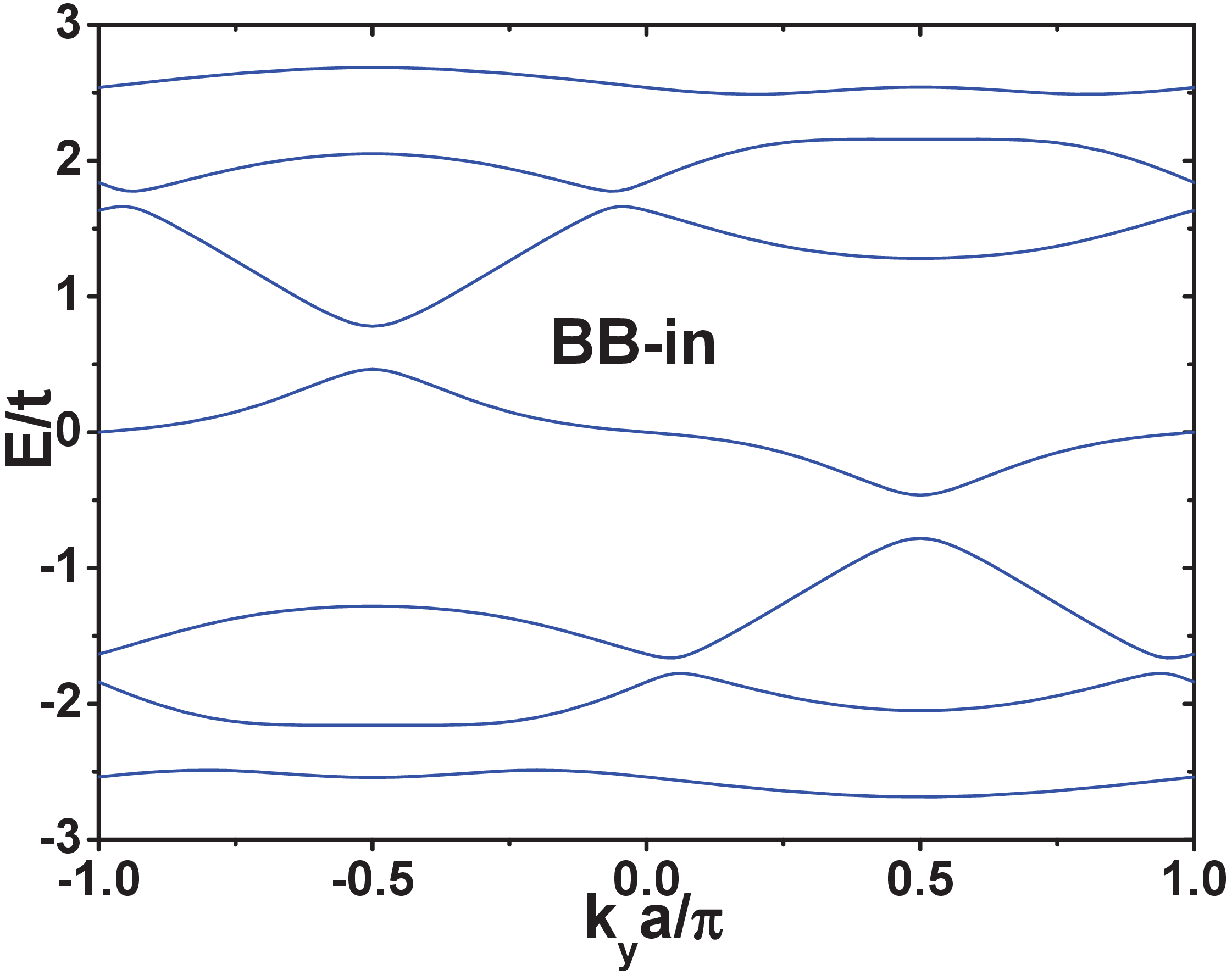}\\
\vspace{-0.10cm}
\hspace{-2.95cm} {\textbf{(e)}} \hspace{3.8cm}{\textbf{(f)}}\\
\hspace{0cm}\includegraphics[width=4.2cm,height=3.5cm]{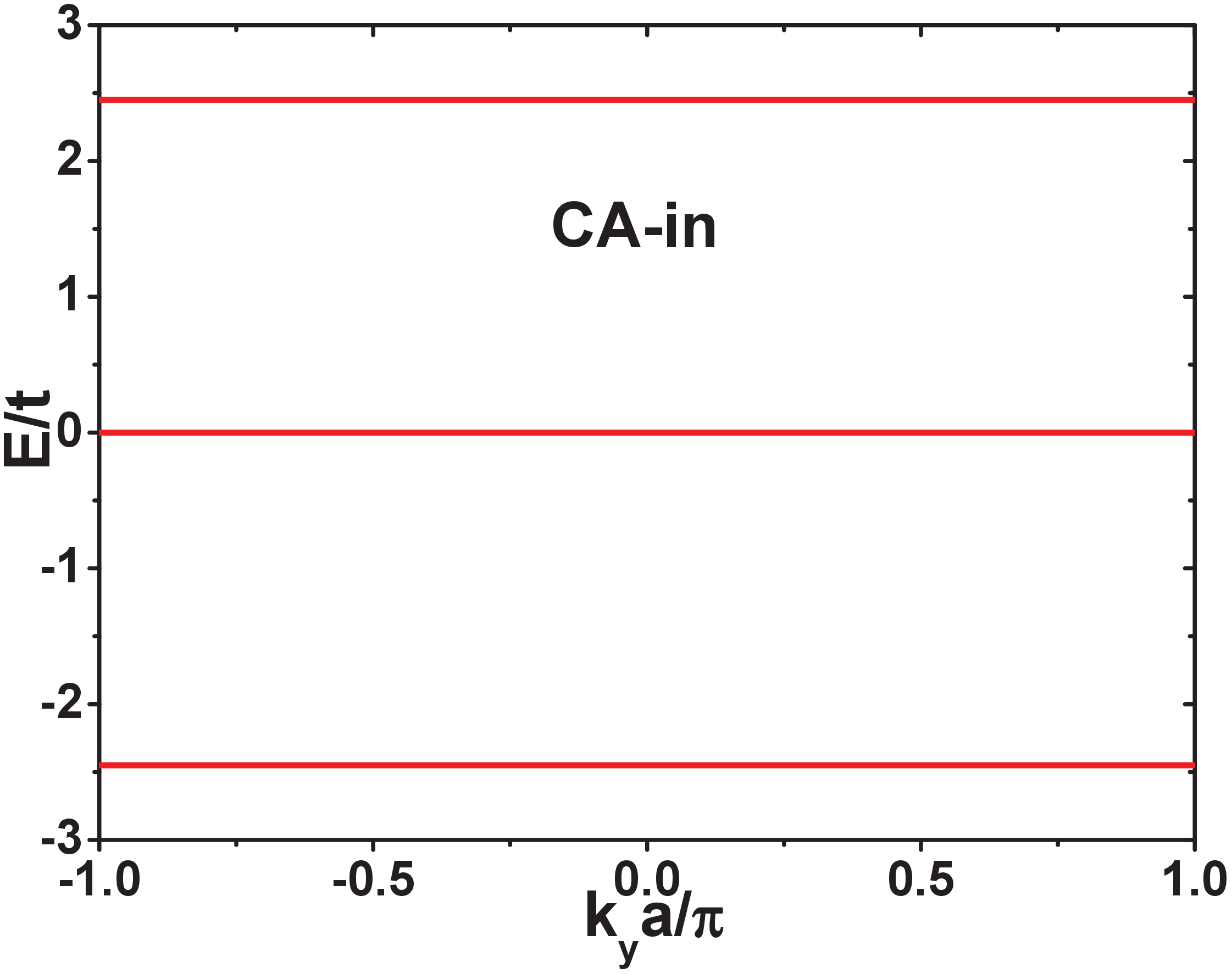}
\includegraphics[width=4.2cm,height=3.5cm]{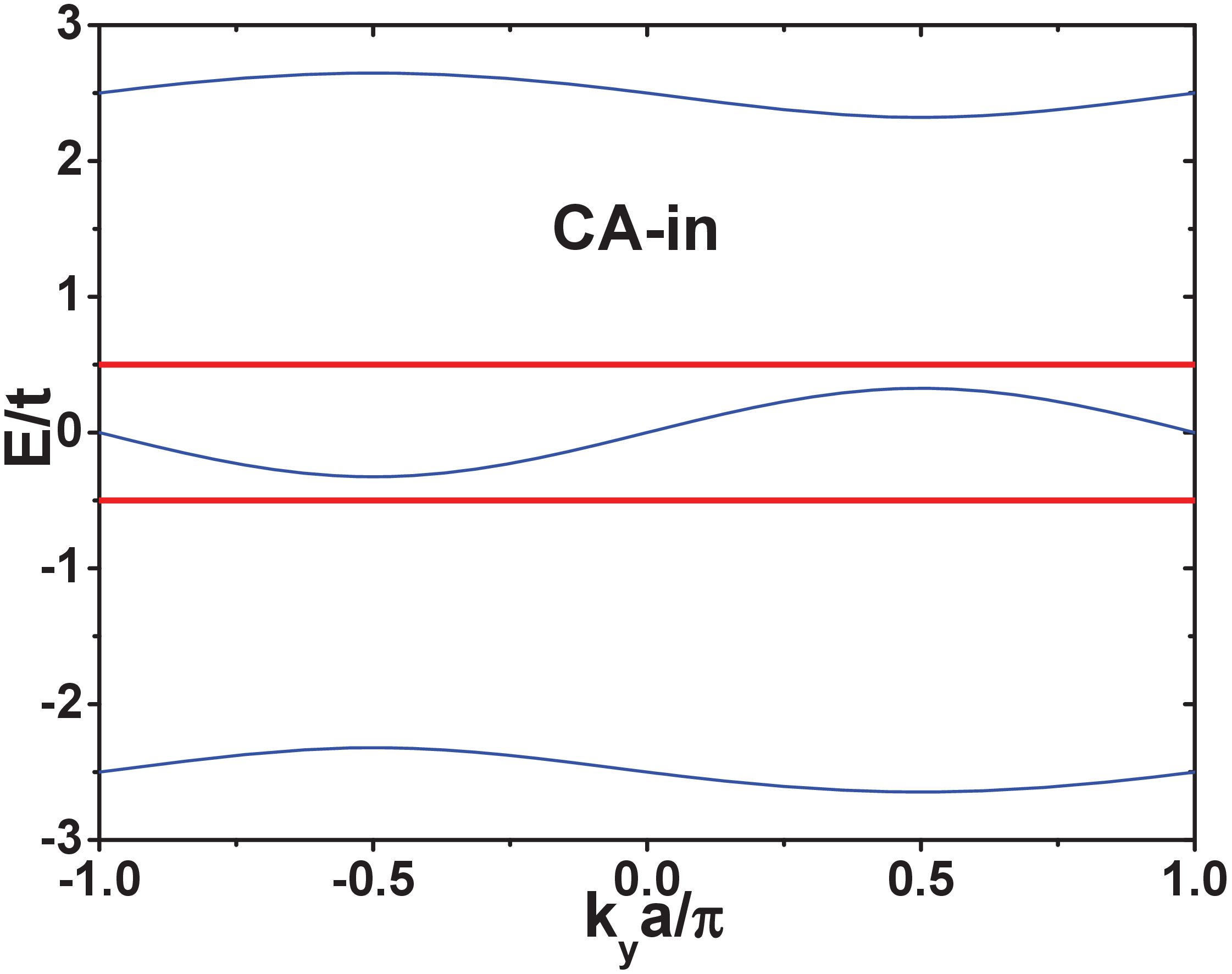} \\
\vspace{-0.10cm}
\hspace{-2.95cm} {\textbf{(g)}} \hspace{3.8cm}{\textbf{(h)}}\\
\hspace{0cm}\includegraphics[width=4.2cm,height=3.5cm]{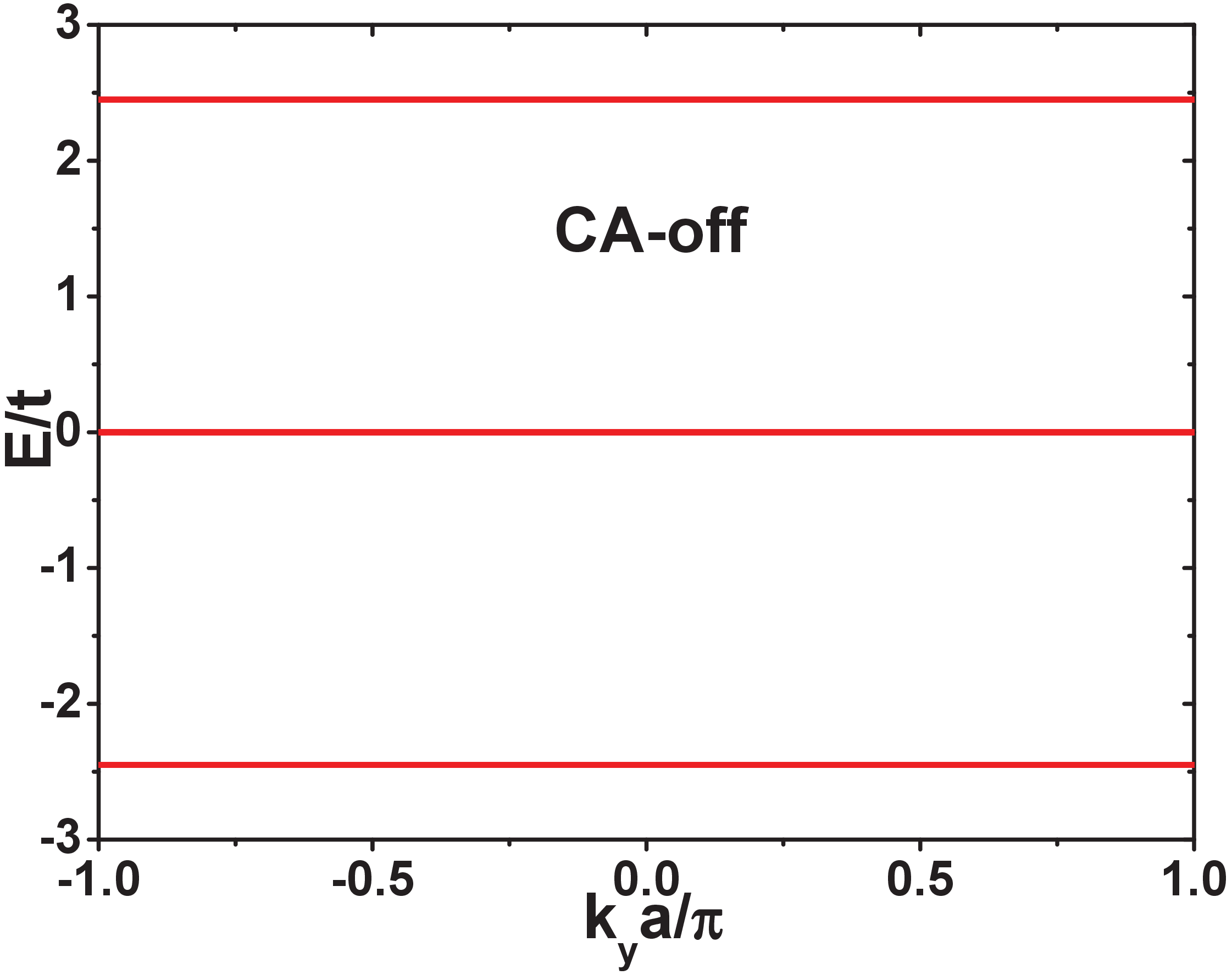}
\includegraphics[width=4.2cm,height=3.5cm]{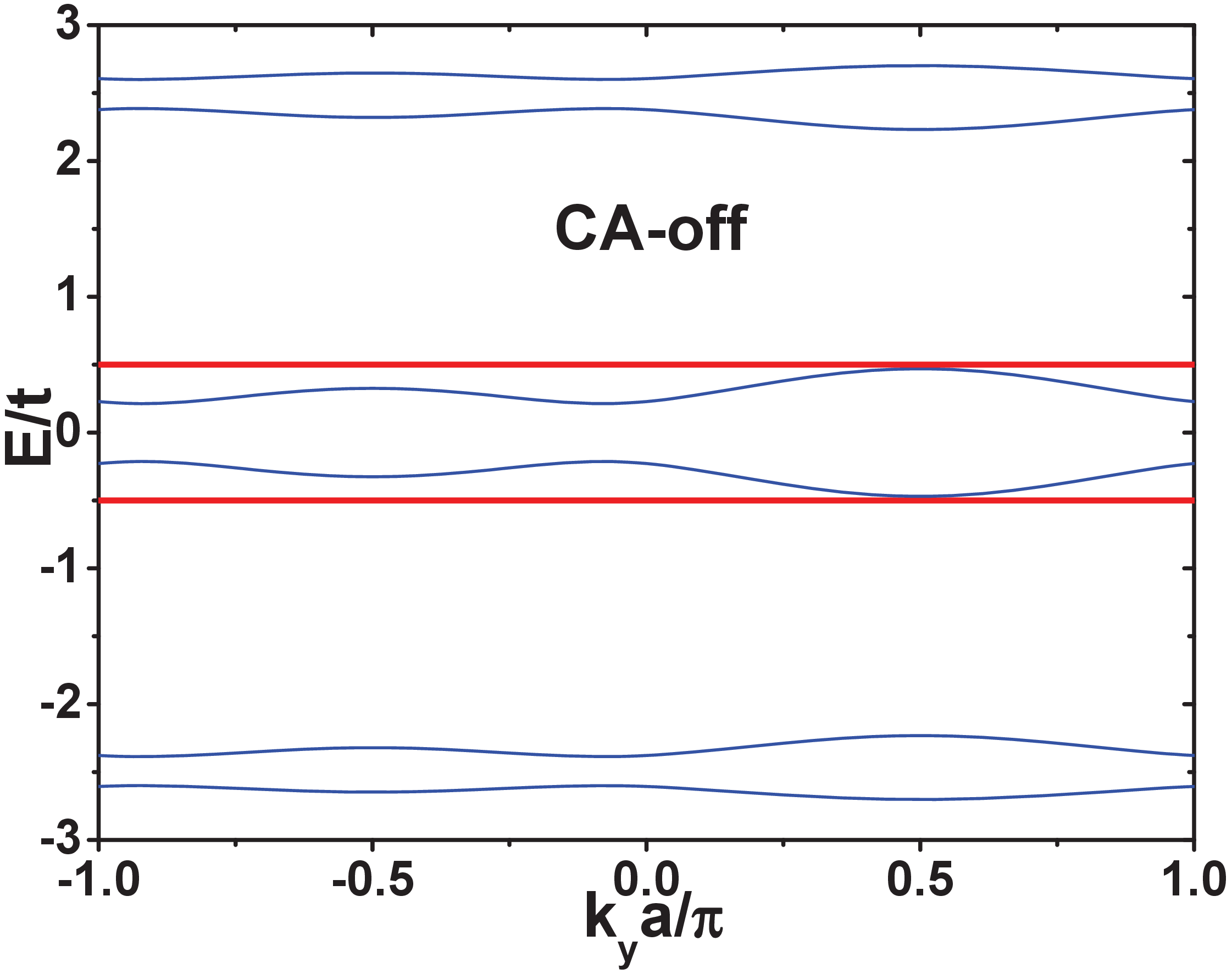} \\
\vspace{-0.10cm}
\hspace{-2.95cm} {\textbf{(i)}} \hspace{3.8cm}{\textbf{(j)}}\\
\hspace{0cm}\includegraphics[width=4.2cm,height=3.5cm]{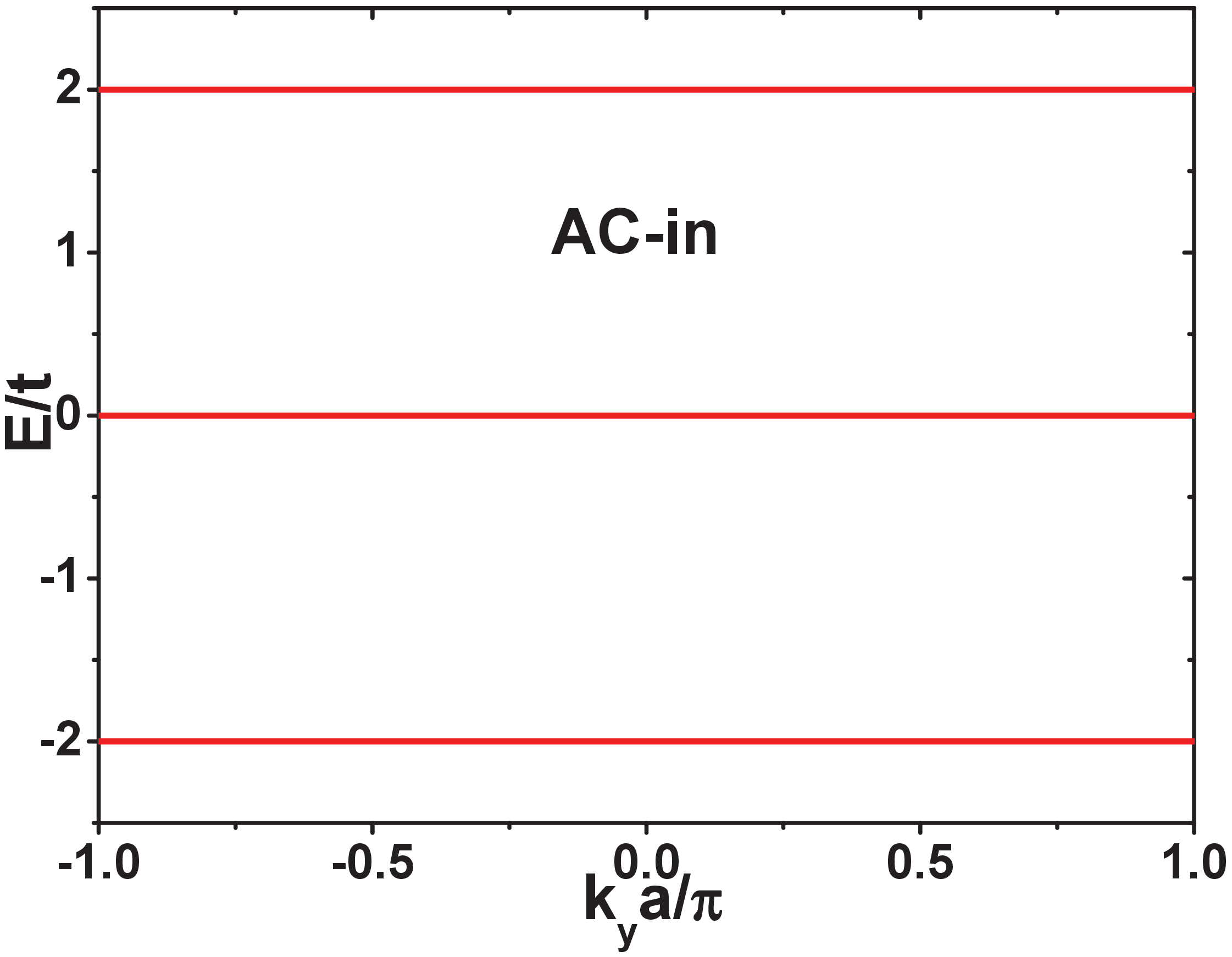}
\includegraphics[width=4.2cm,height=3.5cm]{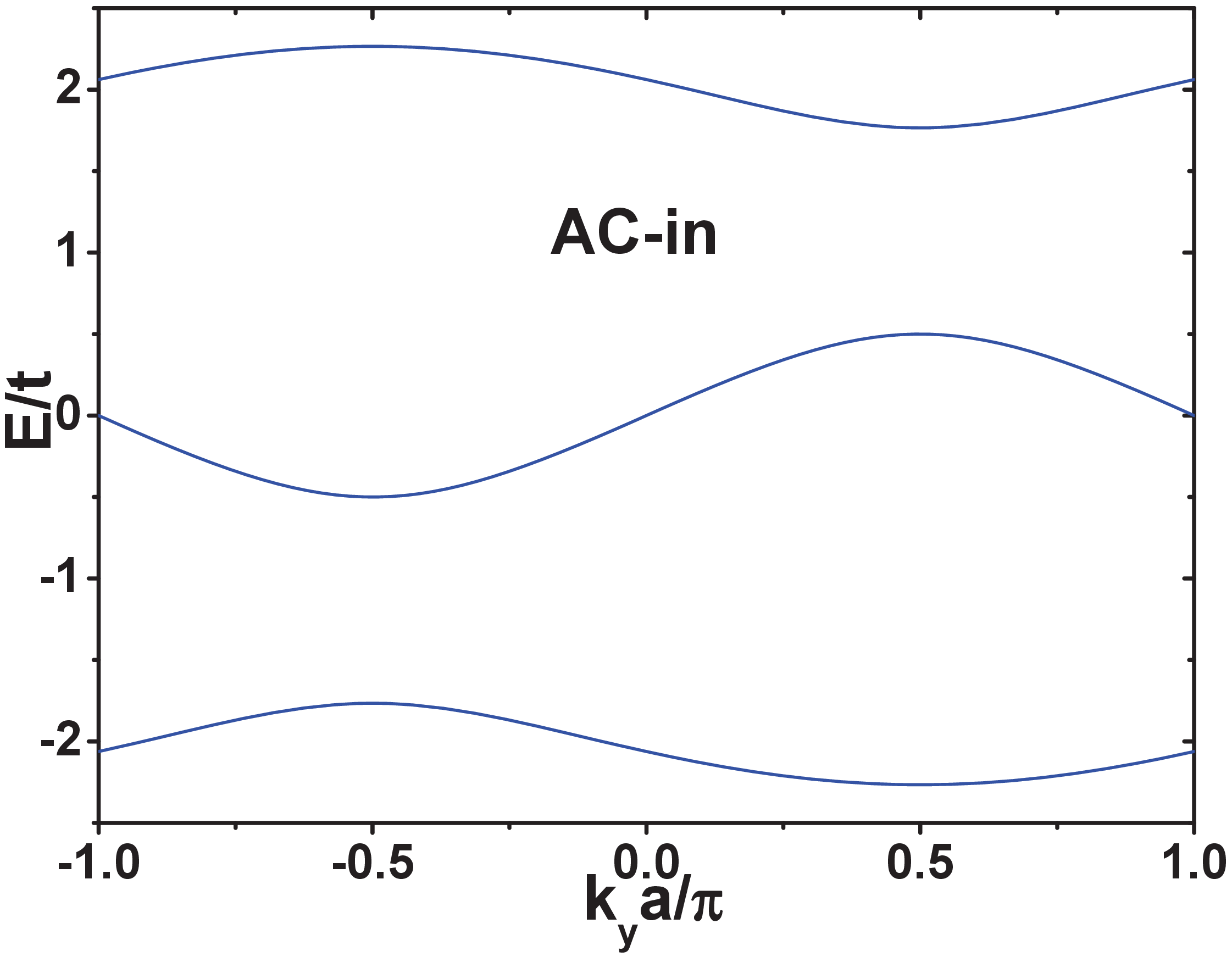}
\caption{Band structures of the five minimal zigzag dice lattice ribbons, in a magnetic field $f=\frac{1}{2}$, for the unbiased and symmetrically biased dice models. The left column (a, c, e, g, i) and the right column (b, d, f, h, j) are separately the band structures for $\Delta=0$ and for $\Delta=0.5t$, of the five ribbons defined in Fig.2 in sequence. The flat bands are highlighted in bold and plotted in red.}
\end{figure}

Comparing the minimal AC-in, minimal CA-in, and minimal CA-off ribbons with other minimal zigzag dice lattice ribbons, they are the only cases in which all the B sublattice sites contained therein coordinate with an even number of rim sublattice sites. All the other minimal zigzag dice lattice ribbons contain B sublattice sites with odd coordination numbers. When we consider wider zigzag dice lattice ribbons, we see that the CA-in and CA-off ribbons of any widths always have only evenly coordinated B sublattice sites, namely all the B sites are fully coordinated with six rim sublattice sites. We therefore speculate that the CA-in and CA-off ribbons of any width have completely pinched energy spectrum, at $0$ and $\pm\sqrt{6}t$, for $f=1/2$ and $\Delta=0$. Explicit numerical calculations for many widths of the CA-in and CA-off ribbons all agree with this conjecture. We therefore have found a series of q-1D structures that have fully pinched spectra in a proper magnetic field.

In the following, we study the dynamical evolutions from localized initial states in the minimal CA-in and CA-off ribbons with fully pinched energy spectra. As we will see, they lead to compact AB cages. Besides AB cages known for the bulk dice lattice, we also identify new compact AB cages unique to these pair of minimal zigzag ribbons. We also construct the CLS from the eigenvectors of the flat bands in the fully-pinched spectrum, from which we uncover interesting qualitative changes brought to the flat band states by the magnetic field.

\subsection{AB cages of the minimal CA-in and CA-off ribbons}

We follow the procedure of Vidal et al \cite{vidal98} and start from a wave packet localized at a single site of the unit cell as the initial state $|0\rangle$. By acting repeatedly the full real-space Hamiltonian $\hat{H}$ of the ribbon, for $f=1/2$ and $\Delta=0$, we get a series of normalized states $|n\rangle$ that satisfy the three-term recurrence relation
\begin{equation}
\hat{H}|n\rangle=b_{n}|n-1\rangle+a_{n}|n\rangle+b_{n+1}|n+1\rangle.
\end{equation}
$n=0,1,2,...$, and $b_{0}=0$. Because the lattice is bipartite, the diagonal terms vanish and $a_{n}=0$. Taking $b_{n+1}>0$ for all $n$, the states generated in the iteration are uniquely determined by normalization. If for some $n$ the corresponding $b_{n+1}=0$, the iteration leads to a chain of $n+1$ states, which form a closed subspace of the full Hamiltonian. In the basis of $(|0\rangle,|1\rangle,...,|n\rangle)$, the Hamiltonian matrix for this subspace consists of two identical subdiagonals with $n$ elements $b_{i}$ ($i=1,...,n$). Diagonalizing this Hamiltonian gives the eigenvalues $E_{\alpha}$ and eigenvectors $|\psi_{\alpha}\rangle$ ($\alpha=1,2,...,n+1$) of this subspace. Representing the initial state $|0\rangle$ as a linear combination of $|\psi_{\alpha}\rangle$ ($\alpha=1,2,...,n+1$), it is easy to see that the dynamical evolution will lead to a pulsating motion within a compact cluster of sites, which defines the AB cage of this initial state.


\subsubsection{AB cages of the minimal CA-in ribbon}

For the minimal CA-in ribbon of Fig.2(c), we firstly consider the initial state at the B$_{1}$ site of the unit cell, which we denote as
\begin{equation}
|0\rangle_{1}=|B_{1},0\rangle.
\end{equation}
The number after the comma is the $y$ coordinate of the site. The iteration gives $b_{1}=\sqrt{6}|t|$ and $b_{2}=0$. The other state of the generated two-dimensional subspace is
\begin{eqnarray}
&&|1\rangle_{1}=\frac{\text{sgn}(t)}{\sqrt{6}}[|C_{1},0\rangle+|A_{2},0\rangle
+e^{-i\gamma_{1}}|A_{1},\frac{a}{2}\rangle       \notag \\
&&+e^{i\gamma_{1}}|A_{1},-\frac{a}{2}\rangle-e^{i\gamma_{1}}|C_{2},\frac{a}{2}\rangle
-e^{-i\gamma_{1}}|C_{2},-\frac{a}{2}\rangle],
\end{eqnarray}
where $\gamma_{1}=\pi/4$, $\text{sgn}(t)=t/|t|$ gives the sign of $t$.
In the basis of Eq.(25) for $|0\rangle_{1}$ and Eq.(26) for $|1\rangle_{1}$, the $2\times2$ model turns out to have two eigenvalues $\pm\sqrt{6}|t|$. Taking the corresponding eigenvectors $|\psi_{\alpha}\rangle_{1}$ ($\alpha=\pm$) as
\begin{equation}
|\psi_{\alpha}\rangle_{1}=\frac{1}{\sqrt{2}}(\alpha|0\rangle_{1}+|1\rangle_{1}),
\end{equation}
the initial state is represented as
\begin{equation}
|0\rangle_{1}=\frac{1}{\sqrt{2}}(|\psi_{+}\rangle_{1}-|\psi_{-}\rangle_{1}).
\end{equation}
The dynamical evolution initiating from $|0\rangle_{1}$ is obtained by acting the time evolution operator $\exp(-i\hat{H}t'/\hbar)$ on it ($t'$ denotes the time), which gives a pulsating periodic motion of period $h/(\sqrt{6}|t|)$ within the seven-site star surrounding the initial $B_{1}$ site. This defines the AB cage for the B$_{1}$ site.

For the initial state at the C$_{1}$ site of the unit cell,
\begin{equation}
|0\rangle_{2}=|C_{1},0\rangle,
\end{equation}
the iteration gives $b_{1}=|t|$, $b_{2}=\sqrt{5}|t|$, and $b_{3}=0$. The other two basis states of the generated three-dimensional subspace are
\begin{equation}
|1\rangle_{2}=\text{sgn}(t)|B_{1},0\rangle,
\end{equation}
\begin{eqnarray}
|2\rangle_{2}&=&\frac{1}{\sqrt{5}}[|A_{2},0\rangle+e^{-i\gamma_{1}}|A_{1},\frac{a}{2}\rangle
+e^{i\gamma_{1}}|A_{1},-\frac{a}{2}\rangle     \notag \\
&&-e^{i\gamma_{1}}|C_{2},\frac{a}{2}\rangle
-e^{-i\gamma_{1}}|C_{2},-\frac{a}{2}\rangle].
\end{eqnarray}
In the three basis states of Eqs.(29)-(31), the $3\times3$ model has eigenvalues $E_{0}=0$ and $E_{\pm}=\pm\sqrt{6}|t|$. We take the three eigenvectors as ($\alpha=\pm$)
\begin{equation}
\begin{cases}
|\psi_{0}\rangle_{2}=\sqrt{\frac{5}{6}}|0\rangle_{2}-\sqrt{\frac{1}{6}}|2\rangle_{2}, \\
|\psi_{\alpha}\rangle_{2}=\frac{\alpha}{2\sqrt{3}}|0\rangle_{2}+\frac{1}{\sqrt{2}}|1\rangle_{2} +\frac{\sqrt{5}\alpha}{2\sqrt{3}}|2\rangle_{2}.
\end{cases}
\end{equation}
The above initial state is represented as
\begin{equation}
|0\rangle_{2}=\frac{1}{2\sqrt{3}}(\sqrt{10}|\psi_{0}\rangle_{2}+|\psi_{+}\rangle_{2}-|\psi_{-}\rangle_{2}).
\end{equation}
The time evolution of the state also gives a compact pulsating state, which defines the AB cage for the C$_{1}$ site. While it is still restricted to the seven-site cluster centering at the B$_{1}$ site of the unit cell, the time evolution is different from the AB cage of the B$_{1}$ site. Note that $|\psi_{\alpha}\rangle_{2}$ of Eq.(32) are the same eigenvectors as $|\psi_{\alpha}\rangle_{1}$ of Eq.(27), up to a sign factor.

For the initial state at the A$_{1}$ site of the unit cell,
\begin{equation}
|0\rangle_{3}=|A_{1},\frac{a}{2}\rangle,
\end{equation}
the iteration gives $b_{1}=\sqrt{2}|t|$, $b_{2}=2|t|$, and $b_{3}=0$.
The two new basis states of the generated three-dimensional subspace are
\begin{equation}
|1\rangle_{3}=\frac{\text{sgn}(t)}{\sqrt{2}}(e^{i\gamma_{1}}|B_{1},0\rangle+e^{-i\gamma_{1}}|B_{1},a\rangle),
\end{equation}
\begin{eqnarray}
|2\rangle_{3}&=&\frac{1}{2\sqrt{2}}[e^{i\gamma_{1}}(|C_{1},0\rangle+|A_{2},0\rangle)   \notag \\ &&+e^{-i\gamma_{1}}(|C_{1},a\rangle+|A_{2},a\rangle)+i(|A_{1},-\frac{a}{2}\rangle-|A_{1},\frac{3a}{2}\rangle)    \notag \\
&&-|C_{2},-\frac{a}{2}\rangle-|C_{2},\frac{3a}{2}\rangle].
\end{eqnarray}
In the basis of Eqs.(34)-(36), the $3\times3$ model has eigenvalues $E_{0}=0$ and $E_{\pm}=\pm\sqrt{6}|t|$. We take the eigenvectors as ($\alpha=\pm$)
\begin{equation}
\begin{cases}
|\psi_{0}\rangle_{3}=\sqrt{\frac{2}{3}}|0\rangle_{3}-\sqrt{\frac{1}{3}}|2\rangle_{3},  \\
|\psi_{\alpha}\rangle_{3}=\frac{\alpha}{\sqrt{6}}|0\rangle_{3}+\frac{1}{\sqrt{2}}|1\rangle_{3} +\frac{\alpha}{\sqrt{3}}|2\rangle_{3}.
\end{cases}
\end{equation}
In terms of these eigenstates, the above initial state is
\begin{equation}
|0\rangle_{3}=\frac{1}{\sqrt{6}}(2|\psi_{0}\rangle_{3}+|\psi_{+}\rangle_{3}-|\psi_{-}\rangle_{3}).
\end{equation}
Starting from an A$_{1}$ site, the dynamical evolution of the wave packet covers a larger area that amounts to a mergence of two seven-site stars centering at the two NN B$_{1}$ sites connecting to this A$_{1}$ site, which leads to a cluster containing 12 sites.
Because of the symmetry of the lattice and the model, the dynamical evolution of an initial state at A$_{2}$ (C$_{2}$) is equivalent to the evolution initiating from a localized wave packet at C$_{1}$ (A$_{1}$). Therefore, the AB cages for all sites of the CA-in ribbon are bounded and compact for $f=1/2$ and $\Delta=0$.

\subsubsection{AB cages of the minimal CA-off ribbon}

Straightforward analysis shows that all the AB cages for the minimal CA-off ribbon are also bounded and compact for $f=1/2$ and $\Delta=0$. For completeness, we list in what follows the states relevant to the evolutions initiating from the C$_{1}$, A$_{1}$, B$_{1}$, and C$_{2}$ sites of the unit cell of Fig. 2(d). The evolutions initiating from the remaining four sites of the unit cell follow by direct mapping in terms of the symmetry of the lattice and the model.

For the initial state at the C$_{1}$ site of the unit cell of Fig. 2(d)
\begin{equation}
|0\rangle_{4}=|C_{1},0\rangle,
\end{equation}
the iteration leads to $b_{1}=|t|$, $b_{2}=\sqrt{5}|t|$, and $b_{3}=0$. The two new basis states of the generated three-dimensional subspace are identical to Eqs.(30) and (31) for the minimal CA-in ribbon. The AB cage for this initial state is therefore equivalent to the AB cage described by Eq.(33) for the initial state of Eq.(29) for the minimal CA-in ribbon. For the initial state at the A$_{1}$ site of the unit cell of Fig.2(d)
\begin{equation}
|0\rangle_{5}=|A_{1},\frac{a}{2}\rangle,
\end{equation}
the three-term iteration, the generated closed subspace, the eigenstates, and the representation of the initial state in terms of the eigenstates, are all equivalent to the case for the initial state Eq.(34) for the minimal CA-in ribbon. For the initial state at the B$_{1}$ site of the unit cell of Fig.2(d)
\begin{equation}
|0\rangle_{6}=|B_{1},0\rangle,
\end{equation}
the three-term iteration, the generated closed subspace, the eigenstates, and the representation of the initial state in terms of the eigenstates, are all equivalent to the case for the initial state Eq.(25) for the minimal CA-in ribbon.

For an initial state at the C$_{2}$ site of the unit cell of Fig. 2(d),
\begin{equation}
|0\rangle_{7}=|C_{2},\frac{a}{2}\rangle,
\end{equation}
the iteration leads to $b_{1}=b_{2}=\sqrt{3}|t|$, $b_{3}=0$. The other two basis states of the generated three-dimensional subspace are
\begin{equation}
|1\rangle_{7}=\frac{\text{sgn}(t)}{\sqrt{3}}(|B_{2},\frac{a}{2}\rangle-e^{-i\gamma_{1}}|B_{1},0\rangle -e^{i\gamma_{1}}|B_{1},a\rangle),
\end{equation}
\begin{eqnarray}
|2\rangle_{7}&=&\frac{1}{3}(|A_{3},\frac{a}{2}\rangle+e^{-i\gamma_{1}}|C_{3},a\rangle +e^{i\gamma_{1}}|C_{3},0\rangle    \notag \\
&&-e^{-i\gamma_{1}}|C_{1},0\rangle-e^{i\gamma_{1}}|C_{1},a\rangle   \\
&&-|A_{1},-\frac{a}{2}\rangle-|A_{1},\frac{3a}{2}\rangle -i|C_{2},-\frac{a}{2}\rangle+i|C_{2},\frac{3a}{2}\rangle).   \notag
\end{eqnarray}
In the basis of Eqs.(42)-(44), the $3\times3$ model has the eigenvalues $E_{0}=0$ and $E_{\pm}=\pm\sqrt{6}|t|$. We take the three eigenvectors as ($\alpha=\pm$)
\begin{equation}
\begin{cases}
|\psi_{0}\rangle_{7}=\frac{1}{\sqrt{2}}(|0\rangle_{7}-|2\rangle_{7}),  \\
|\psi_{\alpha}\rangle_{7}=\frac{\alpha}{2}|0\rangle_{7}+\frac{1}{\sqrt{2}}|1\rangle_{7}+\frac{\alpha}{2}|2\rangle_{7}.
\end{cases}
\end{equation}
The above initial state is therefore represented as
\begin{equation}
|0\rangle_{7}=\frac{1}{2}(\sqrt{2}|\psi_{0}\rangle_{7}+|\psi_{+}\rangle_{7}-|\psi_{-}\rangle_{7}).
\end{equation}
Again, we get a compact AB cage from this initial state.

\subsubsection{Comparison to known results and generalization to wider CA ribbons}

According to Vidal et al \cite{vidal98,vidal00}, there are two types of AB cages both for the bulk dice lattice and for the minimal AC-in ribbon (i.e., the diamond chain lattice), which are related separately to the dynamical evolutions of states initiating from a hub sublattice site and from a rim sublattice site.
From the above analysis, there are three types of AB cages in the dynamical evolutions in the minimal CA-in ribbon, related separately to the initial states on the C$_{1}$-type (or A$_{2}$-type), A$_{1}$-type (or C$_{2}$-type), and B$_{1}$-type of sites of the ribbon lattice. For the minimal CA-off ribbon, there are four types of AB cages related separately to the initial states on the C$_{1}$-type (or A$_{3}$-type) and A$_{1}$-type (or C$_{3}$-type) of sites on the edge, $B_{1}$-type (or B$_{2}$-type) fully coordinated B sublattice sites, and C$_{2}$-type (or A$_{2}$-type) fully coordinated rim sublattice sites. While the minimal CA-in and CA-off ribbons have fully pinched spectrum identical to the bulk dice lattice, the additional AB cages for the edge sites distinguish the two new minimal zigzag ribbons from the bulk dice lattice.

As the widths of the CA-in or CA-off ribbons increase, no other types of sites emerge. Together with the previous result that all CA-in and CA-off ribbons have fully pinched spectra for $f=1/2$ and $\Delta=0$, all the CA-in and CA-off ribbons realize complete AB caging phase with only compact AB cages, for the parameters $f=1/2$ and $\Delta=0$. The experimental setups for exploring physics of the AB cages are therefore greatly expanded.

\subsection{CLS in minimal CA-in and CA-off ribbons}

In comparison to the analysis in Sec.IIIC for the CLS in the flat bands in zero magnetic field, it is interesting to study the CLS in the various flat bands in Fig. 5 for $f=\frac{1}{2}$. The compact eigenstates constructed in the previous section in studying the AB cages are eigenstates of the flat bands with the corresponding energy. They are natural CLS of the corresponding bands. On the other hand, in terms of analytical eigenvectors for the various flat bands, we can explicitly construct the corresponding CLS. In this way, we clarify the qualitative changes brought to the flat band states by the magnetic field.

\begin{figure}\label{fig6} \centering
\hspace{-2.0cm}{\textbf{(a)}} \hspace{3.0cm}{\textbf{(b)}}  \hspace{2.3cm}{\textbf{(c)}} \\
\hspace{-0.5cm}\includegraphics[width=3cm,height=2.446cm]{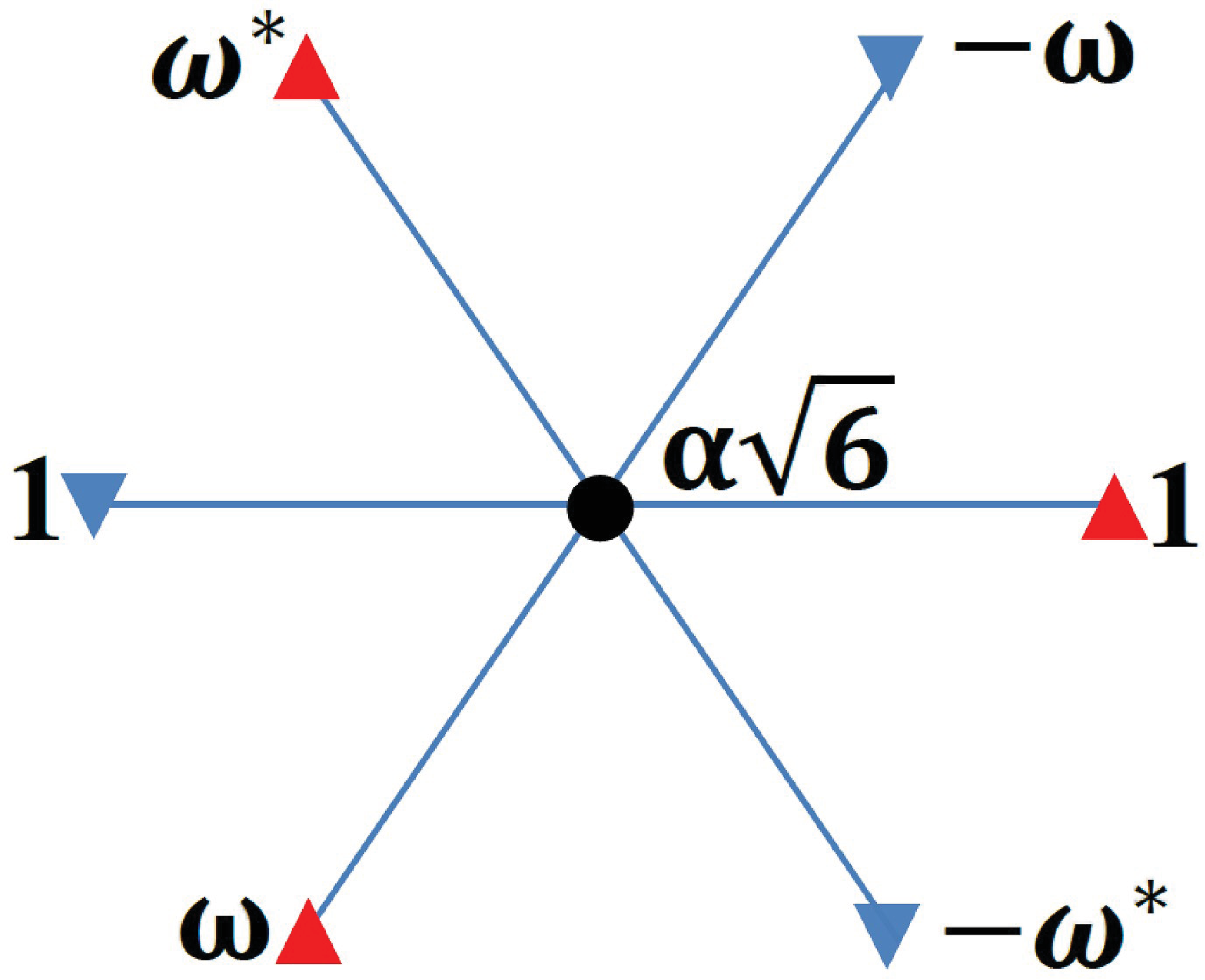}
\hspace{0.3cm}\includegraphics[width=2.3cm,height=1.691cm]{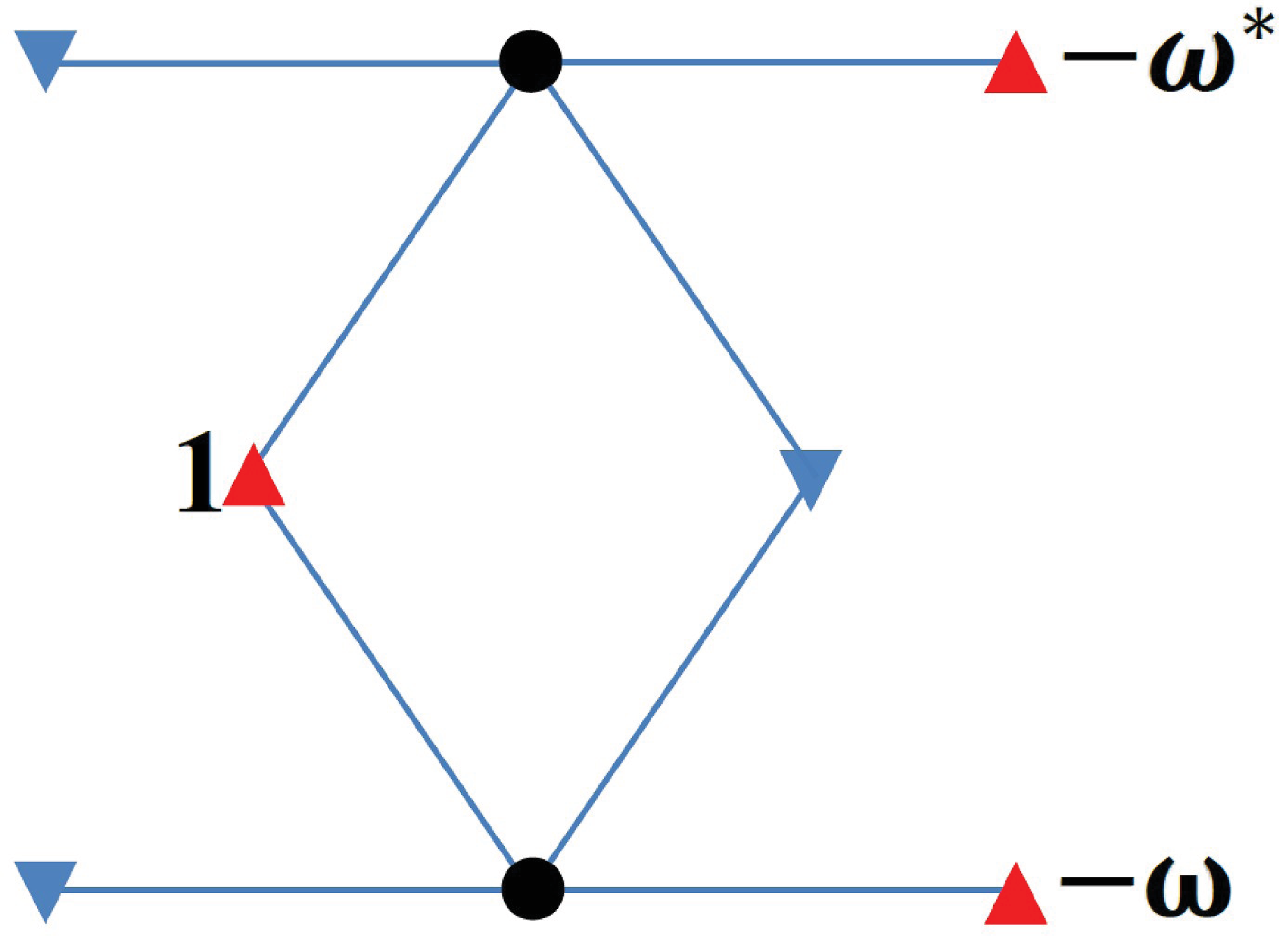}
\hspace{0.3cm}\includegraphics[width=2.3cm,height=1.787cm]{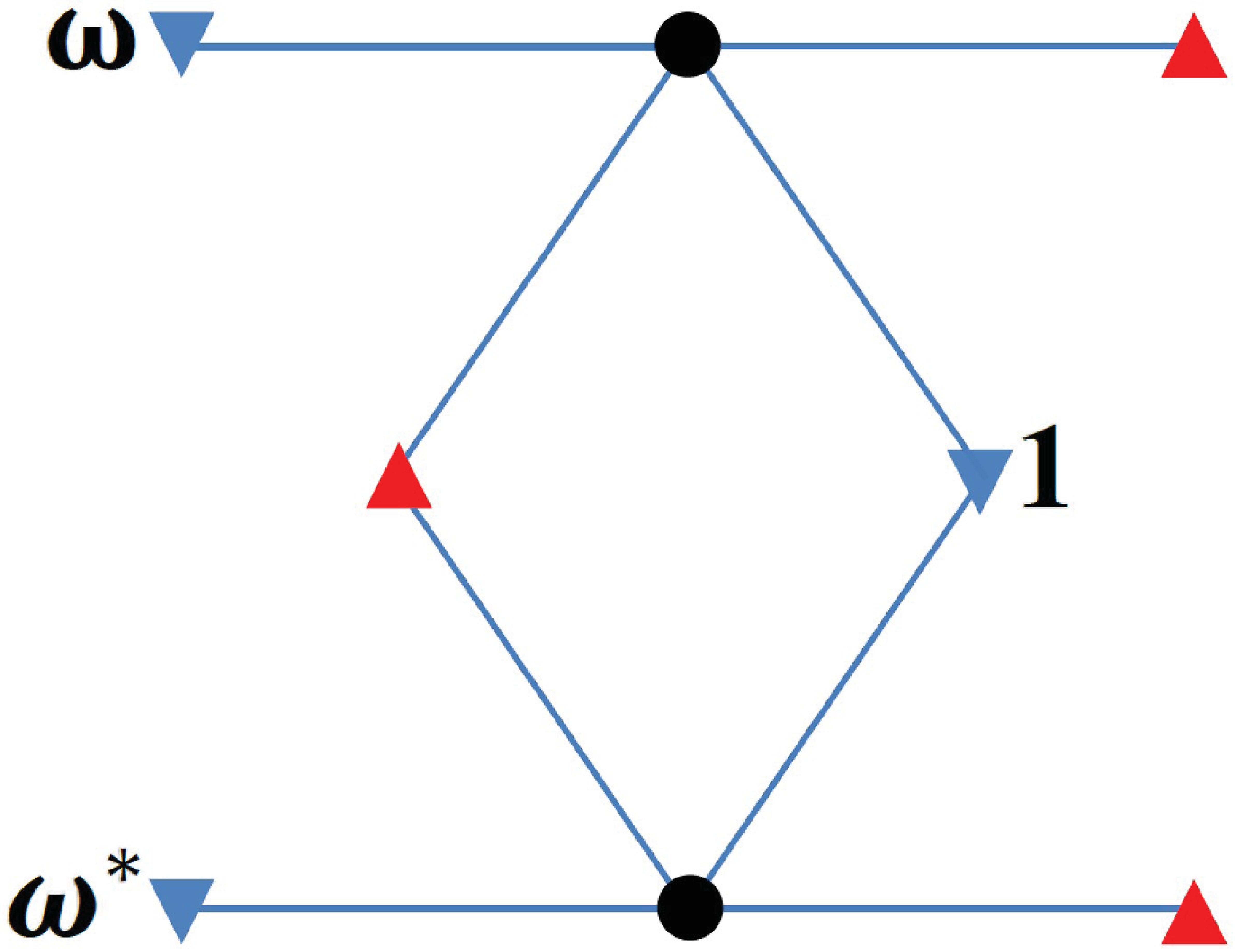}\\
\vspace{0.50cm}
\hspace{-8.0cm} {\textbf{(d)}}\\
\hspace{0.3cm}\includegraphics[width=7.0cm,height=3.319cm]{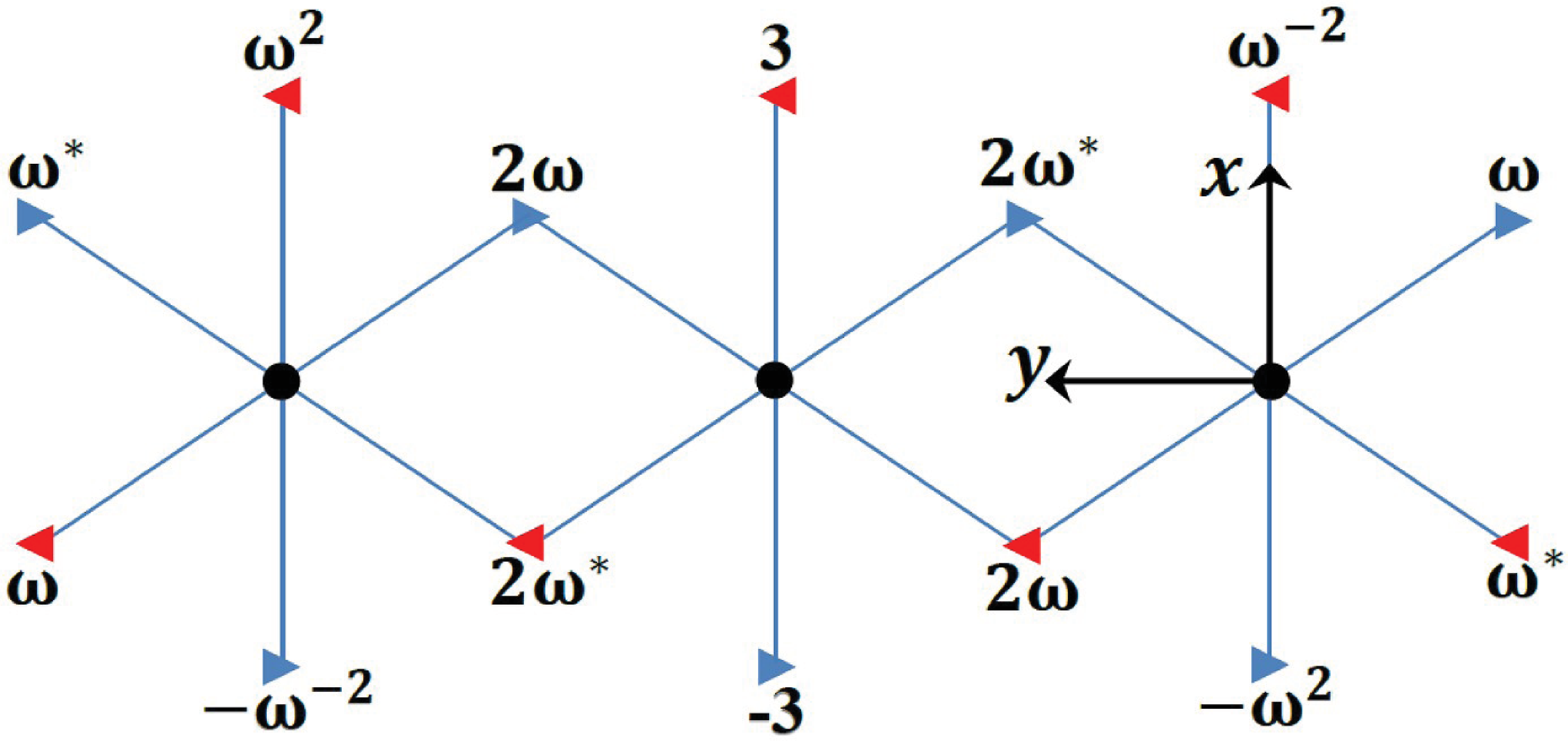}\\
\vspace{0.50cm}
\hspace{-3.6cm} {\textbf{(e)}} \hspace{4.2cm}{\textbf{(f)}}\\
\hspace{0cm}\includegraphics[width=4.0cm,height=3.773cm]{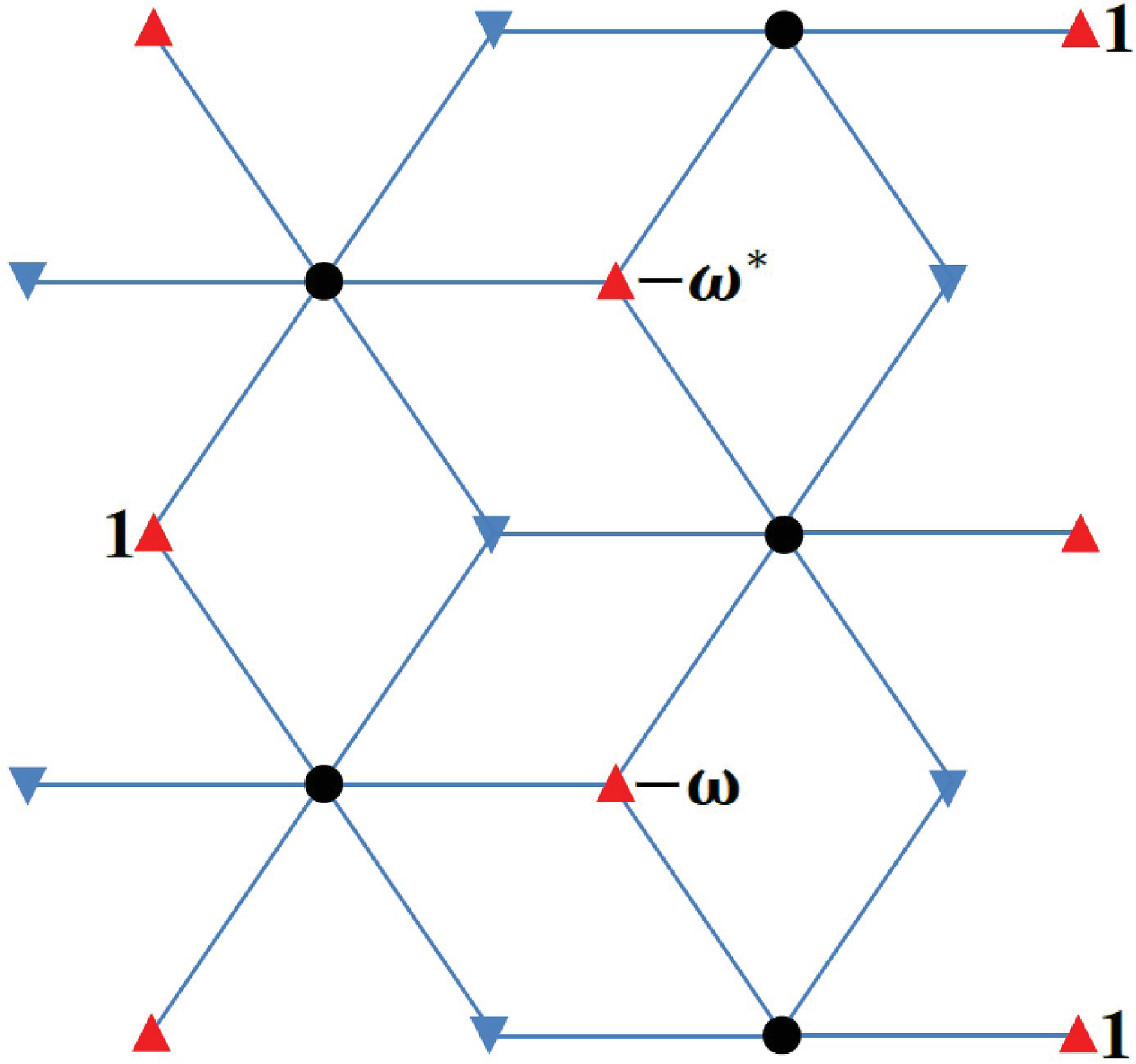}
\hspace{0.5cm}\includegraphics[width=4.0cm,height=3.706cm]{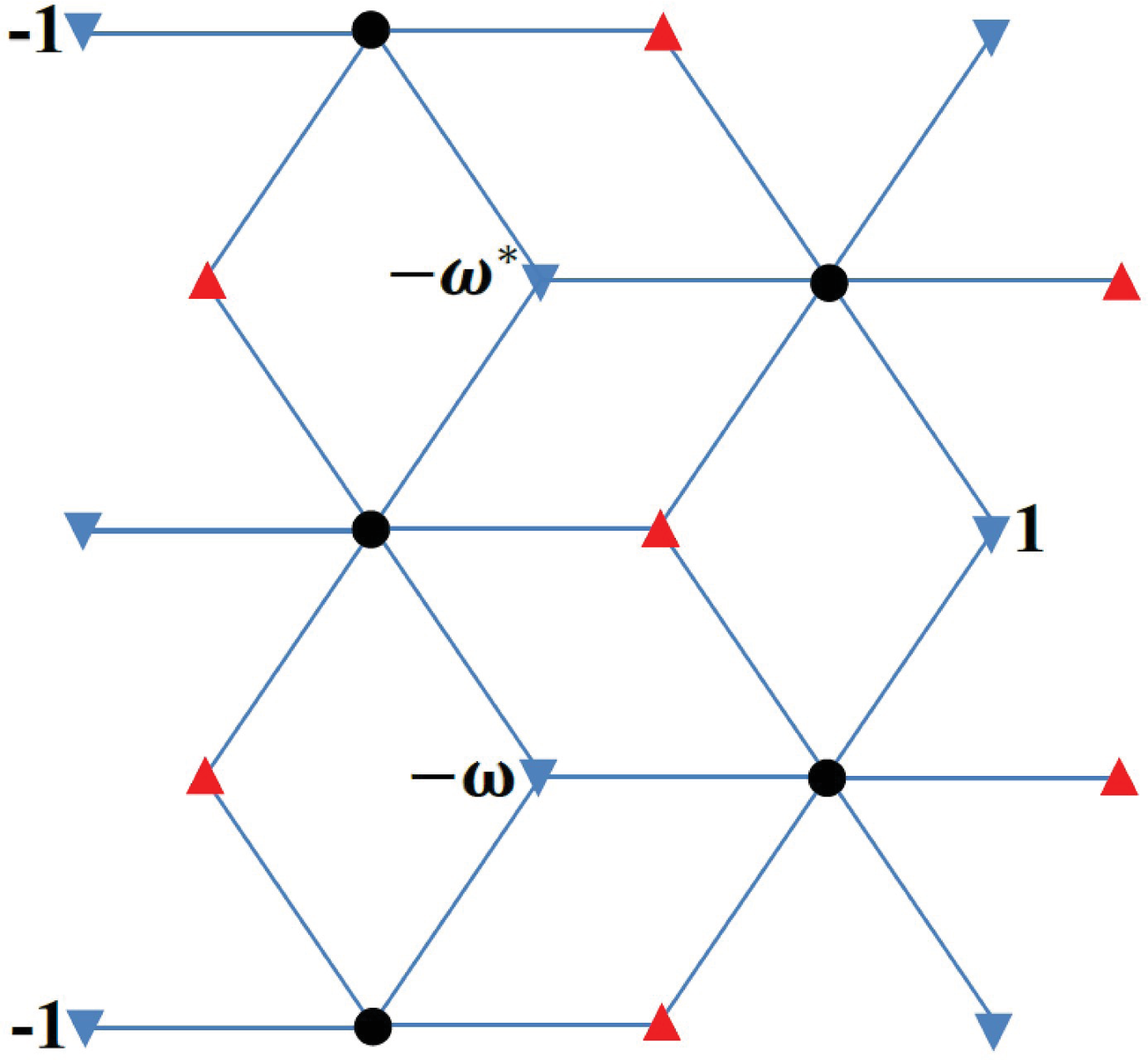}
\vspace{-0.10cm}
\caption{The distinct CLS for flat bands of the minimal zigzag dice lattice ribbons in a nonzero magnetic field of $f=\frac{1}{2}$, up to normalization factors. These CLS separately correspond to (a) the two nonzero energy (i.e., $E_{\alpha}=\alpha\sqrt{6}t$, $\alpha=\pm$) flat bands of the minimal CA-in ribbon, (b) [(c)] the $E=\Delta$ [$E=-\Delta$] flat band of the minimal CA-in ribbon, (d) the third $E=0$ flat band (for $\Delta=0$) of the minimal CA-in ribbon that corresponds to the $E=0$ band of the bulk lattice, and (e) [(f)] the $E=\Delta$ [$E=-\Delta$] flat band of the minimal CA-off ribbon. The nonzero number beside a site represents the amplitude of the CLS on that site. The sites of the ribbon lattice not shown and the sites without a number beside it contribute zero weight to the CLS. $\omega=e^{i\pi/4}=\frac{1}{\sqrt{2}}(1+i)$ and $\omega^{\ast}$ its complex conjugate. The lattice and the coordinate system have rotated 90$^{\circ}$ counterclockwise for (d).}
\end{figure}

For the minimal CA-in ribbon, the eigenvectors for the nondegenerate flat band at $E_{\alpha}(k)=\alpha\sqrt{6}t$ ($\alpha=\pm$) are
\begin{equation}
N_{\alpha}\begin{pmatrix} t, & t_{1}', & \alpha\sqrt{6}t, & t_{2}', & t \\ \end{pmatrix}^{\text{T}}.
\end{equation}
$N_{\alpha}$ ($\alpha=\pm$) is the normalization factor. $t_{1}'=2t\cos(\frac{ka}{2}+\frac{\pi}{4})$ and $t_{2}'=2t\cos(\frac{ka}{2}+\frac{3\pi}{4})$. The corresponding CLS, up to a normalization factor, is shown in Fig.6(a). We note that, upon replacing $\text{sgn}(t)$ of Eq.(26) by $1$, the local eigenstate defined by Eq.(27) has exactly the same form as Fig.6(a).

The zero-energy flat band of the minimal CA-in ribbon is triply degenerate. From Fig.5(f) and Eq.(A26), the triply degenerate zero-energy flat band for $\Delta=0$ contains two flat bands corresponding to the $\pm\Delta$ flat bands for a general $\Delta$. Solving the eigenequation for these two eigenstates, we get the eigenvectors that have weights only on the A or C sublattice sites. The eigenvector for the $E=\Delta$ flat band is
\begin{equation}
\frac{1}{\sqrt{t^{2}+t_{1}'^{2}}}\begin{pmatrix} 0, & -t, & 0, & 0, & t_{1}' \\ \end{pmatrix}^{\text{T}}.
\end{equation}
The eigenvector for the $E=-\Delta$ flat band is
\begin{equation}
\frac{1}{\sqrt{t^{2}+t_{2}'^{2}}}\begin{pmatrix} t_{2}', & 0, & 0, & -t, & 0 \\ \end{pmatrix}^{\text{T}}.
\end{equation}
They are clearly also the zero-energy eigenstates of the model at $\Delta=0$.
Comparing Eqs.(48) and (49) to Eqs.(13) and (14), the magnetic field only moderately modifies the eigenvectors through introducing a phase factor to the $k$-dependent components of the eigenvectors. The corresponding CLS, as shown in Figs.6(b) and 6(c), also slightly differ from the CLS shown in Figs.4(b) and 4(c) for the $\pm\Delta$ flat bands in zero magnetic field.
The third zero-energy eigenstate orthogonal to the two states listed above is found to be
\begin{equation}
N_{0}[t(t^{2}+t_{1}'^{2}),-t_{1}'(t^{2}+t_{2}'^{2}),0,t_{2}'(t^{2}+t_{1}'^{2}),-t(t^{2}+t_{2}'^{2})]^{\text{T}}.
\end{equation}
The factors outside the four parentheses follow closely the components of Eq.(12) for $\Delta=0$. However, the terms within the four parentheses strongly modify the corresponding CLS. Compared to the CLS of Fig.4(a) for Eq.(12), which is a 7-site star centering at a single fully coordinated B site, the CLS shown in Fig.6(d) for Eq.(50) is more extended. Its spatial extension covers three NN 7-site stars along the $y$ axis. This is the first qualitative changes brought to the states of the zero-energy flat bands by the magnetic field. We note that, according to Eqs.(A20) and (A21), the eigenvectors of the three flat bands in Fig.5(c) for the minimal BB-in ribbon are the same as Eqs.(47) and (50), by properly identifying the corresponding sites within the unit cells.

For the minimal CA-off ribbon, the eigenvectors for the twofold degenerate flat band at $E_{\alpha}=\alpha\sqrt{6}t$ ($\alpha=\pm$) may be taken as
\begin{equation}
\frac{1}{2\sqrt{3}|t|}\begin{pmatrix} t, & t_{1}', & E_{\alpha}, & t_{2}', & t, & 0, & 0, & 0 \\ \end{pmatrix}^{\text{T}},
\end{equation}
and
\begin{equation}
\frac{1}{2\sqrt{3}|t|}\begin{pmatrix} 0, & 0, & 0, & t, & -t_{2}', & E_{\alpha}, & t_{1}', & t \\ \end{pmatrix}^{\text{T}}.
\end{equation}
Eq.(51) is identical to Eq.(47), and the corresponding CLS is exactly the same as Fig.6(a). The CLS corresponding to the eigenstate described by Eq.(52) is also a 7-site star, which differs only slightly from Fig.6(a) in the phase factors.

For the fourfold degenerate zero-energy flat bands of the minimal CA-off ribbon, we again firstly derive the eigenvectors for the nondegenerate flat bands at $\Delta$ and $-\Delta$, for nonzero $\Delta$. We find
\begin{equation}
N_{1}\begin{pmatrix} 0, & t^{2}, & 0, & 0, & -tt_{1}', & 0, & 0, & -t_{1}'t_{2}' \\ \end{pmatrix}^{\text{T}}
\end{equation}
for the $E=\Delta$ flat band, and
\begin{equation}
N_{2}\begin{pmatrix} t_{1}'t_{2}', & 0, & 0, & -tt_{1}', & 0, & 0, & t^{2}, & 0 \\ \end{pmatrix}^{\text{T}}
\end{equation}
for the $E=-\Delta$ flat band. $N_{1}$ and $N_{2}$ are normalization factors. Eqs.(53) and (54) are independent of $\Delta$ and separately have weights only on the A and C sublattices. It is easily verified that Eqs.(53) and (54) are zero-energy eigenvectors of the model at $\Delta=0$. They are separately quite similar to Eqs.(16) and (17) for zero magnetic field, up to additional phase factors produced by the Peierls' substitution for the magnetic field. The CLS corresponding to Eqs.(53) and (54) are separately shown in Figs.6(e) and 6(f). While the CLS still have the form of equilateral triangles with a length of $2a$ for each edge, same as those in Figs.4(d) and 4(e), the (unnormalized) amplitude on the middle site of the vertical edge is now 0 instead of 2.

What are the two remaining eigenvectors for the zero-energy flat bands? From Sec.IIIB for the evolution of the zero-energy flat bands from minimal CA-in ribbon to minimal CA-off ribbon, and the general results in a previous work on wide zigzag dice lattice ribbons \cite{hao22}, each chain of fully coordinated B sublattice sites along the $y$ direction contributes a zero-energy flat band that have the same form of eigenvectors, in zero magnetic field. Together with Eq.(A20) for the zero-energy flat band of the minimal BB-in ribbon and Eq.(50) for the minimal CA-in ribbon, it is then natural to expect a similar representation of the eigenstates of the two remaining zero-energy flat bands of the minimal CA-off ribbon in terms of Eq.(50) and Eq.(A20). The ensuing vectors, such as Eq.(50) expanded with three zeros, are unfortunately not eigenvectors of the model. Therefore, the zero-energy flat bands of the minimal CA-off ribbons, leaving out the two bands that will split off to $\pm\Delta$ for nonzero $\Delta$, cannot be represented as eigenvectors associated with the fully coordinated B$_{1}$ and B$_{2}$ sublattice sites, in the same form as Eq.(50) for the minimal CA-in ribbon or Eq.(A20) for the minimal BB-in ribbon. This is a major impact of the magnetic field on the zero-energy flat bands.

The two remaining eigenvectors, which are needed for completeness, can have different choices. One simple choice gives
\begin{equation}
N_{3}'\begin{pmatrix} -t_{1}', & t, & 0, & 0, & 0, & 0, & 0, & 0 \\ \end{pmatrix}^{\text{T}},
\end{equation}
and
\begin{equation}
N_{4}'\begin{pmatrix} 0, & 0, & 0, & 0, & 0, & 0, & t, & -t_{1}' \\ \end{pmatrix}^{\text{T}}.
\end{equation}
In addition, we may hybridize Eqs.(55) and (56) with Eqs.(53) and (54) to eliminate the last three or the leading three components of the eigenvector, and obtain
\begin{equation}
N_{3}''\begin{pmatrix} t_{1}'t_{2}'^{2}, & t^{3}, & 0, & -tt_{1}'t_{2}', & -t^{2}t_{1}', & 0, & 0, & 0 \\ \end{pmatrix}^{\text{T}},
\end{equation}
and
\begin{equation}
N_{4}''\begin{pmatrix} 0, & 0, & 0, & -t^{2}t_{1}', & tt_{1}'t_{2}', & 0, & t^{3}, & t_{1}'t_{2}'^{2} \\ \end{pmatrix}^{\text{T}},
\end{equation}
However, these eigenvectors are nonorthogonal to Eqs.(53) and (54), and they are quite distinct from Eq.(50) for the minimal CA-in ribbon and Eq.(A20) for the minimal BB-in ribbon. In particular, we again do not see a systematic construction of the zero-energy eigenstates in them. Alternatively, we may project out Eqs.(53) and (54) from Eqs.(55) and (56). This leads to
\begin{equation}
N_{3}[-t(t^{2}+t_{1}'^{2}),t_{1}'(t^{2}+t_{2}'^{2}),0,-t_{1}'^{2}t_{2}',t^{3},0,tt_{1}'t_{2}',t^{2}t_{2}']^{\text{T}},
\end{equation}
and
\begin{equation}
N_{4}[-t^{2}t_{2}',-tt_{1}'t_{2}',0,t^{3},t_{1}'^{2}t_{2}',0,t_{1}'(t^{2}+t_{2}'^{2}),-t(t^{2}+t_{1}'^{2})]^{\text{T}}.
\end{equation}
They are mutually orthogonal and orthogonal to Eqs.(53) and (54). Again, they are quite distinct from Eq.(50) and Eq.(A20) and do not only involve the degrees of freedom of a single fully coordinated B sublattice site of the unit cell.

Overall, through the analysis for the CLS, a magnetic field of $f=\frac{1}{2}$ significantly changes the various flat bands. Firstly, it changes the phase factors and the amplitude distributions of the CLS for the flat bands residing solely on the A or C sublattice sites. Secondly, it turns the bands corresponding to the bulk dispersive bands into flat bands by introducing additional phase factors. These correspond to the eigenstates defined by Eqs.(47), (51), and (52). Thirdly, it increases the spatial extension of the CLS for the bulk-like zero-energy flat bands, from Fig.4(a) to Fig.6(d). Fourthly, the bulk-like zero-energy flat bands can no longer be associated to consecutive fully coordinated B sublattice sites of the unit cell, especially not in the manner for zero magnetic field.

\section{discussion and outlook}

In summary, we find novel 1D flat bands and Dirac cones in four minimal zigzag dice lattice ribbons. Firstly, we find all three types of combinations of 1D Dirac cones and flat bands in the low-energy band structures of the four minimal ribbons, including Dirac cones only (the minimal BB-off ribbon), flat bands only (the minimal CA-in and CA-off ribbons), and coexisting Dirac cones and flat bands (the minimal BB-in ribbon). In a perpendicular magnetic field for which there is half a flux quantum through each elementary rhombus, the minimal CA-in and minimal CA-off ribbons enter the AB caging phase, with fully pinched spectra and compact AB cages for arbitrary localized initial states. Secondly, in the band structures of the four minimal ribbons at zero magnetic field, we identify Dirac cones and flat bands corresponding to the edge states of wide ribbons. These constitute the only known edge states that survive the extreme reduction of the ribbon width, to our knowledge. Thirdly, for the Dirac cones and flat bands related to the edge states of wide ribbons, we find interesting connections between them and the geometric structures of the ribbons. These include a BZ folding picture for the two Dirac cones of the minimal BB-off ribbon. In addition, for the flat bands of the minimal CA-in and CA-off ribbons that correspond to the edge states of wide CA-in and CA-off ribbons, we find peculiar triangular CLS and new AB cages associated with initial states on the ribbon edges.

From the above results, the four new narrow zigzag dice lattice ribbons are ideal playgrounds for exploring physics associated with 1D flat bands and Dirac cones. These include the Luttinger liquid phases associated with the Dirac cones \cite{egger97,kane97,yoshioka99}, and various symmetry breaking phases realizable in the flat bands \cite{vidal00,doucot02,hyrkas13,tovmasyan18,cartwright18,tilleke20,roy20,orito21}. In the following parts of this section, we briefly discuss the experimental realizations of the four new minimal zigzag ribbons and the intriguing open theoretical questions related to the Dirac cones and flat bands.

\subsection{Experimental realization}

The dice lattice and the diamond chain lattice (i.e., the minimal AC-in ribbon) have been experimentally fabricated in several physical systems, which may be adapted to fabricate the four new minimal zigzag dice lattice ribbons.

The dice lattice has been realized in superconducting wire networks and normal metal wire networks \cite{abilio99,naud01}.
Such solid-state networks may be created by applying laser direct writing or e-beam lithography to a thin film. The lift-off process, following the e-beam writing \cite{abilio99,naud01} or a nanoimprint lithography process \cite{chou95}, have also been successfully applied to make these nanostructures. By working on a narrow strip of the thin film, a narrow ribbon of the network is obtained. By choosing the widths of the narrow strips in reference to the lattice parameters and the target ribbon lattices, the four minimal zigzag dice lattice ribbons can be fabricated in terms of these methods.

The diamond chain lattice has been fabricated as a Josephson junction array \cite{pop08}, in photonic lattices \cite{mukherjee18,kremer20,caceres22}, in topoelectrical circuits \cite{wang22}, and in a synthetic lattice in the momentum space of ultracold atoms \cite{li22}. To apply the synthetic lattice approach, we have to encode the sites within each unit cell of a minimal ribbon into suitable non-spatial degrees of freedom, such as the spin-orbital freedoms of the atoms or the frequency modes under a periodic driving, and establish the connections among them following the pattern of the target lattice. This has to be constructed in a lattice-by-lattice manner, which we will not go into details. The other approaches mentioned, as we discuss in what follows, can directly be generalized to apply to the four new minimal zigzag dice lattice ribbons.

The Josephson junction diamond chain lattice was made by the e-beam lithography and shadow evaporation technique \cite{pop08}. By making more tiny superconducting islands and connecting them by tunneling barriers, following the patterns of the target minimal ribbons, we can get the corresponding Josephson junction arrays for the four new minimal zigzag dice lattice ribbons.

The photonic lattices are periodic arrays of evanescently coupled optical waveguides, which can be fabricated using ultrafast (e.g., femtosecond) laser writing technique and have proven to be versatile platforms of simulating novel lattice models \cite{ozawa19,liberto19}. Similar to the photonic diamond chain lattice \cite{mukherjee18,kremer20,caceres22}, the photonic lattices of the four minimal zigzag dice lattice ribbons may be fabricated by laser writing a glass wafer. To ensure that the dominant couplings between the waveguides follow the pattern defined by the target ribbon lattice, we may add an additional auxiliary waveguide in the middle of each pair of primary waveguides that are connected by a bond of the ribbon lattice \cite{kremer20}.

The topoelectrical circuit for the diamond chain lattice is constructed via unit by unit assembly \cite{wang22}, which is highly flexible and scalable. As such, the extension of the circuit for the diamond chain lattice to circuits for the four new minimal zigzag dice lattice ribbons is straightforward. The only care that needs to be taken is to ensure the connections among the nodes to follow the patterns of the bonds in the target minimal ribbons.

Besides the above extensions to previously used methods, several other systems may also be used to prepare the four new minimal zigzag dice lattice ribbons. The first system we want to mention is the quantum dot array \cite{ansaloni20,sikdar21}. By etching a two-dimensional electron gas and attaching suitably patterned gates, arrays of quantum dots may be arranged to follow the geometric structures and connections of the four minimal ribbons. This is a highly tunable platform, since the tunneling connections between NN quantum dots and the on-site energies of different sublattices may be tuned independently with suitable gate voltages.

The second approach we mention is the molecular framework. 2D polymers have been used to fabricate the Lieb lattice and the kagome lattice \cite{springer20,jing20,jiang21}. An extremely appealing feature of this method is the existence of elementary molecular building blocks, including nodes with desirable (e.g., two-fold, three-fold, four-fold, and six-fold) rotational symmetries and linear linkers \cite{springer20}, which in principle allows for the design of arbitrary periodic 2D or q-1D networks. For our purpose, the four new minimal zigzag dice lattice ribbons may be fabricated by combining nodes with three-fold rotational symmetry and nodes with six-fold rotational symmetry by linear linkers along the NN bonds of the ribbon lattices.

Finally, there are methods not so easy to implement at present but are valid in principle and may turn out to be feasible later. An example is the thin film on a suitable substrate, grown by controlled deposition or self-assembling, which has been used to fabricate the 2D kagome lattice \cite{mihalyuk22,mihalyuk21}. To apply this method, we have to find a suitable substrate that can guide the growth of the 2D dice lattice, and take narrow strips of the substrate to grow our q-1D minimal ribbons.
Another promising method is the optical lattice \cite{windpassinger13,zhang19,cooper19}. Several proposals for the 2D optical dice lattice are available \cite{rizzi06,bercioux09,hao21epj}. The minimal zigzag dice lattice ribbons may be obtained by applying an additional constraining potential to the 2D optical dice lattice, so that the ultracold atoms load only to the sites belonging to the minimal ribbon of the whole lattice. This constraining potential may be realized by applying a red-detuned traveling-wave Gaussian beam with reshaped beam profile of a line to the 2D optical dice lattice \cite{daniels03}. Lastly, arrays of optical tweezers are emerging as a new flexible platform for simulating novel lattice models \cite{kaufman14,murmann15,spar22}. With the development of the technique, some time later we may be able to make a minimal zigzag dice lattice ribbon in the form of an array of optical tweezers.

\subsection{Theoretical outlook}

The single-body electronic spectra with 1D Dirac cones and (or) flat bands are interesting on their own and deserve further studies. The 1D Dirac cone states in the minimal BB-off ribbon may serve as perfect quantum channels that are immune to disorder \cite{zhu09}. The effects of disorder and impurities on other narrow ribbons are also highly intriguing issues to be explored further \cite{bodyfelt14,roy20,orito21}. Another important question unexplored here is the topological properties of the minimal zigzag ribbons. The diamond chain lattice in the AB caging phase was shown to be a so-called square-root topological insulator \cite{kremer20}. The potential topological properties of the four new minimal zigzag ribbons are unknown presently. In addition, when supplemented with proper spin-orbit coupling terms, the dice model was known to describe various interesting topological phases \cite{wang11,dey20,soni20,soni21,wang21}. The corresponding topological properties of the q-1D narrow zigzag dice lattice ribbons are interesting issues of future studies.
Finally, while we have identified Dirac cones (for the minimal BB-off and minimal BB-in ribbons) and flat bands (for the minimal CA-in and minimal CA-off ribbons) corresponding directly to the edge states of wide ribbons, it is unclear where else to find such edge states that are robust to the extreme reduction of the ribbon width. Here, the ribbons with robust edge states in the form of Dirac cones (i.e., the BB ribbons) or flat bands (i.e., the CA ribbons) may be regarded as the merging of basic building blocks (e.g., the minimal BB-off ribbon for the BB ribbons, the minimal CA-in ribbon for the CA ribbons) via their common hinge orbitals. This connection between the geometric structures and the spectrum seems to underlie the extreme robustness of the edge states. But it is unknown whether more fundamental physics distinguish such edge states from other more common edge states that are strongly modified as the ribbon width decreases to the minimum. The above questions are also interesting theoretical issues for the future.

The many-body correlation effects are highly enhanced in a low-dimensional system. Correlated 1D systems with Dirac cones and (or) flat bands are under intensive studies \cite{vidal00,doucot02,hyrkas13,tovmasyan18,cartwright18,tilleke20,roy20,orito21,bischoff17,tada19}. For all the distinct types of band structures with 1D Dirac cones and (or) flat bands studied in this work, we expect there to be great opportunities of finding fascinating many-body physics.

Similar to the minimal BB-off ribbon, the single-wall armchair carbon nanotubes have a pair of Dirac cones in the low-energy part of the band structures \cite{ajiki93,kane97m}. The Luttinger liquids in the single-wall armchair carbon nanotubes were studied in terms of Bosonization method \cite{egger97,kane97,yoshioka99}. Because the circumference of the nanotubes are commonly much larger than the lattice constant, the backscattering term of the correlation is tiny and may be treated as a perturbation \cite{kane97}. In the minimal BB-off ribbon, however, the ribbon width is comparable to the lattice constant and the backscattering term is no longer perturbative. The properties of the Luttinger liquids in the minimal BB-off ribbon with two Dirac cones and a significant backscattering term is therefore a highly nontrivial and intriguing issue to be explored.

Systems with flat bands at the chemical potential are genuine strongly correlated systems, even for infinitesimally small interaction strengths. Due to the presence of powerful analytical and efficient numerical techniques for these problems \cite{giamarchibook}, 1D correlated models with flat bands in the band structures have attracted widespread interest \cite{vidal00,doucot02,hyrkas13,tovmasyan18,cartwright18,tilleke20,roy20,orito21}. Here, we have three kinds of flat bands. The first kind contains the minimal CA-in and CA-off ribbons in zero magnetic field, where the flat bands are isolated from dispersive bands. The minimal BB-in ribbon for $f=1/2$ and $\Delta=0$ [Fig.5(c)] and the minimal AC-in ribbon for $f=0$ and $\Delta\ne0$ [Fig.3(j)] belong also to this case. The second kind is the fully pinched spectra of the minimal CA-in, CA-off, and AC-in ribbons, for $f=1/2$ and $\Delta=0$. The third kind is the coexisting flat band and Dirac cone in the minimal BB-in ribbon and the minimal AC-in ribbon (for $\Delta=0$), in zero magnetic field. The many-body physics in these different cases are highly intriguing subjects to study.

We briefly mention a straightforward implication of the present studies as regards the many-body physics. The representation of the flat bands in terms of CLS facilitates the construction of certain kinds of exact many-body ground states. Namely, when the filling in the lowest flat band is smaller than a critical density, a many-body state may be constructed by placing the CLS on nonoverlapping clusters, so that there is no interaction between the different CLS and the energy of the state is the direct summation of the energies of the different CLS. This scheme has been applied to several flat band systems \cite{wu07,gulacsi07,derzhko09,derzhko10}. It may also help to understand the physics for more general densities that have to resort to numerical calculations.

\begin{acknowledgments}
L.H. gratefully acknowledges the comments and suggestions of all the referees for this paper.
\end{acknowledgments}\index{}

\begin{appendix}

\section{models of the minimal zigzag dice lattice ribbons in the $k$ space}

In this Appendix, we introduce the model Hamiltonians of the five minimal zigzag dice lattice ribbons defined in Fig.2, in the $k_{y}$ space and in the presence of a tunable perpendicular magnetic field. As shown in Section II, a magnetic field is introduced through the Peierls' substitution, which gives a set of phase factors to the hopping amplitudes for non-horizontal bonds. The magnetic field is measured in terms of the reduced flux $f=\phi/\phi_{0}$, where $\phi$ is the magnetic flux through an elementary rhombus of the lattices and $\phi_{0}$ is the flux quantum. In the Laudau gauge that we consider, the translational invariance along the $y$ direction is retained for an arbitrary magnetic field. So $k_{y}$ is a good quantum number in the 1D BZ, $k_{y}\in(-\pi/a,\pi/a]$, for arbitrary $f$. Since the wave vector has only a single component, we usually denote $k_{y}$ by $k$ for simplicity.

\subsection{The minimal BB-off ribbon}

As shown in Fig.2(a), the minimal BB-off ribbon has four $y$-chains of lattice sites. It may be seen as consisting of two parallel zigzag chains coupled along the transverse (i.e., $x$) direction. The Hamiltonian in the $k$ space is
\begin{equation}
h(k)=\begin{pmatrix} 0 & t & t_{1}' & 0  \\
                     t & \Delta & 0 & t_{2}' \\
                     t_{1}' & 0 & -\Delta & t \\
                     0 & t_{2}' & t & 0
               \end{pmatrix}.
\end{equation}
The basis is taken as $(b_{1k}^{\dagger},a_{1k}^{\dagger},c_{1k}^{\dagger},b_{2k}^{\dagger})$. $a_{ik}^{\dagger}$, $b_{ik}^{\dagger}$, and $c_{ik}^{\dagger}$ separately create a spinless electron of wave vector $k$ on the $i$th sites of the A, B, and C sublattices in the unit cell of Fig.2(a). $t_{1}'=2t\cos(\frac{ka}{2}+\gamma_{1})$ and $t_{2}'=2t\cos(\frac{ka}{2}+\gamma_{2})$. $\gamma_{1}=\frac{\pi}{2}f$ and $\gamma_{2}=\frac{5\pi}{2}f$.
The eigenequation for the four bands are
\begin{eqnarray}
&&\text{det}[h(k)-\lambda I_{4}]=\lambda^{4}-(2t^{2}+t_{1}'^{2}+t_{2}'^{2}+\Delta^{2})\lambda^{2}   \notag \\ &&+\Delta(t_{1}'^{2}-t_{2}'^{2})\lambda+(t^{2}-t_{1}'t_{2}')^{2}=0.
\end{eqnarray}
$I_{4}$ is the $4\times4$ unit matrix.

In zero magnetic field, we have $f=0$ and $\gamma_1=\gamma_2=0$, so that $t_{1}'=t_{2}'=t'=2t\cos\frac{ka}{2}$. The dispersions of the four bands are
\begin{equation}
E_{\alpha\beta}(k)=\alpha\left|\sqrt{t^{2}+\frac{\Delta^{2}}{4}}+\beta\sqrt{t'^{2}+\frac{\Delta^{2}}{4}}\right|,
\end{equation}
where $\alpha=\pm$ and $\beta=\pm$. The two low-energy bands $E_{--}(k)$ and $E_{+-}(k)$ touch at the two Dirac points $k=\pm\frac{2\pi}{3a}$.

In a magnetic field giving half a flux quantum through each elementary rhombus, we have $f=\frac{1}{2}$, $\gamma_{1}=\frac{\pi}{4}$, and $\gamma_{2}=\frac{5\pi}{4}$. In this case, $t_{2}'=-t_{1}'$, the dispersions of the four bands are
\begin{equation}
E_{\alpha\beta}(k)=\alpha\left|\sqrt{t^{2}+t_{1}'^{2}+\frac{\Delta^{2}}{4}}+\beta\frac{|\Delta|}{2}\right|,
\end{equation}
where $\alpha=\pm$ and $\beta=\pm$. There are neither Dirac cones nor flat bands.

\subsection{The minimal BB-in ribbon}

The minimal BB-in ribbon shown in Fig.2(b) contains 7 $y$-chains. There are 7 sites in the unit cell. In the $k$ space and in the basis of $(b_{1k}^{\dagger},a_{1k}^{\dagger},c_{1k}^{\dagger},b_{2k}^{\dagger} ,a_{2k}^{\dagger},c_{2k}^{\dagger},b_{3k}^{\dagger})$, the Hamiltonian is written as
\begin{equation}
h(k)=\begin{pmatrix} 0 & t & t_{1}' & 0 & 0 & 0 & 0  \\
                     t & \Delta & 0 & t_{2}' & 0 & 0 & 0  \\
                     t_{1}' & 0 & -\Delta & t & 0 & 0 & 0  \\
                     0 & t_{2}' & t & 0 & t & t_{3}' & 0  \\
                     0 & 0 & 0 & t & \Delta & 0 & t_{4}'  \\
                     0 & 0 & 0 & t_{3}' & 0 & -\Delta & t \\
                     0 & 0 & 0 & 0 & t_{4}' & t & 0
               \end{pmatrix}.
\end{equation}
$t_{i}'=2t\cos(\frac{ka}{2}+\gamma_{i})$, $i=1,2,3,4$. The four phases are $\gamma_{1}=\frac{\pi}{2}f$, $\gamma_{2}=\frac{5\pi}{2}f$, $\gamma_{3}=\frac{7\pi}{2}f$, and $\gamma_{4}=\frac{11\pi}{2}f$.

In zero magnetic field, $\gamma_{i}=0$ ($i=1,2,3,4$), $t_{i}'=2t\cos(\frac{ka}{2})=t'$. The eigenequation for the seven bands are
\begin{equation}
\text{det}[h(k)-\lambda I_{7}]=-\lambda(\lambda^{6}+a\lambda^{4}+b\lambda^{2}+c)=0,
\end{equation}
where $I_{7}$ is the seventh-order unit matrix, and
\begin{equation}
\begin{cases}
a=-2(2t^{2}+2t'^{2}+\Delta^{2}),  \\
b=(2t^{2}+\Delta^{2})^{2}+(2t'^{2}+\Delta^{2})^{2}+(t^{2}+t'^{2})^{2}-\Delta^{4},  \\
c=-(t^{2}-t'^{2})^{2}(2t^{2}+2t'^{2}+\Delta^{2}).
\end{cases}
\end{equation}

Eq.(A6) is a multiplication of $\lambda$ and a cubic function of $\lambda^{2}$. The factor $\lambda$ gives the zero-energy flat band, whose eigenvector is found to be
\begin{equation}
\frac{1}{\sqrt{2t^{2}+2t'^{2}+\Delta^{2}}}\begin{pmatrix} 0, & -t', & t, & \Delta, & -t, & t', & 0 \\ \end{pmatrix}^{\text{T}}.
\end{equation}
It is related to the B$_{2}$ sublattice sites in the middle of the ribbon and the NN rim sublattice sites connecting to them.
This agrees with the eigenvectors of the zero-energy flat bands of wide zigzag dice lattice ribbons \cite{hao22}.

The remaining six bands of the minimal BB-in ribbon are determined by the cubic equation of $\lambda^{2}$ of Eq.(A6). When $t'^{2}=t^{2}$, we have $c=0$ and can take out another $\lambda^{2}$ factor, which gives two further zero-energy modes. These account for the two Dirac points at $ka=\pm\frac{2}{3}\pi$.
Away from the Dirac points, the energies of the six nonzero energy bands are determined as roots of the cubic function of $\lambda^{2}$. Because the Hermitian matrix Eq.(A5) has only real roots, the cubic function of $\lambda^{2}$ must have three real positive roots. We introduce two further parameters,
\begin{equation}
\begin{cases}
p=\frac{3b-a^{2}}{3}, \\
q=\frac{2a^{3}+27c-9ab}{27}.
\end{cases}
\end{equation}
In terms of Cardano's method, the three positive roots for $\lambda^{2}$ are
\begin{equation}
\begin{cases}
\lambda_{1}^{2}=u+v-\frac{a}{3},  \\
\lambda_{2}^{2}=\omega' u+\omega'^{2}v-\frac{a}{3},  \\
\lambda_{3}^{2}=\omega'^{2}u +\omega' v-\frac{a}{3},
\end{cases}
\end{equation}
where $\omega'=\exp(i\frac{2\pi}{3})$, and
\begin{equation}
\begin{cases}
u=\sqrt[3]{-\frac{q}{2}+\sqrt{(\frac{q}{2})^{2}+(\frac{p}{3})^{3}}}, \\
v=\sqrt[3]{-\frac{q}{2}-\sqrt{(\frac{q}{2})^{2}+(\frac{p}{3})^{3}}}.
\end{cases}
\end{equation}
The six dispersive bands have dispersions
\begin{equation}
\begin{cases}
E_{1\alpha}=\alpha\sqrt{u+v-\frac{a}{3}},  \\
E_{2\alpha}=\alpha\sqrt{\omega' u+\omega'^{2}v-\frac{a}{3}},  \\
E_{3\alpha}=\alpha\sqrt{\omega'^{2}u +\omega' v-\frac{a}{3}},
\end{cases}
\end{equation}
where $\alpha=\pm$.

We consider the unbiased case with $\Delta=0$ to determine the sequences of the various bands. In this case,
\begin{equation}
\begin{cases}
-\frac{q}{2}=\frac{1}{27}(t^{2}+t'^{2})(t^{4}+t'^{4}+38t^{2}t'^{2}),   \\
\Big(\frac{q}{2}\Big)^{2}+\Big(\frac{p}{3}\Big)^{3}=-\frac{16t^{4}t'^{4}}{27}(t^{4}+t'^{4}+34t^{2}t'^{2}).
\end{cases}
\end{equation}
We take
\begin{eqnarray}
u&=&v^{\ast}=\sqrt[3]{\Big(-\frac{q}{2}\Big)+i\sqrt{\Big|\Big(\frac{q}{2}\Big)^{2}+\Big(\frac{p}{3}\Big)^{3}\Big|}}  \notag \\
&=&\sqrt[3]{Ae^{i\phi}}=\sqrt[3]{A}e^{i\phi/3}=u_{r}+iu_{i},
\end{eqnarray}
where $A>0$, $u_{r}$ and $u_{i}$ are the real and imaginary parts of $u$. In terms of $u_{r}$ and $u_{i}$, we have
\begin{equation}
\begin{cases}
\lambda_{1}^{2}=2u_{r}+\frac{4}{3}(t^{2}+t'^{2}),  \\
\lambda_{2}^{2}=-u_{r}-\sqrt{3}u_{i}+\frac{4}{3}(t^{2}+t'^{2}),  \\
\lambda_{3}^{2}=-u_{r}+\sqrt{3}u_{i}+\frac{4}{3}(t^{2}+t'^{2}).
\end{cases}
\end{equation}
Since $A>0$, $A\cos\phi>0$, and $A\sin\phi\ge0$, we have $0\le\phi<\frac{\pi}{2}$. Therefore, $0\le\frac{\phi}{3}<\frac{\pi}{6}$ and $0\le\sqrt{3}u_{i}<u_{r}$. As a result,
\begin{equation}
\lambda_{2}^{2}\le\lambda_{3}^{2}<\lambda_{1}^{2}.
\end{equation}
In conclusion, $E_{2+}$ and $E_{2-}$ are the two lowest-energy bands constituting the two 1D Dirac cones. $E_{3\alpha}$ ($\alpha=\pm$) have higher energy than the Dirac cone bands and touch the Dirac cone bands at the BZ boundary for $\Delta=0$. $E_{1\alpha}$ ($\alpha=\pm$) are the two highest energy bands.

In a magnetic field of $f=\frac{1}{2}$, $\gamma_{1}=\frac{\pi}{4}$, $\gamma_{2}=\frac{5\pi}{4}$, $\gamma_{3}=\frac{7\pi}{4}$, and $\gamma_{4}=\frac{11\pi}{4}$. The four hoping terms are related through $t_{2}'=-t_{1}'$, $t_{4}'=-t_{3}'$, and $t_{1}'^{2}+t_{3}'^{2}=4t^{2}$. The eigenequation for the energy spectrum becomes
\begin{equation}
-\lambda^{7}+c_{5}\lambda^{5}+c_{3}\lambda^{3}+c_{2}\lambda^{2}+c_{1}\lambda+c_{0}=0,
\end{equation}
where
\begin{equation}
\begin{cases}
c_{5}=2(6t^{2}+\Delta^{2}),  \\
c_{3}=-(6t^{2}+\Delta^{2})^{2}-(5t^{4}+t_{1}'^{2}t_{3}'^{2}),  \\
c_{2}=2\Delta t^{2}(t_{1}'^{2}-t_{3}'^{2}),  \\
c_{1}=\Delta^{2}(29t^{4}-3t_{1}'^{2}t_{3}'^{2})+6t^{2}(5t^{4}+t_{1}'^{2}t_{3}'^{2}),  \\
c_{0}=\Delta(7t^{4}-t_{1}'^{2}t_{3}'^{2})(t_{3}'^{2}-t_{1}'^{2}).  \\
\end{cases}
\end{equation}
For $\Delta\ne0$, the above equation does not have explicit analytical solutions.

When $\Delta=0$, Eq.(A17) reduces to
\begin{equation}
\lambda(\lambda^{2}-6t^{2})(\lambda^{4}-6t^{2}\lambda^{2}+5t^{4}+t_{1}'^{2}t_{3}'^{2})=0.
\end{equation}
It has three flat bands at $0$, $-\sqrt{6}t$, and $\sqrt{6}t$, which coincide with the three flat bands of the bulk dice lattice. The eigenvector for the zero-energy flat band is
\begin{equation}
\frac{1}{\sqrt{2(5t^{4}+t_{1}'^{2}t_{3}'^{2})}}\begin{pmatrix} 0 \\ t_{1}'(t^{2}+t_{3}'^{2}) \\ -t(t^{2}+t_{3}'^{2}) \\ 0 \\ t(t^{2}+t_{1}'^{2}) \\ t_{3}'(t^{2}+t_{1}'^{2}) \\ 0 \end{pmatrix}.
\end{equation}
The eigenvectors for the flat bands at $\pm\sqrt{6}t$ are
\begin{equation}
\frac{1}{2\sqrt{3}|t|}\begin{pmatrix} 0, & -t_{1}', & t, & \pm\sqrt{6}t, & t, & t_{3}', & 0 \\ \end{pmatrix}^{\text{T}}.
\end{equation}
The last factor of the eigenequation (A19) gives four additional dispersive bands,
\begin{equation}
\alpha|t|\sqrt{3+2\beta\big|\cos(ka+\frac{\pi}{2})\big|},
\end{equation}
where $\alpha=\pm$ and $\beta=\pm$. At $ka=0$ and $ka=\pi$, $\cos(ka+\frac{\pi}{2})=0$, the two positive (negative) energy bands become degenerate. These lead to the Dirac points in the dispersive bands of Fig.5(c).

\subsection{The minimal CA-in ribbon}

The minimal CA-in ribbon shown in Fig.2(c) has five $y$-chains, and the unit cell contains five sites. In the $k$ space and with the basis $(c_{1k}^{\dagger},a_{1k}^{\dagger},b_{1k}^{\dagger},c_{2k}^{\dagger},a_{2k}^{\dagger})$, the Hamiltonian is
\begin{equation}
h(k)=\begin{pmatrix} -\Delta & 0 & t & 0 & 0  \\
                     0 & \Delta & t_{1}' & 0 & 0   \\
                     t & t_{1}' & 0 & t_{2}' & t   \\
                     0 & 0 & t_{2}' & -\Delta & 0   \\
                     0 & 0 & t & 0 & \Delta  \\
               \end{pmatrix}.
\end{equation}
$t_{i}'=2t\cos(\frac{ka}{2}+\gamma_{i})$ ($i=1,2$), $\gamma_{1}=\frac{\pi}{2}f$, and $\gamma_{2}=\frac{3\pi}{2}f$.
The eigenequation, $\text{det}[h(k)-\lambda I_{5}]$, for the five bands is
\begin{equation}
(\lambda^{2}-\Delta^{2})[-\lambda^{3} +(2t^{2}+t_{1}'^{2}+t_{2}'^{2}+\Delta^{2})\lambda+\Delta(t_{1}'^{2}-t_{2}'^{2})]=0.
\end{equation}
$I_{5}$ is the fifth-order unit matrix.

In zero magnetic field, $\gamma_{1}=\gamma_{2}=f=0$ and $t_{1}'=t_{2}'=2t\cos\frac{ka}{2}=t'$. The eigenequation for the five bands becomes
\begin{equation}
\lambda(\lambda^{2}-\Delta^{2})(\lambda^{2}-\Delta^{2}-2t^{2}-2t'^{2})=0.
\end{equation}
The eigenequation leads to a zero-energy flat band, two flat bands separately at $\Delta$ and $-\Delta$, and two dispersive bands of higher energy that are well separated from the three low-energy flat bands.

In a perpendicular magnetic field which gives $f=\frac{1}{2}$, we have $\gamma_{1}=\frac{\pi}{4}$ and $\gamma_{2}=\frac{3\pi}{4}$. The eigenequation for the five bands in this magnetic field is
\begin{equation}
(\lambda^{2}-\Delta^{2})[-\lambda^{3}+(6t^{2}+\Delta^{2})\lambda+\Delta(t_{1}'^{2}-t_{2}'^{2})]=0.
\end{equation}
For a general nonzero $\Delta$, the spectrum contains two flat bands separately at $\Delta$ and $-\Delta$, and three dispersive bands determined by solving a cubic function.
For $\Delta=0$, the spectrum contains only flat bands and are fully pinched at three energies, 0 and $\pm\sqrt{6}t$. The zero-energy flat band is three-fold degenerate, the flat bands at $\sqrt{6}t$ and $-\sqrt{6}t$ are nondegenerate.

\subsection{The minimal CA-off ribbon}

The minimal CA-off ribbon shown in Fig.2(d) has eight $y$-chains, and the unit cell contains eight sites. In the $k$ space and with the basis $(c_{1k}^{\dagger},a_{1k}^{\dagger},b_{1k}^{\dagger},c_{2k}^{\dagger}, a_{2k}^{\dagger},b_{2k}^{\dagger},c_{3k}^{\dagger},a_{3k}^{\dagger})$, the Hamiltonian is
\begin{equation}
h(k)=\begin{pmatrix} -\Delta & 0 & t & 0 & 0 & 0 & 0 & 0  \\
                     0 & \Delta & t_{1}' & 0  & 0 & 0 & 0 & 0  \\
                     t & t_{1}' & 0 & t_{2}' & t & 0 & 0 & 0  \\
                     0 & 0 & t_{2}' & -\Delta & 0 & t & 0 & 0  \\
                     0 & 0 & t & 0 & \Delta & t_{3}' & 0 & 0  \\
                     0 & 0 & 0 & t & t_{3}' & 0 & t_{4}' & t  \\
                     0 & 0 & 0 & 0 & 0 & t_{4}' & -\Delta & 0  \\
                     0 & 0 & 0 & 0 & 0 & t & 0 & \Delta  \\
               \end{pmatrix}.
\end{equation}
$t_{i}'=2t\cos(\frac{ka}{2}+\gamma_{i})$ ($i=1,2,3,4$), $\gamma_{1}=\frac{\pi}{2}f$, $\gamma_{2}=\frac{3\pi}{2}f$, $\gamma_{3}=\frac{7\pi}{2}f$, and $\gamma_{4}=\frac{9\pi}{2}f$.

In zero magnetic field, $f=0$, and $t_{i}'=t'=2t\cos\frac{ka}{2}$ ($i=1,2,3,4$). The eigenequation, $\text{det}[h(k)-\lambda I_{8}]$, for the eight bands is
\begin{equation}
\lambda^{2}(\lambda^{2}-\Delta^{2})[(\lambda^{2}-\Delta^{2}-2t^{2}-2t'^{2})^{2}-4t^{2}t'^{2}]=0,
\end{equation}
where $I_{8}$ is the eighth-order unit matrix. There are two zero-energy flat bands, a flat band at $\Delta$ and another flat band at $-\Delta$. The remaining four dispersive bands are well separated from the four low-energy flat bands.

In a magnetic field for $f=\frac{1}{2}$, $\gamma_{1}=\frac{\pi}{4}$, $\gamma_{2}=\frac{3\pi}{4}$, $\gamma_{3}=\frac{7\pi}{4}$, and $\gamma_{4}=\frac{9\pi}{4}$. $t_{4}'=t_{1}'$ and $t_{3}'=-t_{2}'$. The eigenequation is
\begin{eqnarray}
&&(\lambda^{2}-\Delta^{2})[\lambda^{2}(\lambda^{2}-6t^{2}-\Delta^{2})^{2}    \notag \\
&&  -(t_{1}'^{2}-t_{2}'^{2})^{2}\Delta^{2}-4t^{2}t_{2}'^{2}\Delta^{2}]=0.
\end{eqnarray}
For a general nonzero $\Delta$, the spectrum contains two flat bands at $\Delta$ and $-\Delta$ and six dispersive bands determined by a cubic function of $\lambda^{2}$. For $\Delta=0$, the eight bands collapse into three sets of flat bands at $0$ (fourfold degenerate), $\sqrt{6}t$ (twofold degenerate), and $-\sqrt{6}t$ (twofold degenerate).

\subsection{The minimal AC-in ribbon}

The minimal AC-in ribbon contains only three $y$-chains, as shown in Fig.2(e), which is the thinnest among the five minimal ribbons considered. This lattice is also known as the diamond lattice, diamond chain lattice, or rhombic chain lattice, and is widely studied in the literature as a prototypical 1D system with flat bands \cite{vidal00,doucot02,hyrkas13,tovmasyan18,cartwright18,tilleke20,huda20}. It is discussed in this work for the purposes of completeness and comparison. In the $k$ space and with the basis $(a_{1k}^{\dagger},b_{1k}^{\dagger},c_{1k}^{\dagger})$, the Hamiltonian is
\begin{equation}
h(k)=\begin{pmatrix} \Delta & t_{1}' & 0 \\ t_{1}' & 0 & t_{2}' \\ 0 & t_{2}' & -\Delta \\ \end{pmatrix}.
\end{equation}
$t_{i}'=2t\cos(\frac{ka}{2}+\gamma_{i})$ ($i=1,2$), $\gamma_{1}=\frac{\pi}{2}f$ and $\gamma_{2}=\frac{3\pi}{2}f$. The eigenequation for the three bands is
\begin{equation}
\text{det}[h(k)-\lambda I_{3}]=-\lambda^{3}+(t_{1}'^{2}+t_{2}'^{2}+\Delta^{2})\lambda+\Delta(t_{1}'^{2}-t_{2}'^{2})=0,
\end{equation}
where $I_{3}$ is the third-order unit matrix.

In zero magnetic field, $f=0$ and $t_{1}'=t_{2}'=2t\cos\frac{ka}{2}=t'$, the eigenequation becomes
\begin{equation}
\lambda(\lambda^{2}-2t'^{2}-\Delta^{2})=0.
\end{equation}
It gives a zero-energy flat band and a pair of dispersive bands. The three bands are isolated from each other for nonzero $\Delta$. At $\Delta=0$, the three bands touch at the BZ boundary $ka=\pi$. The two dispersive bands connect linearly and give a single Dirac cone.

In a magnetic field of $f=\frac{1}{2}$, $\gamma_{1}=\frac{\pi}{4}$ and $\gamma_{2}=\frac{3\pi}{4}$. The eigenequation becomes
\begin{equation}
\lambda^{3}-(4t^{2}+\Delta^{2})\lambda+4\Delta t^{2}\sin(ka)=0.
\end{equation}
When $\Delta=0$, the spectrum is fully pinched and contain three flat bands at $0$, $2t$, and $-2t$. When $\Delta\ne0$, all three bands become dispersive.

\end{appendix}






\end{document}